\documentclass[letterpaper,11pt]{article}
\usepackage{authblk}

\usepackage[margin=1in]{geometry}
\usepackage{setspace}
\usepackage{titlesec}
\usepackage{tocloft}
\titlespacing*{\section}{0pt}{2.0ex plus 1ex minus .2ex}{1.5ex plus .2ex}
\titlespacing*{\subsection}{0pt}{1.8ex plus 0.8ex minus .2ex}{1.3ex plus .2ex}
\usepackage[T1]{fontenc}
\usepackage{times}
\usepackage{xcolor}
\usepackage{hyperref}
\usepackage[centertags,fleqn]{amsmath}
\usepackage{amssymb,amsfonts,xspace}
\usepackage{booktabs}
\usepackage{multirow}
\usepackage{subfig}
\usepackage{tikz}
\usetikzlibrary{fit,matrix,shapes,petri}
\usetikzlibrary{arrows,positioning,automata}
\usepackage[style=authoryear,natbib=true,backend=bibtex,maxbibnames=15,maxcitenames=15]{biblatex}

\usepackage{amsthm}
\usepackage{mathtools}
\usepackage{thmtools}
\usepackage{thm-restate}
\usepackage{longtable}

\declaretheorem[name=Theorem,numberwithin=section]{theorem}
\declaretheorem[name=Lemma,sibling=theorem]{lemma}
\declaretheorem[name=Proposition,sibling=theorem]{proposition}
\declaretheorem[name=Corollary,sibling=theorem]{corollary}
\declaretheorem[name=Conjecture,sibling=theorem]{conjecture}

\newtheorem{observation}[theorem]{Observation}
\newtheorem{definition}[theorem]{Definition}
\newtheorem{example}[theorem]{Example}

\DeclarePairedDelimiter\parentheses{\lparen}{\rparen}
\DeclarePairedDelimiter\brackets{\lbrack}{\rbrack}
\DeclarePairedDelimiter\vbars{\lvert}{\rvert}
\DeclarePairedDelimiter\bracketparenthesis{\lbrack}{\rparen}

\newcommand{\comp}[1]{D\parentheses*{#1}}
\newcommand{\proto}{\mathcal{P}}
\newcommand{\icomp}{\circ}
\newcommand{\gcomp}[2]{#1 \icomp #2}
\newcommand{\goodOver}[2]{#1[#2]}
\newcommand{\basedecomp}[2]{\brackets{#1}_{#2}}
\newcommand{\extractmatrix}[3]{\varepsilon(#1,#2,#3)}
\newcommand{\floor}[1]{\left\lfloor#1\right\rfloor}
\newcommand{\ceil}[1]{\left\lceil#1\right\rceil}
\newcommand{\suchthat}{\;\ifnum\currentgrouptype=16 \middle\fi|\;}
\newcommand{\card}[1]{\vbars{#1}}
\newcommand{\rangeintegers}[1]{\bracketparenthesis{#1}}
\newcommand{\remainder}[2]{#1 \bmod #2}
\newcommand{\transpose}[1]{#1^{\intercal}}
\newcommand{\overOp}[2]{#1{\left\langle{#2}\right\rangle}}
\newcommand{\bracket}[4]{{#1}\brackets*{#2,#3,#4}}
\newcommand{\yval}[1]{y_{#1}}
\newcommand{\cstTopMultS}{B}
\newcommand{\symTotGames}{p}
\newcommand{\symGamePartSize}{\ell}
\newcommand{\symNumInstances}{\ell}
\newcommand{\indexOne}{\gamma}

\newcommand{\numberT}{T}

\title{Refuting the Direct Sum Conjecture for Total Functions in Deterministic Communication Complexity}
\author[1]{Simon Mackenzie}
\author[2]{Abdallah Saffidine}
\affil[1]{\texttt{simon.william.mackenzie@gmail.com}}
\affil[2]{ \texttt{abdallah.saffidine@gmail.com}}
\date{}

\begin{document}
\maketitle

\begin{abstract}
  In communication complexity the input of a function $f:X\times Y\rightarrow Z$ is distributed between two players Alice and Bob.
  If Alice knows only $x\in X$ and Bob only $y\in Y$, how much information must Alice and Bob share to be able to elicit the value of $f(x,y)$?

  Do we need $\symNumInstances$ more resources to solve $\symNumInstances$ instances of a problem?
  This question is the direct sum question and has been studied in many computational models.
  In this paper we focus on the case of 2-party deterministic communication complexity and give a counterexample to the direct sum conjecture in its strongest form.
  To do so we exhibit a family of functions for which the complexity of solving $\symNumInstances$ instances is less than $(1 -\epsilon )\symNumInstances$ times the complexity of solving one instance for some small enough $\epsilon>0$.

  We use a customised method in the analysis of our family of total functions, showing that one can force the alternation of rounds between players.

  This idea allows us to exploit the integrality of the complexity measure to create an increasing gap between the complexity of solving the instances independently with that of solving them together.
\end{abstract}

\newpage

\section{Introduction}

Communication complexity, introduced by \citet{yao1979some}, studies
computational problems in a distributed model where inputs are split between
two players or more. In this paradigm, two or more players, each privy to
separate pieces of information, collaborate to compute a function based on
their collective inputs. The measure of complexity is the amount of
information exchanged to achieve this goal, providing insights into the
minimum communication necessary.

While the basic model is simple, numerous variants arise by tweaking aspects
of the communication process, the nature of the function or relation under
consideration, and the resources allowed for the players. The functions
studied can range from total functions to partial relations, as well as
multi-party functions where the input is split in a certain way among more
than two players (\citet{rao2020communication}, \citet{Kushilevitz1997}).

Communication complexity also encompasses a variety of complexity measures and
computational models. Beyond deterministic protocols, one often considers
randomised or non-deterministic ones. The complexity measure used may also
vary from the standard communication complexity measure consisting of number
of bits exchanged, a prominent such example is information complexity
(\citet{braverman2012interactive}, \citet{chakrabarti2001informational}). In
this paper, we restrict our attention to the case of deterministic protocols,
total functions, 2 players, and the number of bits exchanged as a measure of
complexity.

One significant question within this field is that of the direct sum problem,
a topic that has garnered substantial interest due to its theoretical
implications. The issue revolves around whether solving $\symNumInstances$
instances of a problem requires $\symNumInstances$ times the resources of
solving a single instance. Specifically, we study whether computing a
function $f$ on $\symNumInstances$ instances can be easier than paying
$\symNumInstances$ times the cost of a single instance. We provide a
counterexample that challenges this conjecture in its most stringent form.

\subsection{Motivation}

Aside from being of independent interest, the computational model used in
communication complexity has proven to be highly applicable to other
settings. In particular, it has proven fruitful in proving results regarding
other computational models. This relevance stems from its capacity to
abstract essential features of computational models, particularly focusing on
the communication aspect within the model of study, which is often a
bottleneck. A remarkable application of communication complexity is in the
realm of circuit complexity, providing techniques to derive lower bounds
which have remained elusive for many years
(\citet{arora2009computational}). Moreover, the applications of the model
include many more computational models and problems, including streaming
algorithms, VLSI design, faster SAT algorithms, decision trees and turing
machines (\citet{roughgarden2016communication}, \citet{Kushilevitz1997}).

The direct sum question, aside from being of independent interest and
pervasive in computer science, was originally motivated in the context of
communication complexity by the work of \citet{karchmer1995super} connecting
it to circuit lower bounds. By building on previous work
(\citet{karchmer1988monotone}) characterising circuit depth in terms of the
communication complexity of certain relations, the authors showed that a
direct sum theorem on such relations would imply that $NC \subsetneq P$. In
the same paper they successfully apply the approach for monotone circuits.

There are different formulations of the direct sum conjecture, we will refute
the following found in \citet{AmbainisBGKT2000} and more recently in
\citet{hazan2017two}:
\begin{equation}
  \comp{f^{\symNumInstances}} = \symNumInstances(\comp{f} - O(1)).
  \label{eq:DS}
\end{equation}
There are many variants of the question of how multiple instances of a
function affect the complexity of the problem. In \citet{Kushilevitz1997},
they ask if $\comp{f,g}$ is smaller than $\comp{f}+\comp{g}$ for families of
functions $f$ and $g$. Other ways of framing the question might ask for a
weaker form of the conjecture. For example the problem is stated as
$\comp{f^{\symNumInstances}}\geq \Omega(\symNumInstances.\comp{f})$ in some
sources (\citet{braverman2022communication}).

However a counter-example to the conjecture in its strongest form has thus
far remained elusive and is of significance in understanding better the
original variant needed for $NC^1\neq NC^2$ which allows only for an additive
term (\citet{AmbainisBGKT2000}). A counterexample to the conjecture in this
form also provides the first example of a family of functions where the
number of bits saved actually grows as the functions gets larger.

The direct sum question is also of independent interest beyond the original
motivation linked to circuit lower bounds, as demonstrated by the considerable
amount of research on the various variants of the question
(\citet{meir2018direct}, \citet{nachum2018direct}, \citet{jain2008optimal},
\citet{molinaro2013beating}).

\subsection{Related Work}

Much work has been done on the variants of the direct sum question in
communication complexity. However, a comprehensive understanding of the
question is far from reached. The best lower bound found so far for the
deterministic case is
\[
  D(f^{\symNumInstances}) \geq
  \symNumInstances.\left(\sqrt{\frac{D(f)}{2}}-\log{n}-O(1)\right)
\]
(\citet{feder1995amortized}). This is a consequence of a stronger direct sum
result that they prove on non-deterministic protocols on relations, together
with a result from \citet{aho1983notions} which relates the nondeterministic
communication complexity of a function to its deterministic communication
complexity. In the same paper it is shown that for partial functions or
randomised protocols significant savings can be made, specifically that there
exist functions with $O(\log{n})$ complexity but constant amortised
complexity. Importantly they do not make progress on the case where the
function is \textbf{deterministic} and \textbf{total}, which is the case we
address in this paper. For the two-round deterministic total model,
\citet{karchmer1995fractional} showed that essentially
$\comp{f^{\symNumInstances}} = \symNumInstances \comp{f}$ for any $f$, giving
a strong direct sum theorem under those restrictions. In the context of
deterministic protocols on relations, a function where a small amount of
savings, namely an additive factor of $\log{n}$, is possible is described in
\citet{rao2020communication}.

More recently there has been much interest in a new kind of
information-theoretic technique used in the study of communication complexity
(\citet{chakrabarti2001informational}, \citet{jain2003direct},
\citet{bar2004information}). These have proven very useful in showing many
forms of direct-sum theorems. Information complexity might be a natural
complexity measure with which to study the direct sum question in
communication complexity (\citet{gavinsky2017toward},
\citet{braverman2012interactive}, \citet{braverman2022communication}).

The way we build our counter-example in this paper gives further weight to the
assertion in \citet{braverman2012interactive} that the integrality of the
communication complexity measure causes it to lose some nice properties that
may hold for its continuous relaxation.

\subsection{Overview of the Paper}\label{section:overview}

\paragraph{Organisation.} We exhibit a family of \textbf{total} functions for
which multiple instances can be solved with a small saving over treating the
instances independently. Section~2 introduces the background and formal
statement of the problem. Section~3 presents the construction and the
framework used to analyse it. Section~4 proves the lower bound, deferring the
detailed proof of the iterated partition argument to
Appendix~\ref{appendix:lower-bound-details}. The main idea is that by forcing
the interaction between the players we can make an arbitrary number of bits be
used wastefully. This exploits the integrality of the complexity measure and
shows how it complicates clean direct-sum statements. Section~5 proves the
corresponding upper bound when $\symNumInstances\geq 178$, combines the two
bounds to refute Equation~\eqref{eq:DS}, and then derives a stronger
quantitative consequence.

\paragraph{Discussion.}

The central contribution of the paper is the refutation of the direct sum
conjecture in deterministic communication complexity. However, we would like
to highlight that we use a novel method to establish a lower bound on
communication complexity for our specific construction, and we believe that
the combination of the \nameref{def:overOperation} with the accompanying
analysis could be refined to derive further results. The
\nameref{def:overOperation} is a natural construction which, combined with the
framework for analysing it developed in
Sections \ref{section:CoreLemmas} and \ref{section:LB}, can be used to build
functions for which desired behaviour on optimal protocols can be forced. The
operation and analysis given in this paper is specialised to the problem at
hand, but the core concept extends beyond this setting and variants adapted to
different situations can be defined. One particular example of this would be
relaxing the notion of independence in the \nameref{def:overOperation} so as
to obtain a polynomial size construction.

Aside from applying the idea to new problems, a more straightforward
extension of the work presented here would be to make progress on Open Problem
4.1 in \citet{KushilevitzN1997}, which asks how $\comp{f,g}$ relates to
$\comp{f}+\comp{g}$. Specifically, for the family of functions $f$ presented
in this paper, there should exist a non-trivial family of functions $g$ such
that $\comp{f^{177},g}=\comp{f^{177}}$.

A more complex endeavour would be to improve the constant factor of
communication saved. For example, if one could improve the proof of the lower
bound so that it would hold for $B=3$, the constant obtained would be under
$0.8$, meaning that more than $\frac{1}{5}$ of the communication could be
saved. It might also be possible to adapt the construction to go beyond
constant-factor savings.

These considerations suggest interesting further work related to the paper,
extending and refining both the result and the technique.
\section{Two-Party Communication and the Direct Sum Conjecture}

In this section we will introduce the concepts we will be using to discuss
the problem. Importantly, we will interchangeably use the concepts of
communication game, two-argument boolean function, and boolean matrix, using
whichever formalism seems more natural in the context.

\subsection{Two-Party Communication Complexity}\label{subsec:commModel}

\begin{definition}\label{definition:Protocol}
  A \emph{protocol} $\proto$ over domain $X \times Y$ with range $Z$ is a
  binary tree where each internal node $v$ is labelled either by a function
  $a_v: X \rightarrow \{0, 1\}$ or by a function $b_v : Y \rightarrow \{0, 1\}$,
  and each leaf is labelled with an element $z \in Z$. The \emph{value} of the
  protocol $\proto$ on input $(x, y)$ is the label of the leaf reached by
  starting from the root, and walking on the tree. At each internal node $v$
  labelled by $a_v$, we walk left if $a_v(x) = 0$ and right if $a_v(x) = 1$,
  and at each internal node labelled by $b_v$, we walk left if $b_v(y) = 0$
  and right if $b_v(y) = 1$. The \emph{cost} of the protocol $\proto$ is the
  depth of the tree.
\end{definition}

\begin{definition}\label{definition:PartialProtocol}
  We can extend the definition of a protocol to \emph{partial protocols} by
  allowing the value of the leaves to be $X_v\times Y_v$ where $X_v\subseteq X$
  and $Y_v \subseteq Y$. We require that the union of $X_v$ over all leaves
  $v$ equals $X$. Likewise for $Y$. Finally, for two leaves $v$ and $v'$,
  $X_v \cap X_v'=\emptyset$ and likewise for $Y$.
\end{definition}

For a function $f : X \times Y \rightarrow Z$, we say that $\proto$ computes
$f$ if the value of $\proto$ on each input $(x, y)$ is $f(x,y)$.
Alternatively we will use the term `communication complexity of a game'. In
this case we are referring to the communication game played on the matrix
representing the function of interest. We define $\proto_v$ to be the protocol
over domain $R_v$ corresponding to the binary tree rooted at $v$, and $f_v$ to
be the corresponding function.

\begin{definition}
  We define the \emph{matrix of a game} by the boolean function corresponding
  to the function $f : X \times Y \rightarrow Z$ that the two agents are
  communicating.
\end{definition}

Abusing the notation, we will sometimes use the matrix $M$ of a game as an
argument to the function $D(M)$ to denote the communication complexity of the
function represented by the matrix. Furthermore we will often use matrix,
function and game interchangeably throughout the paper.

\begin{definition}\label{def:CC}
  For a function $f:X\times Y \rightarrow \{0,1\}$ we denote by $\comp{f}$ the
  \emph{communication complexity} of $f$, which consists of the minimum depth
  over all protocols computing $f$.

  We extend this definition to sets of functions. For a set of functions
  $\Phi$, we denote by $\comp{\Phi}$ the \emph{communication complexity} of
  $\Phi$ and define it as
  \[
    \comp{\Phi}=\min_{f \in \Phi} \comp{f}.
  \]
\end{definition}

The following observation is important to keep in mind for our proof.

\begin{proposition}[Complexity Invariant to Transposition]
  For any matrix $M$, the complexity of its transposed matrix is the same:
  $\comp{M^*} = \comp{M}$.
\end{proposition}

\subsection{Direct Sum in Deterministic Communication Complexity}

Direct sum theorems are pervasive in computer science. The general form of
such statements is that one cannot save a significant amount of resources when
jointly resolving several instances of a given problem rather than addressing
the instances separately.

The direct sum conjecture, as stated formally by \citet{AmbainisBGKT2000}, is

\begin{conjecture}[Direct Sum Conjecture]
  Let $(f_n)$ be a family of functions where for all $n$, we have
  $f_n: \{0, 1\}^n \times \{0, 1\}^n \rightarrow \{0, 1\}$; then
  $\comp{f^\symNumInstances} = \symNumInstances(\comp{f} - O(1))$.
  Formally,
  \[
    \exists N, L, C \in \mathbb{N}, \forall n \geq N,
    \forall \symNumInstances \geq L,
    \comp{f_n^\symNumInstances} \geq \symNumInstances(\comp{f_n} - C).
  \]
\end{conjecture}

The result of this paper is Theorem~\ref{th:main}: we disprove the above form
of the direct sum conjecture. We do so by exhibiting a family of functions for
which we can prove that parallel savings can be made to grow as a function of
$n$.

\begin{theorem}\label{th:main}
  There exists a family $(f_n)$ of total functions satisfying
  \[
    \comp{f} = \frac{\comp{f^\symNumInstances}}{\symNumInstances} + \omega(1).
  \]
  Formally,
  \[
    \forall N, L, C \in \mathbb{N}, \exists n \geq N,
    \exists \symNumInstances \geq L,
    \comp{f_n} > \frac{\comp{f_n^\symNumInstances}}{\symNumInstances}+ C.
  \]
\end{theorem}

The family of functions exhibited also yields a further quantitative
comparison which will be developed later in the upper-bound section.
\section{The Counterexample}\label{section:CoreLemmas}

\subsection{Preliminary Definitions}\label{subsec:PreliminaryDefinitions}

A central concept for our construction is the \nameref{def:overOperation}.
Conceptually, this is a binary operation which takes two functions and returns
a new function built in a way that the row player needs to separate the output
function into copies of the argument functions before any communication from
the column player can take place. The way this is achieved is by concatenating
the two matrices corresponding to the functions along the rows, duplicating the
columns and maximally decorrelating the columns of the second function from
those of the first function. Therefore, in our resulting matrix, we end up
with a column built from the concatenation of each possible combination of
columns of the two matrices. We define the operation formally as such:

\begin{definition}[Binary Interlacing Operation on Functions]\label{def:OverBinary}
  Let $f: X_1 \times Y_1 \rightarrow \{0, 1\}$ and
  $g: X_2 \times Y_2 \rightarrow \{0, 1\}$ be functions. Define the sets
  $X' = X_1 \sqcup X_2$ and $Y' = Y_1 \times Y_2$. We then define the function
  \[
    f\circ g : X' \times Y' \rightarrow \{0, 1\}
  \]
  as follows:
  \begin{align*}
    \text{For } x \in X' \text{ with } x = (i, j), \text{ and } y \in Y'
    \text{ with } y = (y_1, y_2), \quad
    f\circ g(x, y)
    =
    \begin{cases}
      f(j,y_1) & \text{if } i = 1, \\
      g(j,y_2) & \text{if } i =2.
    \end{cases}
  \end{align*}
\end{definition}

We will not be using the \nameref{def:overOperation} to combine different
functions, so it is more convenient to define the power with respect to the
\nameref{def:overOperation} and work with that.

\begin{definition}[Interlacing Operation on Function]\label{def:funcOverOp}
  Let $f: X \times Y \rightarrow \{0, 1\}$ be a function. Define the sets
  $X' = \bigsqcup_{i \in [k]} X$ and $Y' = Y^k$. We then define the function
  \[
    \overOp{f}{k} : X' \times Y' \rightarrow \{0, 1\}
  \]
  as follows:
  \begin{align*}
    \text{For } x \in X' \text{ with } x = (i, j), \text{ and } y \in Y'
    \text{ with } y = (y_1, \ldots, y_k), \quad
    \goodOver{f}{k}(x, y) = f(j, y_i).
  \end{align*}
\end{definition}

An alternative definition of the same operation framed for matrices is also
given. We use this definition in Section~\ref{section:CoreLemmas} where we
found working with matrices more convenient.

\begin{definition}[Interlacing Operation]\label{def:overOperation}
  Let $A = (a_{i,j})$ be an $m \times n$ matrix and $\symTotGames$ a positive
  integer.

  We define $\overOp{A}{\symTotGames}$ as the
  $m\symTotGames \times n^\symTotGames$ matrix $(b_{i,j})$ such that for all
  $i \in \rangeintegers{m\symTotGames}\footnote{In this paper we start matrix
  indices at $0$.}$ and $j \in \rangeintegers{n^\symTotGames}$ we have
  $b_{i,j} = a_{i',j'}$ where $i' = \remainder{i}{m}$ and
  $j' = \remainder{\floor{\frac{j}{n^{\indexOne}}}}{n}$ for
  $\indexOne = \floor{\frac{i}{m}}$. The value $\indexOne$ is called the
  \emph{component}.
\end{definition}

\begin{example}
  \[
    \gcomp{A}{B} =
    \left[\begin{array}{@{}c@{}}A \otimes X_{n_b} \\ X_{n_a} \otimes B\end{array}\right]
    =
    \left[\begin{array}{@{}*{11}{c@{}}}
      a_{1,1} & a_{1,1} & \dots & a_{1,1} & a_{1,2} & \dots & a_{1,2} & \dots & a_{1,n_a} & \dots & a_{1,n_a} \\
      a_{2,1} & a_{2,1} & \dots & a_{2,1} & a_{2,2} & \dots & a_{2,2} & \dots & a_{2,n_a} & \dots & a_{2,n_a} \\
      \vdots & \vdots & \ddots & \vdots & \vdots & \ddots & \vdots & \ddots & \vdots& \ddots & \vdots\\
      a_{m_a,1} & a_{m_a,1} & \dots & a_{m_a,1} & a_{m_a,2} & \dots & a_{m_a,2} & \dots & a_{m_a,n_a} & \dots & a_{m_a,n_a}\\
      b_{1,1} & b_{1,2} & \dots & b_{1,n_b} & b_{1,1} & \dots & b_{1,n_b} & \dots & b_{1,1} & \dots & b_{1,n_b} \\
      b_{2,1} & b_{2,2} & \dots & b_{2,n_b} & b_{2,1} & \dots & b_{2,n_b} & \dots & b_{2,1} & \dots & b_{2,n_b} \\
      \vdots & \vdots & \ddots & \vdots & \vdots & \ddots & \vdots & \ddots & \vdots & \ddots & \vdots\\
      b_{m_b,1} & b_{m_b,2} & \dots & b_{m_b,n_b} & b_{m_b,1} & \dots & b_{m_b,n_b} & \dots & b_{m_b,1} & \dots & b_{m_b,n_b}
    \end{array}\right]
  \]
\end{example}

\begin{example}
  Let $A=\{\begin{bmatrix}1 & 0\end{bmatrix}\}$.
  The matrix $\overOp{A}{2}$ is
  $\{\gcomp{\begin{bmatrix}1 & 0\end{bmatrix}}{\begin{bmatrix}1 & 0\end{bmatrix}}\}$
  and can be seen as:
  \begin{align*}
    \overOp{A}{2}
    &=
    \left\{\left[\begin{array}{*{4}{c}}
      1 & 1 & 0 & 0\\
      1 & 0 & 1 & 0
    \end{array}\right]\right\}
    &&
    \overOp{A}{3}
    =
    \left\{\left[\begin{array}{*{16}{c}}
      1 & 1 & 1 & 1 & 0 & 0 & 0 & 0 \\
      1 & 1 & 0 & 0 & 1 & 1 & 0 & 0 \\
      1 & 0 & 1 & 0 & 1 & 0 & 1 & 0
    \end{array}\right]\right\}
  \end{align*}
  \begin{align*}
    \overOp{A}{4}
    &=
    \left\{\left[\begin{array}{*{16}{c}}
      1 & 1 & 1 & 1 & 1 & 1 & 1 & 1 & 0 & 0 & 0 & 0 & 0 & 0 & 0 & 0 \\
      1 & 1 & 1 & 1 & 0 & 0 & 0 & 0 & 1 & 1 & 1 & 1 & 0 & 0 & 0 & 0 \\
      1 & 1 & 0 & 0 & 1 & 1 & 0 & 0 & 1 & 1 & 0 & 0 & 1 & 1 & 0 & 0 \\
      1 & 0 & 1 & 0 & 1 & 0 & 1 & 0 & 1 & 0 & 1 & 0 & 1 & 0 & 1 & 0
    \end{array}\right]\right\}
  \end{align*}
\end{example}

\quad

As already discussed in Section \ref{section:overview}, we believe the idea of
composing games in such a way is a canonical construction in communication
complexity and that such constructions can be adapted to prove further results.
The operation, along with the analysis developed in Sections
\ref{section:CoreLemmas} and \ref{section:LB}, offer a method to build
examples of functions where one can prove that the optimal protocols must
behave in a prescribed way. Though what is presented in this paper is a
specific adaptation of the technique to the direct sum problem, we strongly
believe that variants of the operator and analysis can be leveraged to derive
other results of interest.

Another crucial notion we will use is the idea of subgames and supergames in
Definition \ref{def:Subgames}. The notion of subgame is useful because if $A$
is a subgame of $B$ then a protocol resolving $B$ also resolves $A$. The
notion therefore allows us to lower bound the communication complexity of $B$
by that of $A$. Conceptually, $A$ is a subgame of $B$ if a copy of $A$ is
present in $B$. One way to formalise this is:

\begin{definition}\label{def:Subgames}
  We say $\phi'$ is a \emph{subgame} of $\phi$, denoted $\phi'\sqsubseteq \phi$,
  if we can select a number of rows and columns from the matrix representing
  $\phi$ and rearrange their intersection into $\phi'$. A \emph{supergame} is
  the dual notion of a subgame.
\end{definition}

We can extend the subgame relation to sets of games. Conceptually a set of
games $A$ is a subgame of a set of games $B$ if all games in $B$ have a
subgame in $A$. This concept is used heavily in our lower bound proof, where
we keep track of sets of subgames with desirable properties.

\begin{definition}\label{def:SetSubgames}
  We say that a set $\Phi'$ is a subgame of set $\Phi$ if all matrices in
  $\Phi$ have a subgame in $\Phi'$. That is, $\Phi' \sqsubseteq \Phi$ iff
  $\forall \phi \in \Phi, \exists \phi' \in \Phi', \phi' \sqsubseteq \phi$.
\end{definition}

\begin{proposition}[Subgames Are Easier]\label{prop:subgames}
  If set $\Phi'$ and set $\Phi$ satisfy $\Phi'\sqsubseteq \Phi$ then
  $\comp{\Phi'}\leq \comp{\Phi}$.
\end{proposition}
\begin{proof}
  Since the complexity of a set of games is defined to be the minimum
  complexity amongst all games in the set and since a set of games $\Phi'$ is
  a subgame of a set of games $\Phi$ if for any game $t$ in $\Phi$ there is a
  game $s$ in $\Phi'$ such that $s\sqsubseteq t$, then taking the game $t^{*}$
  of minimum complexity in $\Phi$ we can find $s^{*}$ in $\Phi'$ such that
  $s^{*} \sqsubseteq t^{*}$. Since $s^{*}$ is a restriction of $t^{*}$ on a
  combinatorial rectangle, we have $\comp{s^{*}}\leq \comp{t^{*}}$, implying
  $\comp{\Phi'}\leq \comp{\Phi}$.
\end{proof}

\subsection{The Construction}\label{subsec:construction}

\begin{figure}
  \begin{center}
    \begin{tabular}[b]{l l l l}
      \begin{tabular}{l}
        \subfloat[$\overOp{\phi_0}{1}$]{  \begin{tikzpicture}[yscale=1,xscale=1,rotate=0]
 \begin{scope}[xscale=1,yscale=1,shift={(0,0)},rotate=0]
 \begin{scope}[opacity=1]
 \draw[fill=gray] (0,0) rectangle (1,1);
 \draw[fill=white] (1,0) rectangle (2,1);
 \end{scope}
 \begin{scope}[opacity=0]
 \draw[fill=gray] (0,1) rectangle (0.5,2);
 \draw[fill=white] (0.5,1) rectangle (1,2);
 \draw[fill=gray] (1,1) rectangle (1.5,2);
 \draw[fill=white] (1.5,1) rectangle (2,2);
 \end{scope}
 \begin{scope}[opacity=0]
 \begin{scope}[shift={(0,1)},xscale=0.5]
 \draw[fill=gray] (0,1) rectangle (0.5,2);
 \draw[fill=white] (0.5,1) rectangle (1,2);
 \draw[fill=gray] (1,1) rectangle (1.5,2);
 \draw[fill=white] (1.5,1) rectangle (2,2);
 \end{scope}
 \begin{scope}[shift={(1,1)},xscale=0.5]
 \draw[fill=gray] (0,1) rectangle (0.5,2);
 \draw[fill=white] (0.5,1) rectangle (1,2);
 \draw[fill=gray] (1,1) rectangle (1.5,2);
 \draw[fill=white] (1.5,1) rectangle (2,2);
 \end{scope}
 \end{scope}
 \end{scope}
 \end{tikzpicture}}
        \\
        \subfloat[$\overOp{\phi_0}{2}$]{  \begin{tikzpicture}[yscale=1,xscale=1,rotate=0]
 \begin{scope}[xscale=1,yscale=1,shift={(0,0)},rotate=0]
 \begin{scope}[opacity=1]
 \draw[fill=gray] (0,0) rectangle (1,1);
 \draw[fill=white] (1,0) rectangle (2,1);
 \end{scope}
 \begin{scope}[opacity=1]
 \draw[fill=gray] (0,1) rectangle (0.5,2);
 \draw[fill=white] (0.5,1) rectangle (1,2);
 \draw[fill=gray] (1,1) rectangle (1.5,2);
 \draw[fill=white] (1.5,1) rectangle (2,2);
 \end{scope}
 \begin{scope}[opacity=0]
 \begin{scope}[shift={(0,1)},xscale=0.5]
 \draw[fill=gray] (0,1) rectangle (0.5,2);
 \draw[fill=white] (0.5,1) rectangle (1,2);
 \draw[fill=gray] (1,1) rectangle (1.5,2);
 \draw[fill=white] (1.5,1) rectangle (2,2);
 \end{scope}
 \begin{scope}[shift={(1,1)},xscale=0.5]
 \draw[fill=gray] (0,1) rectangle (0.5,2);
 \draw[fill=white] (0.5,1) rectangle (1,2);
 \draw[fill=gray] (1,1) rectangle (1.5,2);
 \draw[fill=white] (1.5,1) rectangle (2,2);
 \end{scope}
 \end{scope}
 \end{scope}
 \end{tikzpicture} }
        \\
        \subfloat[$\overOp{\phi_0}{3}$]{  \begin{tikzpicture}[yscale=1,xscale=1,rotate=0]
 \begin{scope}[xscale=1,yscale=1,shift={(0,0)},rotate=0]
 \begin{scope}[opacity=1]
 \draw[fill=gray] (0,0) rectangle (1,1);
 \draw[fill=white] (1,0) rectangle (2,1);
 \end{scope}
 \begin{scope}[opacity=1]
 \draw[fill=gray] (0,1) rectangle (0.5,2);
 \draw[fill=white] (0.5,1) rectangle (1,2);
 \draw[fill=gray] (1,1) rectangle (1.5,2);
 \draw[fill=white] (1.5,1) rectangle (2,2);
 \end{scope}
 \begin{scope}[opacity=1]
 \begin{scope}[shift={(0,1)},xscale=0.5]
 \draw[fill=gray] (0,1) rectangle (0.5,2);
 \draw[fill=white] (0.5,1) rectangle (1,2);
 \draw[fill=gray] (1,1) rectangle (1.5,2);
 \draw[fill=white] (1.5,1) rectangle (2,2);
 \end{scope}
 \begin{scope}[shift={(1,1)},xscale=0.5]
 \draw[fill=gray] (0,1) rectangle (0.5,2);
 \draw[fill=white] (0.5,1) rectangle (1,2);
 \draw[fill=gray] (1,1) rectangle (1.5,2);
 \draw[fill=white] (1.5,1) rectangle (2,2);
 \end{scope}
 \end{scope}
 \end{scope}
 \end{tikzpicture} }
      \end{tabular}
      &
      \begin{tabular}{l}
        \subfloat[$\overOp{\phi_1}{1}$]{  \begin{tikzpicture}[yscale=1,xscale=1,rotate=90]
\begin{scope}
 \begin{scope}[xscale=1,yscale=1,shift={(0,0)},rotate=0]
 \begin{scope}[opacity=1]
 \draw[fill=gray] (0,0) rectangle (1,1);
 \draw[fill=white] (1,0) rectangle (2,1);
 \end{scope}
 \begin{scope}[opacity=1]
 \draw[fill=gray] (0,1) rectangle (0.5,2);
 \draw[fill=white] (0.5,1) rectangle (1,2);
 \draw[fill=gray] (1,1) rectangle (1.5,2);
 \draw[fill=white] (1.5,1) rectangle (2,2);
 \end{scope}
 \begin{scope}[opacity=1]
 \begin{scope}[shift={(0,1)},xscale=0.5]
 \draw[fill=gray] (0,1) rectangle (0.5,2);
 \draw[fill=white] (0.5,1) rectangle (1,2);
 \draw[fill=gray] (1,1) rectangle (1.5,2);
 \draw[fill=white] (1.5,1) rectangle (2,2);
 \end{scope}
 \begin{scope}[shift={(1,1)},xscale=0.5]
 \draw[fill=gray] (0,1) rectangle (0.5,2);
 \draw[fill=white] (0.5,1) rectangle (1,2);
 \draw[fill=gray] (1,1) rectangle (1.5,2);
 \draw[fill=white] (1.5,1) rectangle (2,2);
 \end{scope}
 \end{scope}
 \end{scope}
 \begin{scope}[opacity=0,shift={(0,0)}]
 \begin{scope}[xscale=1,yscale=1/3,shift={(2,0)},rotate=0]
 \begin{scope}
 \draw[fill=gray] (0,0) rectangle (1,1);
 \draw[fill=white] (1,0) rectangle (2,1);
 \end{scope}
 \begin{scope}
 \draw[fill=gray] (0,1) rectangle (0.5,2);
 \draw[fill=white] (0.5,1) rectangle (1,2);
 \draw[fill=gray] (1,1) rectangle (1.5,2);
 \draw[fill=white] (1.5,1) rectangle (2,2);
 \end{scope}
 \begin{scope}
 \begin{scope}[shift={(0,1)},xscale=0.5]
 \draw[fill=gray] (0,1) rectangle (0.5,2);
 \draw[fill=white] (0.5,1) rectangle (1,2);
 \draw[fill=gray] (1,1) rectangle (1.5,2);
 \draw[fill=white] (1.5,1) rectangle (2,2);
 \end{scope}
 \begin{scope}[shift={(1,1)},xscale=0.5]
 \draw[fill=gray] (0,1) rectangle (0.5,2);
 \draw[fill=white] (0.5,1) rectangle (1,2);
 \draw[fill=gray] (1,1) rectangle (1.5,2);
 \draw[fill=white] (1.5,1) rectangle (2,2);
 \end{scope}
 \end{scope}
 \end{scope}
\begin{scope}[xscale=1,yscale=1/3,shift={(2,3)},rotate=0]
 \begin{scope}
 \draw[fill=gray] (0,0) rectangle (1,1);
 \draw[fill=white] (1,0) rectangle (2,1);
 \end{scope}
 \begin{scope}
 \draw[fill=gray] (0,1) rectangle (0.5,2);
 \draw[fill=white] (0.5,1) rectangle (1,2);
 \draw[fill=gray] (1,1) rectangle (1.5,2);
 \draw[fill=white] (1.5,1) rectangle (2,2);
 \end{scope}
 \begin{scope}
 \begin{scope}[shift={(0,1)},xscale=0.5]
 \draw[fill=gray] (0,1) rectangle (0.5,2);
 \draw[fill=white] (0.5,1) rectangle (1,2);
 \draw[fill=gray] (1,1) rectangle (1.5,2);
 \draw[fill=white] (1.5,1) rectangle (2,2);
 \end{scope}
 \begin{scope}[shift={(1,1)},xscale=0.5]
 \draw[fill=gray] (0,1) rectangle (0.5,2);
 \draw[fill=white] (0.5,1) rectangle (1,2);
 \draw[fill=gray] (1,1) rectangle (1.5,2);
 \draw[fill=white] (1.5,1) rectangle (2,2);
 \end{scope}
 \end{scope}
 \end{scope}
 \begin{scope}[xscale=1,yscale=1/3,shift={(2,6)},rotate=0]
 \begin{scope}
 \draw[fill=gray] (0,0) rectangle (1,1);
 \draw[fill=white] (1,0) rectangle (2,1);
 \end{scope}
 \begin{scope}
 \draw[fill=gray] (0,1) rectangle (0.5,2);
 \draw[fill=white] (0.5,1) rectangle (1,2);
 \draw[fill=gray] (1,1) rectangle (1.5,2);
 \draw[fill=white] (1.5,1) rectangle (2,2);
 \end{scope}
 \begin{scope}
 \begin{scope}[shift={(0,1)},xscale=0.5]
 \draw[fill=gray] (0,1) rectangle (0.5,2);
 \draw[fill=white] (0.5,1) rectangle (1,2);
 \draw[fill=gray] (1,1) rectangle (1.5,2);
 \draw[fill=white] (1.5,1) rectangle (2,2);
 \end{scope}
 \begin{scope}[shift={(1,1)},xscale=0.5]
 \draw[fill=gray] (0,1) rectangle (0.5,2);
 \draw[fill=white] (0.5,1) rectangle (1,2);
 \draw[fill=gray] (1,1) rectangle (1.5,2);
 \draw[fill=white] (1.5,1) rectangle (2,2);
 \end{scope}
 \end{scope}
 \end{scope}
 \end{scope}
 \begin{scope}[opacity=0]
 \begin{scope}[shift={(2,0)},yscale=1/3]
 \begin{scope}[xscale=1,yscale=1/3,shift={(2,0)},rotate=0]
 \begin{scope}
 \draw[fill=gray] (0,0) rectangle (1,1);
 \draw[fill=white] (1,0) rectangle (2,1);
 \end{scope}
 \begin{scope}
 \draw[fill=gray] (0,1) rectangle (0.5,2);
 \draw[fill=white] (0.5,1) rectangle (1,2);
 \draw[fill=gray] (1,1) rectangle (1.5,2);
 \draw[fill=white] (1.5,1) rectangle (2,2);
 \end{scope}
 \begin{scope}
 \begin{scope}[shift={(0,1)},xscale=0.5]
 \draw[fill=gray] (0,1) rectangle (0.5,2);
 \draw[fill=white] (0.5,1) rectangle (1,2);
 \draw[fill=gray] (1,1) rectangle (1.5,2);
 \draw[fill=white] (1.5,1) rectangle (2,2);
 \end{scope}
 \begin{scope}[shift={(1,1)},xscale=0.5]
 \draw[fill=gray] (0,1) rectangle (0.5,2);
 \draw[fill=white] (0.5,1) rectangle (1,2);
 \draw[fill=gray] (1,1) rectangle (1.5,2);
 \draw[fill=white] (1.5,1) rectangle (2,2);
 \end{scope}
 \end{scope}
 \end{scope}
\begin{scope}[xscale=1,yscale=1/3,shift={(2,3)},rotate=0]
 \begin{scope}
 \draw[fill=gray] (0,0) rectangle (1,1);
 \draw[fill=white] (1,0) rectangle (2,1);
 \end{scope}
 \begin{scope}
 \draw[fill=gray] (0,1) rectangle (0.5,2);
 \draw[fill=white] (0.5,1) rectangle (1,2);
 \draw[fill=gray] (1,1) rectangle (1.5,2);
 \draw[fill=white] (1.5,1) rectangle (2,2);
 \end{scope}
 \begin{scope}
 \begin{scope}[shift={(0,1)},xscale=0.5]
 \draw[fill=gray] (0,1) rectangle (0.5,2);
 \draw[fill=white] (0.5,1) rectangle (1,2);
 \draw[fill=gray] (1,1) rectangle (1.5,2);
 \draw[fill=white] (1.5,1) rectangle (2,2);
 \end{scope}
 \begin{scope}[shift={(1,1)},xscale=0.5]
 \draw[fill=gray] (0,1) rectangle (0.5,2);
 \draw[fill=white] (0.5,1) rectangle (1,2);
 \draw[fill=gray] (1,1) rectangle (1.5,2);
 \draw[fill=white] (1.5,1) rectangle (2,2);
 \end{scope}
 \end{scope}
 \end{scope}
 \begin{scope}[xscale=1,yscale=1/3,shift={(2,6)},rotate=0]
 \begin{scope}
 \draw[fill=gray] (0,0) rectangle (1,1);
 \draw[fill=white] (1,0) rectangle (2,1);
 \end{scope}
 \begin{scope}
 \draw[fill=gray] (0,1) rectangle (0.5,2);
 \draw[fill=white] (0.5,1) rectangle (1,2);
 \draw[fill=gray] (1,1) rectangle (1.5,2);
 \draw[fill=white] (1.5,1) rectangle (2,2);
 \end{scope}
 \begin{scope}
 \begin{scope}[shift={(0,1)},xscale=0.5]
 \draw[fill=gray] (0,1) rectangle (0.5,2);
 \draw[fill=white] (0.5,1) rectangle (1,2);
 \draw[fill=gray] (1,1) rectangle (1.5,2);
 \draw[fill=white] (1.5,1) rectangle (2,2);
 \end{scope}
 \begin{scope}[shift={(1,1)},xscale=0.5]
 \draw[fill=gray] (0,1) rectangle (0.5,2);
 \draw[fill=white] (0.5,1) rectangle (1,2);
 \draw[fill=gray] (1,1) rectangle (1.5,2);
 \draw[fill=white] (1.5,1) rectangle (2,2);
 \end{scope}
 \end{scope}
 \end{scope}
 \end{scope}
  \begin{scope}[shift={(2,1)},yscale=1/3]
 \begin{scope}[xscale=1,yscale=1/3,shift={(2,0)},rotate=0]
 \begin{scope}
 \draw[fill=gray] (0,0) rectangle (1,1);
 \draw[fill=white] (1,0) rectangle (2,1);
 \end{scope}
 \begin{scope}
 \draw[fill=gray] (0,1) rectangle (0.5,2);
 \draw[fill=white] (0.5,1) rectangle (1,2);
 \draw[fill=gray] (1,1) rectangle (1.5,2);
 \draw[fill=white] (1.5,1) rectangle (2,2);
 \end{scope}
 \begin{scope}
 \begin{scope}[shift={(0,1)},xscale=0.5]
 \draw[fill=gray] (0,1) rectangle (0.5,2);
 \draw[fill=white] (0.5,1) rectangle (1,2);
 \draw[fill=gray] (1,1) rectangle (1.5,2);
 \draw[fill=white] (1.5,1) rectangle (2,2);
 \end{scope}
 \begin{scope}[shift={(1,1)},xscale=0.5]
 \draw[fill=gray] (0,1) rectangle (0.5,2);
 \draw[fill=white] (0.5,1) rectangle (1,2);
 \draw[fill=gray] (1,1) rectangle (1.5,2);
 \draw[fill=white] (1.5,1) rectangle (2,2);
 \end{scope}
 \end{scope}
 \end{scope}
\begin{scope}[xscale=1,yscale=1/3,shift={(2,3)},rotate=0]
 \begin{scope}
 \draw[fill=gray] (0,0) rectangle (1,1);
 \draw[fill=white] (1,0) rectangle (2,1);
 \end{scope}
 \begin{scope}
 \draw[fill=gray] (0,1) rectangle (0.5,2);
 \draw[fill=white] (0.5,1) rectangle (1,2);
 \draw[fill=gray] (1,1) rectangle (1.5,2);
 \draw[fill=white] (1.5,1) rectangle (2,2);
 \end{scope}
 \begin{scope}
 \begin{scope}[shift={(0,1)},xscale=0.5]
 \draw[fill=gray] (0,1) rectangle (0.5,2);
 \draw[fill=white] (0.5,1) rectangle (1,2);
 \draw[fill=gray] (1,1) rectangle (1.5,2);
 \draw[fill=white] (1.5,1) rectangle (2,2);
 \end{scope}
 \begin{scope}[shift={(1,1)},xscale=0.5]
 \draw[fill=gray] (0,1) rectangle (0.5,2);
 \draw[fill=white] (0.5,1) rectangle (1,2);
 \draw[fill=gray] (1,1) rectangle (1.5,2);
 \draw[fill=white] (1.5,1) rectangle (2,2);
 \end{scope}
 \end{scope}
 \end{scope}
 \begin{scope}[xscale=1,yscale=1/3,shift={(2,6)},rotate=0]
 \begin{scope}
 \draw[fill=gray] (0,0) rectangle (1,1);
 \draw[fill=white] (1,0) rectangle (2,1);
 \end{scope}
 \begin{scope}
 \draw[fill=gray] (0,1) rectangle (0.5,2);
 \draw[fill=white] (0.5,1) rectangle (1,2);
 \draw[fill=gray] (1,1) rectangle (1.5,2);
 \draw[fill=white] (1.5,1) rectangle (2,2);
 \end{scope}
 \begin{scope}
 \begin{scope}[shift={(0,1)},xscale=0.5]
 \draw[fill=gray] (0,1) rectangle (0.5,2);
 \draw[fill=white] (0.5,1) rectangle (1,2);
 \draw[fill=gray] (1,1) rectangle (1.5,2);
 \draw[fill=white] (1.5,1) rectangle (2,2);
 \end{scope}
 \begin{scope}[shift={(1,1)},xscale=0.5]
 \draw[fill=gray] (0,1) rectangle (0.5,2);
 \draw[fill=white] (0.5,1) rectangle (1,2);
 \draw[fill=gray] (1,1) rectangle (1.5,2);
 \draw[fill=white] (1.5,1) rectangle (2,2);
 \end{scope}
 \end{scope}
 \end{scope}
 \end{scope}
  \begin{scope}[shift={(2,2)},yscale=1/3]
 \begin{scope}[xscale=1,yscale=1/3,shift={(2,0)},rotate=0]
 \begin{scope}
 \draw[fill=gray] (0,0) rectangle (1,1);
 \draw[fill=white] (1,0) rectangle (2,1);
 \end{scope}
 \begin{scope}
 \draw[fill=gray] (0,1) rectangle (0.5,2);
 \draw[fill=white] (0.5,1) rectangle (1,2);
 \draw[fill=gray] (1,1) rectangle (1.5,2);
 \draw[fill=white] (1.5,1) rectangle (2,2);
 \end{scope}
 \begin{scope}
 \begin{scope}[shift={(0,1)},xscale=0.5]
 \draw[fill=gray] (0,1) rectangle (0.5,2);
 \draw[fill=white] (0.5,1) rectangle (1,2);
 \draw[fill=gray] (1,1) rectangle (1.5,2);
 \draw[fill=white] (1.5,1) rectangle (2,2);
 \end{scope}
 \begin{scope}[shift={(1,1)},xscale=0.5]
 \draw[fill=gray] (0,1) rectangle (0.5,2);
 \draw[fill=white] (0.5,1) rectangle (1,2);
 \draw[fill=gray] (1,1) rectangle (1.5,2);
 \draw[fill=white] (1.5,1) rectangle (2,2);
 \end{scope}
 \end{scope}
 \end{scope}
\begin{scope}[xscale=1,yscale=1/3,shift={(2,3)},rotate=0]
 \begin{scope}
 \draw[fill=gray] (0,0) rectangle (1,1);
 \draw[fill=white] (1,0) rectangle (2,1);
 \end{scope}
 \begin{scope}
 \draw[fill=gray] (0,1) rectangle (0.5,2);
 \draw[fill=white] (0.5,1) rectangle (1,2);
 \draw[fill=gray] (1,1) rectangle (1.5,2);
 \draw[fill=white] (1.5,1) rectangle (2,2);
 \end{scope}
 \begin{scope}
 \begin{scope}[shift={(0,1)},xscale=0.5]
 \draw[fill=gray] (0,1) rectangle (0.5,2);
 \draw[fill=white] (0.5,1) rectangle (1,2);
 \draw[fill=gray] (1,1) rectangle (1.5,2);
 \draw[fill=white] (1.5,1) rectangle (2,2);
 \end{scope}
 \begin{scope}[shift={(1,1)},xscale=0.5]
 \draw[fill=gray] (0,1) rectangle (0.5,2);
 \draw[fill=white] (0.5,1) rectangle (1,2);
 \draw[fill=gray] (1,1) rectangle (1.5,2);
 \draw[fill=white] (1.5,1) rectangle (2,2);
 \end{scope}
 \end{scope}
 \end{scope}
 \begin{scope}[xscale=1,yscale=1/3,shift={(2,6)},rotate=0]
 \begin{scope}
 \draw[fill=gray] (0,0) rectangle (1,1);
 \draw[fill=white] (1,0) rectangle (2,1);
 \end{scope}
 \begin{scope}
 \draw[fill=gray] (0,1) rectangle (0.5,2);
 \draw[fill=white] (0.5,1) rectangle (1,2);
 \draw[fill=gray] (1,1) rectangle (1.5,2);
 \draw[fill=white] (1.5,1) rectangle (2,2);
 \end{scope}
 \begin{scope}
 \begin{scope}[shift={(0,1)},xscale=0.5]
 \draw[fill=gray] (0,1) rectangle (0.5,2);
 \draw[fill=white] (0.5,1) rectangle (1,2);
 \draw[fill=gray] (1,1) rectangle (1.5,2);
 \draw[fill=white] (1.5,1) rectangle (2,2);
 \end{scope}
 \begin{scope}[shift={(1,1)},xscale=0.5]
 \draw[fill=gray] (0,1) rectangle (0.5,2);
 \draw[fill=white] (0.5,1) rectangle (1,2);
 \draw[fill=gray] (1,1) rectangle (1.5,2);
 \draw[fill=white] (1.5,1) rectangle (2,2);
 \end{scope}
 \end{scope}
 \end{scope}
 \end{scope}
 \end{scope}
 \end{scope}
\end{tikzpicture}}
      \end{tabular}
      &
      \begin{tabular}{l}
        \subfloat[$\overOp{\phi_1}{2}$]{   \begin{tikzpicture}[yscale=1,xscale=1,rotate=90]
\begin{scope}
 \begin{scope}[xscale=1,yscale=1,shift={(0,0)},rotate=0]
 \begin{scope}[opacity=1]
 \draw[fill=gray] (0,0) rectangle (1,1);
 \draw[fill=white] (1,0) rectangle (2,1);
 \end{scope}
 \begin{scope}[opacity=1]
 \draw[fill=gray] (0,1) rectangle (0.5,2);
 \draw[fill=white] (0.5,1) rectangle (1,2);
 \draw[fill=gray] (1,1) rectangle (1.5,2);
 \draw[fill=white] (1.5,1) rectangle (2,2);
 \end{scope}
 \begin{scope}[opacity=1]
 \begin{scope}[shift={(0,1)},xscale=0.5]
 \draw[fill=gray] (0,1) rectangle (0.5,2);
 \draw[fill=white] (0.5,1) rectangle (1,2);
 \draw[fill=gray] (1,1) rectangle (1.5,2);
 \draw[fill=white] (1.5,1) rectangle (2,2);
 \end{scope}
 \begin{scope}[shift={(1,1)},xscale=0.5]
 \draw[fill=gray] (0,1) rectangle (0.5,2);
 \draw[fill=white] (0.5,1) rectangle (1,2);
 \draw[fill=gray] (1,1) rectangle (1.5,2);
 \draw[fill=white] (1.5,1) rectangle (2,2);
 \end{scope}
 \end{scope}
 \end{scope}
 \begin{scope}[opacity=1,shift={(0,0)}]
 \begin{scope}[xscale=1,yscale=1/3,shift={(2,0)},rotate=0]
 \begin{scope}
 \draw[fill=gray] (0,0) rectangle (1,1);
 \draw[fill=white] (1,0) rectangle (2,1);
 \end{scope}
 \begin{scope}
 \draw[fill=gray] (0,1) rectangle (0.5,2);
 \draw[fill=white] (0.5,1) rectangle (1,2);
 \draw[fill=gray] (1,1) rectangle (1.5,2);
 \draw[fill=white] (1.5,1) rectangle (2,2);
 \end{scope}
 \begin{scope}
 \begin{scope}[shift={(0,1)},xscale=0.5]
 \draw[fill=gray] (0,1) rectangle (0.5,2);
 \draw[fill=white] (0.5,1) rectangle (1,2);
 \draw[fill=gray] (1,1) rectangle (1.5,2);
 \draw[fill=white] (1.5,1) rectangle (2,2);
 \end{scope}
 \begin{scope}[shift={(1,1)},xscale=0.5]
 \draw[fill=gray] (0,1) rectangle (0.5,2);
 \draw[fill=white] (0.5,1) rectangle (1,2);
 \draw[fill=gray] (1,1) rectangle (1.5,2);
 \draw[fill=white] (1.5,1) rectangle (2,2);
 \end{scope}
 \end{scope}
 \end{scope}
\begin{scope}[xscale=1,yscale=1/3,shift={(2,3)},rotate=0]
 \begin{scope}
 \draw[fill=gray] (0,0) rectangle (1,1);
 \draw[fill=white] (1,0) rectangle (2,1);
 \end{scope}
 \begin{scope}
 \draw[fill=gray] (0,1) rectangle (0.5,2);
 \draw[fill=white] (0.5,1) rectangle (1,2);
 \draw[fill=gray] (1,1) rectangle (1.5,2);
 \draw[fill=white] (1.5,1) rectangle (2,2);
 \end{scope}
 \begin{scope}
 \begin{scope}[shift={(0,1)},xscale=0.5]
 \draw[fill=gray] (0,1) rectangle (0.5,2);
 \draw[fill=white] (0.5,1) rectangle (1,2);
 \draw[fill=gray] (1,1) rectangle (1.5,2);
 \draw[fill=white] (1.5,1) rectangle (2,2);
 \end{scope}
 \begin{scope}[shift={(1,1)},xscale=0.5]
 \draw[fill=gray] (0,1) rectangle (0.5,2);
 \draw[fill=white] (0.5,1) rectangle (1,2);
 \draw[fill=gray] (1,1) rectangle (1.5,2);
 \draw[fill=white] (1.5,1) rectangle (2,2);
 \end{scope}
 \end{scope}
 \end{scope}
 \begin{scope}[xscale=1,yscale=1/3,shift={(2,6)},rotate=0]
 \begin{scope}
 \draw[fill=gray] (0,0) rectangle (1,1);
 \draw[fill=white] (1,0) rectangle (2,1);
 \end{scope}
 \begin{scope}
 \draw[fill=gray] (0,1) rectangle (0.5,2);
 \draw[fill=white] (0.5,1) rectangle (1,2);
 \draw[fill=gray] (1,1) rectangle (1.5,2);
 \draw[fill=white] (1.5,1) rectangle (2,2);
 \end{scope}
 \begin{scope}
 \begin{scope}[shift={(0,1)},xscale=0.5]
 \draw[fill=gray] (0,1) rectangle (0.5,2);
 \draw[fill=white] (0.5,1) rectangle (1,2);
 \draw[fill=gray] (1,1) rectangle (1.5,2);
 \draw[fill=white] (1.5,1) rectangle (2,2);
 \end{scope}
 \begin{scope}[shift={(1,1)},xscale=0.5]
 \draw[fill=gray] (0,1) rectangle (0.5,2);
 \draw[fill=white] (0.5,1) rectangle (1,2);
 \draw[fill=gray] (1,1) rectangle (1.5,2);
 \draw[fill=white] (1.5,1) rectangle (2,2);
 \end{scope}
 \end{scope}
 \end{scope}
 \end{scope}
 \begin{scope}[opacity=0]
 \begin{scope}[shift={(2,0)},yscale=1/3]
 \begin{scope}[xscale=1,yscale=1/3,shift={(2,0)},rotate=0]
 \begin{scope}
 \draw[fill=gray] (0,0) rectangle (1,1);
 \draw[fill=white] (1,0) rectangle (2,1);
 \end{scope}
 \begin{scope}
 \draw[fill=gray] (0,1) rectangle (0.5,2);
 \draw[fill=white] (0.5,1) rectangle (1,2);
 \draw[fill=gray] (1,1) rectangle (1.5,2);
 \draw[fill=white] (1.5,1) rectangle (2,2);
 \end{scope}
 \begin{scope}
 \begin{scope}[shift={(0,1)},xscale=0.5]
 \draw[fill=gray] (0,1) rectangle (0.5,2);
 \draw[fill=white] (0.5,1) rectangle (1,2);
 \draw[fill=gray] (1,1) rectangle (1.5,2);
 \draw[fill=white] (1.5,1) rectangle (2,2);
 \end{scope}
 \begin{scope}[shift={(1,1)},xscale=0.5]
 \draw[fill=gray] (0,1) rectangle (0.5,2);
 \draw[fill=white] (0.5,1) rectangle (1,2);
 \draw[fill=gray] (1,1) rectangle (1.5,2);
 \draw[fill=white] (1.5,1) rectangle (2,2);
 \end{scope}
 \end{scope}
 \end{scope}
\begin{scope}[xscale=1,yscale=1/3,shift={(2,3)},rotate=0]
 \begin{scope}
 \draw[fill=gray] (0,0) rectangle (1,1);
 \draw[fill=white] (1,0) rectangle (2,1);
 \end{scope}
 \begin{scope}
 \draw[fill=gray] (0,1) rectangle (0.5,2);
 \draw[fill=white] (0.5,1) rectangle (1,2);
 \draw[fill=gray] (1,1) rectangle (1.5,2);
 \draw[fill=white] (1.5,1) rectangle (2,2);
 \end{scope}
 \begin{scope}
 \begin{scope}[shift={(0,1)},xscale=0.5]
 \draw[fill=gray] (0,1) rectangle (0.5,2);
 \draw[fill=white] (0.5,1) rectangle (1,2);
 \draw[fill=gray] (1,1) rectangle (1.5,2);
 \draw[fill=white] (1.5,1) rectangle (2,2);
 \end{scope}
 \begin{scope}[shift={(1,1)},xscale=0.5]
 \draw[fill=gray] (0,1) rectangle (0.5,2);
 \draw[fill=white] (0.5,1) rectangle (1,2);
 \draw[fill=gray] (1,1) rectangle (1.5,2);
 \draw[fill=white] (1.5,1) rectangle (2,2);
 \end{scope}
 \end{scope}
 \end{scope}
 \begin{scope}[xscale=1,yscale=1/3,shift={(2,6)},rotate=0]
 \begin{scope}
 \draw[fill=gray] (0,0) rectangle (1,1);
 \draw[fill=white] (1,0) rectangle (2,1);
 \end{scope}
 \begin{scope}
 \draw[fill=gray] (0,1) rectangle (0.5,2);
 \draw[fill=white] (0.5,1) rectangle (1,2);
 \draw[fill=gray] (1,1) rectangle (1.5,2);
 \draw[fill=white] (1.5,1) rectangle (2,2);
 \end{scope}
 \begin{scope}
 \begin{scope}[shift={(0,1)},xscale=0.5]
 \draw[fill=gray] (0,1) rectangle (0.5,2);
 \draw[fill=white] (0.5,1) rectangle (1,2);
 \draw[fill=gray] (1,1) rectangle (1.5,2);
 \draw[fill=white] (1.5,1) rectangle (2,2);
 \end{scope}
 \begin{scope}[shift={(1,1)},xscale=0.5]
 \draw[fill=gray] (0,1) rectangle (0.5,2);
 \draw[fill=white] (0.5,1) rectangle (1,2);
 \draw[fill=gray] (1,1) rectangle (1.5,2);
 \draw[fill=white] (1.5,1) rectangle (2,2);
 \end{scope}
 \end{scope}
 \end{scope}
 \end{scope}
  \begin{scope}[shift={(2,1)},yscale=1/3]
 \begin{scope}[xscale=1,yscale=1/3,shift={(2,0)},rotate=0]
 \begin{scope}
 \draw[fill=gray] (0,0) rectangle (1,1);
 \draw[fill=white] (1,0) rectangle (2,1);
 \end{scope}
 \begin{scope}
 \draw[fill=gray] (0,1) rectangle (0.5,2);
 \draw[fill=white] (0.5,1) rectangle (1,2);
 \draw[fill=gray] (1,1) rectangle (1.5,2);
 \draw[fill=white] (1.5,1) rectangle (2,2);
 \end{scope}
 \begin{scope}
 \begin{scope}[shift={(0,1)},xscale=0.5]
 \draw[fill=gray] (0,1) rectangle (0.5,2);
 \draw[fill=white] (0.5,1) rectangle (1,2);
 \draw[fill=gray] (1,1) rectangle (1.5,2);
 \draw[fill=white] (1.5,1) rectangle (2,2);
 \end{scope}
 \begin{scope}[shift={(1,1)},xscale=0.5]
 \draw[fill=gray] (0,1) rectangle (0.5,2);
 \draw[fill=white] (0.5,1) rectangle (1,2);
 \draw[fill=gray] (1,1) rectangle (1.5,2);
 \draw[fill=white] (1.5,1) rectangle (2,2);
 \end{scope}
 \end{scope}
 \end{scope}
\begin{scope}[xscale=1,yscale=1/3,shift={(2,3)},rotate=0]
 \begin{scope}
 \draw[fill=gray] (0,0) rectangle (1,1);
 \draw[fill=white] (1,0) rectangle (2,1);
 \end{scope}
 \begin{scope}
 \draw[fill=gray] (0,1) rectangle (0.5,2);
 \draw[fill=white] (0.5,1) rectangle (1,2);
 \draw[fill=gray] (1,1) rectangle (1.5,2);
 \draw[fill=white] (1.5,1) rectangle (2,2);
 \end{scope}
 \begin{scope}
 \begin{scope}[shift={(0,1)},xscale=0.5]
 \draw[fill=gray] (0,1) rectangle (0.5,2);
 \draw[fill=white] (0.5,1) rectangle (1,2);
 \draw[fill=gray] (1,1) rectangle (1.5,2);
 \draw[fill=white] (1.5,1) rectangle (2,2);
 \end{scope}
 \begin{scope}[shift={(1,1)},xscale=0.5]
 \draw[fill=gray] (0,1) rectangle (0.5,2);
 \draw[fill=white] (0.5,1) rectangle (1,2);
 \draw[fill=gray] (1,1) rectangle (1.5,2);
 \draw[fill=white] (1.5,1) rectangle (2,2);
 \end{scope}
 \end{scope}
 \end{scope}
 \begin{scope}[xscale=1,yscale=1/3,shift={(2,6)},rotate=0]
 \begin{scope}
 \draw[fill=gray] (0,0) rectangle (1,1);
 \draw[fill=white] (1,0) rectangle (2,1);
 \end{scope}
 \begin{scope}
 \draw[fill=gray] (0,1) rectangle (0.5,2);
 \draw[fill=white] (0.5,1) rectangle (1,2);
 \draw[fill=gray] (1,1) rectangle (1.5,2);
 \draw[fill=white] (1.5,1) rectangle (2,2);
 \end{scope}
 \begin{scope}
 \begin{scope}[shift={(0,1)},xscale=0.5]
 \draw[fill=gray] (0,1) rectangle (0.5,2);
 \draw[fill=white] (0.5,1) rectangle (1,2);
 \draw[fill=gray] (1,1) rectangle (1.5,2);
 \draw[fill=white] (1.5,1) rectangle (2,2);
 \end{scope}
 \begin{scope}[shift={(1,1)},xscale=0.5]
 \draw[fill=gray] (0,1) rectangle (0.5,2);
 \draw[fill=white] (0.5,1) rectangle (1,2);
 \draw[fill=gray] (1,1) rectangle (1.5,2);
 \draw[fill=white] (1.5,1) rectangle (2,2);
 \end{scope}
 \end{scope}
 \end{scope}
 \end{scope}
  \begin{scope}[shift={(2,2)},yscale=1/3]
 \begin{scope}[xscale=1,yscale=1/3,shift={(2,0)},rotate=0]
 \begin{scope}
 \draw[fill=gray] (0,0) rectangle (1,1);
 \draw[fill=white] (1,0) rectangle (2,1);
 \end{scope}
 \begin{scope}
 \draw[fill=gray] (0,1) rectangle (0.5,2);
 \draw[fill=white] (0.5,1) rectangle (1,2);
 \draw[fill=gray] (1,1) rectangle (1.5,2);
 \draw[fill=white] (1.5,1) rectangle (2,2);
 \end{scope}
 \begin{scope}
 \begin{scope}[shift={(0,1)},xscale=0.5]
 \draw[fill=gray] (0,1) rectangle (0.5,2);
 \draw[fill=white] (0.5,1) rectangle (1,2);
 \draw[fill=gray] (1,1) rectangle (1.5,2);
 \draw[fill=white] (1.5,1) rectangle (2,2);
 \end{scope}
 \begin{scope}[shift={(1,1)},xscale=0.5]
 \draw[fill=gray] (0,1) rectangle (0.5,2);
 \draw[fill=white] (0.5,1) rectangle (1,2);
 \draw[fill=gray] (1,1) rectangle (1.5,2);
 \draw[fill=white] (1.5,1) rectangle (2,2);
 \end{scope}
 \end{scope}
 \end{scope}
\begin{scope}[xscale=1,yscale=1/3,shift={(2,3)},rotate=0]
 \begin{scope}
 \draw[fill=gray] (0,0) rectangle (1,1);
 \draw[fill=white] (1,0) rectangle (2,1);
 \end{scope}
 \begin{scope}
 \draw[fill=gray] (0,1) rectangle (0.5,2);
 \draw[fill=white] (0.5,1) rectangle (1,2);
 \draw[fill=gray] (1,1) rectangle (1.5,2);
 \draw[fill=white] (1.5,1) rectangle (2,2);
 \end{scope}
 \begin{scope}
 \begin{scope}[shift={(0,1)},xscale=0.5]
 \draw[fill=gray] (0,1) rectangle (0.5,2);
 \draw[fill=white] (0.5,1) rectangle (1,2);
 \draw[fill=gray] (1,1) rectangle (1.5,2);
 \draw[fill=white] (1.5,1) rectangle (2,2);
 \end{scope}
 \begin{scope}[shift={(1,1)},xscale=0.5]
 \draw[fill=gray] (0,1) rectangle (0.5,2);
 \draw[fill=white] (0.5,1) rectangle (1,2);
 \draw[fill=gray] (1,1) rectangle (1.5,2);
 \draw[fill=white] (1.5,1) rectangle (2,2);
 \end{scope}
 \end{scope}
 \end{scope}
 \begin{scope}[xscale=1,yscale=1/3,shift={(2,6)},rotate=0]
 \begin{scope}
 \draw[fill=gray] (0,0) rectangle (1,1);
 \draw[fill=white] (1,0) rectangle (2,1);
 \end{scope}
 \begin{scope}
 \draw[fill=gray] (0,1) rectangle (0.5,2);
 \draw[fill=white] (0.5,1) rectangle (1,2);
 \draw[fill=gray] (1,1) rectangle (1.5,2);
 \draw[fill=white] (1.5,1) rectangle (2,2);
 \end{scope}
 \begin{scope}
 \begin{scope}[shift={(0,1)},xscale=0.5]
 \draw[fill=gray] (0,1) rectangle (0.5,2);
 \draw[fill=white] (0.5,1) rectangle (1,2);
 \draw[fill=gray] (1,1) rectangle (1.5,2);
 \draw[fill=white] (1.5,1) rectangle (2,2);
 \end{scope}
 \begin{scope}[shift={(1,1)},xscale=0.5]
 \draw[fill=gray] (0,1) rectangle (0.5,2);
 \draw[fill=white] (0.5,1) rectangle (1,2);
 \draw[fill=gray] (1,1) rectangle (1.5,2);
 \draw[fill=white] (1.5,1) rectangle (2,2);
 \end{scope}
 \end{scope}
 \end{scope}
 \end{scope}
 \end{scope}
 \end{scope}
 \end{tikzpicture}}
      \end{tabular}
      &
      \begin{tabular}{l}
        \subfloat[$\overOp{\phi_1}{3}$]{  \begin{tikzpicture}[yscale=1,xscale=1,rotate=90]
\begin{scope}
 \begin{scope}[xscale=1,yscale=1,shift={(0,0)},rotate=0]
 \begin{scope}[opacity=1]
 \draw[fill=gray] (0,0) rectangle (1,1);
 \draw[fill=white] (1,0) rectangle (2,1);
 \end{scope}
 \begin{scope}[opacity=1]
 \draw[fill=gray] (0,1) rectangle (0.5,2);
 \draw[fill=white] (0.5,1) rectangle (1,2);
 \draw[fill=gray] (1,1) rectangle (1.5,2);
 \draw[fill=white] (1.5,1) rectangle (2,2);
 \end{scope}
 \begin{scope}[opacity=1]
 \begin{scope}[shift={(0,1)},xscale=0.5]
 \draw[fill=gray] (0,1) rectangle (0.5,2);
 \draw[fill=white] (0.5,1) rectangle (1,2);
 \draw[fill=gray] (1,1) rectangle (1.5,2);
 \draw[fill=white] (1.5,1) rectangle (2,2);
 \end{scope}
 \begin{scope}[shift={(1,1)},xscale=0.5]
 \draw[fill=gray] (0,1) rectangle (0.5,2);
 \draw[fill=white] (0.5,1) rectangle (1,2);
 \draw[fill=gray] (1,1) rectangle (1.5,2);
 \draw[fill=white] (1.5,1) rectangle (2,2);
 \end{scope}
 \end{scope}
 \end{scope}
 \begin{scope}[opacity=1,shift={(0,0)}]
 \begin{scope}[xscale=1,yscale=1/3,shift={(2,0)},rotate=0]
 \begin{scope}
 \draw[fill=gray] (0,0) rectangle (1,1);
 \draw[fill=white] (1,0) rectangle (2,1);
 \end{scope}
 \begin{scope}
 \draw[fill=gray] (0,1) rectangle (0.5,2);
 \draw[fill=white] (0.5,1) rectangle (1,2);
 \draw[fill=gray] (1,1) rectangle (1.5,2);
 \draw[fill=white] (1.5,1) rectangle (2,2);
 \end{scope}
 \begin{scope}
 \begin{scope}[shift={(0,1)},xscale=0.5]
 \draw[fill=gray] (0,1) rectangle (0.5,2);
 \draw[fill=white] (0.5,1) rectangle (1,2);
 \draw[fill=gray] (1,1) rectangle (1.5,2);
 \draw[fill=white] (1.5,1) rectangle (2,2);
 \end{scope}
 \begin{scope}[shift={(1,1)},xscale=0.5]
 \draw[fill=gray] (0,1) rectangle (0.5,2);
 \draw[fill=white] (0.5,1) rectangle (1,2);
 \draw[fill=gray] (1,1) rectangle (1.5,2);
 \draw[fill=white] (1.5,1) rectangle (2,2);
 \end{scope}
 \end{scope}
 \end{scope}
\begin{scope}[xscale=1,yscale=1/3,shift={(2,3)},rotate=0]
 \begin{scope}
 \draw[fill=gray] (0,0) rectangle (1,1);
 \draw[fill=white] (1,0) rectangle (2,1);
 \end{scope}
 \begin{scope}
 \draw[fill=gray] (0,1) rectangle (0.5,2);
 \draw[fill=white] (0.5,1) rectangle (1,2);
 \draw[fill=gray] (1,1) rectangle (1.5,2);
 \draw[fill=white] (1.5,1) rectangle (2,2);
 \end{scope}
 \begin{scope}
 \begin{scope}[shift={(0,1)},xscale=0.5]
 \draw[fill=gray] (0,1) rectangle (0.5,2);
 \draw[fill=white] (0.5,1) rectangle (1,2);
 \draw[fill=gray] (1,1) rectangle (1.5,2);
 \draw[fill=white] (1.5,1) rectangle (2,2);
 \end{scope}
 \begin{scope}[shift={(1,1)},xscale=0.5]
 \draw[fill=gray] (0,1) rectangle (0.5,2);
 \draw[fill=white] (0.5,1) rectangle (1,2);
 \draw[fill=gray] (1,1) rectangle (1.5,2);
 \draw[fill=white] (1.5,1) rectangle (2,2);
 \end{scope}
 \end{scope}
 \end{scope}
 \begin{scope}[xscale=1,yscale=1/3,shift={(2,6)},rotate=0]
 \begin{scope}
 \draw[fill=gray] (0,0) rectangle (1,1);
 \draw[fill=white] (1,0) rectangle (2,1);
 \end{scope}
 \begin{scope}
 \draw[fill=gray] (0,1) rectangle (0.5,2);
 \draw[fill=white] (0.5,1) rectangle (1,2);
 \draw[fill=gray] (1,1) rectangle (1.5,2);
 \draw[fill=white] (1.5,1) rectangle (2,2);
 \end{scope}
 \begin{scope}
 \begin{scope}[shift={(0,1)},xscale=0.5]
 \draw[fill=gray] (0,1) rectangle (0.5,2);
 \draw[fill=white] (0.5,1) rectangle (1,2);
 \draw[fill=gray] (1,1) rectangle (1.5,2);
 \draw[fill=white] (1.5,1) rectangle (2,2);
 \end{scope}
 \begin{scope}[shift={(1,1)},xscale=0.5]
 \draw[fill=gray] (0,1) rectangle (0.5,2);
 \draw[fill=white] (0.5,1) rectangle (1,2);
 \draw[fill=gray] (1,1) rectangle (1.5,2);
 \draw[fill=white] (1.5,1) rectangle (2,2);
 \end{scope}
 \end{scope}
 \end{scope}
 \end{scope}
 \begin{scope}[opacity=1]
 \begin{scope}[shift={(2,0)},yscale=1/3]
 \begin{scope}[xscale=1,yscale=1/3,shift={(2,0)},rotate=0]
 \begin{scope}
 \draw[fill=gray] (0,0) rectangle (1,1);
 \draw[fill=white] (1,0) rectangle (2,1);
 \end{scope}
 \begin{scope}
 \draw[fill=gray] (0,1) rectangle (0.5,2);
 \draw[fill=white] (0.5,1) rectangle (1,2);
 \draw[fill=gray] (1,1) rectangle (1.5,2);
 \draw[fill=white] (1.5,1) rectangle (2,2);
 \end{scope}
 \begin{scope}
 \begin{scope}[shift={(0,1)},xscale=0.5]
 \draw[fill=gray] (0,1) rectangle (0.5,2);
 \draw[fill=white] (0.5,1) rectangle (1,2);
 \draw[fill=gray] (1,1) rectangle (1.5,2);
 \draw[fill=white] (1.5,1) rectangle (2,2);
 \end{scope}
 \begin{scope}[shift={(1,1)},xscale=0.5]
 \draw[fill=gray] (0,1) rectangle (0.5,2);
 \draw[fill=white] (0.5,1) rectangle (1,2);
 \draw[fill=gray] (1,1) rectangle (1.5,2);
 \draw[fill=white] (1.5,1) rectangle (2,2);
 \end{scope}
 \end{scope}
 \end{scope}
\begin{scope}[xscale=1,yscale=1/3,shift={(2,3)},rotate=0]
 \begin{scope}
 \draw[fill=gray] (0,0) rectangle (1,1);
 \draw[fill=white] (1,0) rectangle (2,1);
 \end{scope}
 \begin{scope}
 \draw[fill=gray] (0,1) rectangle (0.5,2);
 \draw[fill=white] (0.5,1) rectangle (1,2);
 \draw[fill=gray] (1,1) rectangle (1.5,2);
 \draw[fill=white] (1.5,1) rectangle (2,2);
 \end{scope}
 \begin{scope}
 \begin{scope}[shift={(0,1)},xscale=0.5]
 \draw[fill=gray] (0,1) rectangle (0.5,2);
 \draw[fill=white] (0.5,1) rectangle (1,2);
 \draw[fill=gray] (1,1) rectangle (1.5,2);
 \draw[fill=white] (1.5,1) rectangle (2,2);
 \end{scope}
 \begin{scope}[shift={(1,1)},xscale=0.5]
 \draw[fill=gray] (0,1) rectangle (0.5,2);
 \draw[fill=white] (0.5,1) rectangle (1,2);
 \draw[fill=gray] (1,1) rectangle (1.5,2);
 \draw[fill=white] (1.5,1) rectangle (2,2);
 \end{scope}
 \end{scope}
 \end{scope}
 \begin{scope}[xscale=1,yscale=1/3,shift={(2,6)},rotate=0]
 \begin{scope}
 \draw[fill=gray] (0,0) rectangle (1,1);
 \draw[fill=white] (1,0) rectangle (2,1);
 \end{scope}
 \begin{scope}
 \draw[fill=gray] (0,1) rectangle (0.5,2);
 \draw[fill=white] (0.5,1) rectangle (1,2);
 \draw[fill=gray] (1,1) rectangle (1.5,2);
 \draw[fill=white] (1.5,1) rectangle (2,2);
 \end{scope}
 \begin{scope}
 \begin{scope}[shift={(0,1)},xscale=0.5]
 \draw[fill=gray] (0,1) rectangle (0.5,2);
 \draw[fill=white] (0.5,1) rectangle (1,2);
 \draw[fill=gray] (1,1) rectangle (1.5,2);
 \draw[fill=white] (1.5,1) rectangle (2,2);
 \end{scope}
 \begin{scope}[shift={(1,1)},xscale=0.5]
 \draw[fill=gray] (0,1) rectangle (0.5,2);
 \draw[fill=white] (0.5,1) rectangle (1,2);
 \draw[fill=gray] (1,1) rectangle (1.5,2);
 \draw[fill=white] (1.5,1) rectangle (2,2);
 \end{scope}
 \end{scope}
 \end{scope}
 \end{scope}
  \begin{scope}[shift={(2,1)},yscale=1/3]
 \begin{scope}[xscale=1,yscale=1/3,shift={(2,0)},rotate=0]
 \begin{scope}
 \draw[fill=gray] (0,0) rectangle (1,1);
 \draw[fill=white] (1,0) rectangle (2,1);
 \end{scope}
 \begin{scope}
 \draw[fill=gray] (0,1) rectangle (0.5,2);
 \draw[fill=white] (0.5,1) rectangle (1,2);
 \draw[fill=gray] (1,1) rectangle (1.5,2);
 \draw[fill=white] (1.5,1) rectangle (2,2);
 \end{scope}
 \begin{scope}
 \begin{scope}[shift={(0,1)},xscale=0.5]
 \draw[fill=gray] (0,1) rectangle (0.5,2);
 \draw[fill=white] (0.5,1) rectangle (1,2);
 \draw[fill=gray] (1,1) rectangle (1.5,2);
 \draw[fill=white] (1.5,1) rectangle (2,2);
 \end{scope}
 \begin{scope}[shift={(1,1)},xscale=0.5]
 \draw[fill=gray] (0,1) rectangle (0.5,2);
 \draw[fill=white] (0.5,1) rectangle (1,2);
 \draw[fill=gray] (1,1) rectangle (1.5,2);
 \draw[fill=white] (1.5,1) rectangle (2,2);
 \end{scope}
 \end{scope}
 \end{scope}
\begin{scope}[xscale=1,yscale=1/3,shift={(2,3)},rotate=0]
 \begin{scope}
 \draw[fill=gray] (0,0) rectangle (1,1);
 \draw[fill=white] (1,0) rectangle (2,1);
 \end{scope}
 \begin{scope}
 \draw[fill=gray] (0,1) rectangle (0.5,2);
 \draw[fill=white] (0.5,1) rectangle (1,2);
 \draw[fill=gray] (1,1) rectangle (1.5,2);
 \draw[fill=white] (1.5,1) rectangle (2,2);
 \end{scope}
 \begin{scope}
 \begin{scope}[shift={(0,1)},xscale=0.5]
 \draw[fill=gray] (0,1) rectangle (0.5,2);
 \draw[fill=white] (0.5,1) rectangle (1,2);
 \draw[fill=gray] (1,1) rectangle (1.5,2);
 \draw[fill=white] (1.5,1) rectangle (2,2);
 \end{scope}
 \begin{scope}[shift={(1,1)},xscale=0.5]
 \draw[fill=gray] (0,1) rectangle (0.5,2);
 \draw[fill=white] (0.5,1) rectangle (1,2);
 \draw[fill=gray] (1,1) rectangle (1.5,2);
 \draw[fill=white] (1.5,1) rectangle (2,2);
 \end{scope}
 \end{scope}
 \end{scope}
 \begin{scope}[xscale=1,yscale=1/3,shift={(2,6)},rotate=0]
 \begin{scope}
 \draw[fill=gray] (0,0) rectangle (1,1);
 \draw[fill=white] (1,0) rectangle (2,1);
 \end{scope}
 \begin{scope}
 \draw[fill=gray] (0,1) rectangle (0.5,2);
 \draw[fill=white] (0.5,1) rectangle (1,2);
 \draw[fill=gray] (1,1) rectangle (1.5,2);
 \draw[fill=white] (1.5,1) rectangle (2,2);
 \end{scope}
 \begin{scope}
 \begin{scope}[shift={(0,1)},xscale=0.5]
 \draw[fill=gray] (0,1) rectangle (0.5,2);
 \draw[fill=white] (0.5,1) rectangle (1,2);
 \draw[fill=gray] (1,1) rectangle (1.5,2);
 \draw[fill=white] (1.5,1) rectangle (2,2);
 \end{scope}
 \begin{scope}[shift={(1,1)},xscale=0.5]
 \draw[fill=gray] (0,1) rectangle (0.5,2);
 \draw[fill=white] (0.5,1) rectangle (1,2);
 \draw[fill=gray] (1,1) rectangle (1.5,2);
 \draw[fill=white] (1.5,1) rectangle (2,2);
 \end{scope}
 \end{scope}
 \end{scope}
 \end{scope}
  \begin{scope}[shift={(2,2)},yscale=1/3]
 \begin{scope}[xscale=1,yscale=1/3,shift={(2,0)},rotate=0]
 \begin{scope}
 \draw[fill=gray] (0,0) rectangle (1,1);
 \draw[fill=white] (1,0) rectangle (2,1);
 \end{scope}
 \begin{scope}
 \draw[fill=gray] (0,1) rectangle (0.5,2);
 \draw[fill=white] (0.5,1) rectangle (1,2);
 \draw[fill=gray] (1,1) rectangle (1.5,2);
 \draw[fill=white] (1.5,1) rectangle (2,2);
 \end{scope}
 \begin{scope}
 \begin{scope}[shift={(0,1)},xscale=0.5]
 \draw[fill=gray] (0,1) rectangle (0.5,2);
 \draw[fill=white] (0.5,1) rectangle (1,2);
 \draw[fill=gray] (1,1) rectangle (1.5,2);
 \draw[fill=white] (1.5,1) rectangle (2,2);
 \end{scope}
 \begin{scope}[shift={(1,1)},xscale=0.5]
 \draw[fill=gray] (0,1) rectangle (0.5,2);
 \draw[fill=white] (0.5,1) rectangle (1,2);
 \draw[fill=gray] (1,1) rectangle (1.5,2);
 \draw[fill=white] (1.5,1) rectangle (2,2);
 \end{scope}
 \end{scope}
 \end{scope}
\begin{scope}[xscale=1,yscale=1/3,shift={(2,3)},rotate=0]
 \begin{scope}
 \draw[fill=gray] (0,0) rectangle (1,1);
 \draw[fill=white] (1,0) rectangle (2,1);
 \end{scope}
 \begin{scope}
 \draw[fill=gray] (0,1) rectangle (0.5,2);
 \draw[fill=white] (0.5,1) rectangle (1,2);
 \draw[fill=gray] (1,1) rectangle (1.5,2);
 \draw[fill=white] (1.5,1) rectangle (2,2);
 \end{scope}
 \begin{scope}
 \begin{scope}[shift={(0,1)},xscale=0.5]
 \draw[fill=gray] (0,1) rectangle (0.5,2);
 \draw[fill=white] (0.5,1) rectangle (1,2);
 \draw[fill=gray] (1,1) rectangle (1.5,2);
 \draw[fill=white] (1.5,1) rectangle (2,2);
 \end{scope}
 \begin{scope}[shift={(1,1)},xscale=0.5]
 \draw[fill=gray] (0,1) rectangle (0.5,2);
 \draw[fill=white] (0.5,1) rectangle (1,2);
 \draw[fill=gray] (1,1) rectangle (1.5,2);
 \draw[fill=white] (1.5,1) rectangle (2,2);
 \end{scope}
 \end{scope}
 \end{scope}
 \begin{scope}[xscale=1,yscale=1/3,shift={(2,6)},rotate=0]
 \begin{scope}
 \draw[fill=gray] (0,0) rectangle (1,1);
 \draw[fill=white] (1,0) rectangle (2,1);
 \end{scope}
 \begin{scope}
 \draw[fill=gray] (0,1) rectangle (0.5,2);
 \draw[fill=white] (0.5,1) rectangle (1,2);
 \draw[fill=gray] (1,1) rectangle (1.5,2);
 \draw[fill=white] (1.5,1) rectangle (2,2);
 \end{scope}
 \begin{scope}
 \begin{scope}[shift={(0,1)},xscale=0.5]
 \draw[fill=gray] (0,1) rectangle (0.5,2);
 \draw[fill=white] (0.5,1) rectangle (1,2);
 \draw[fill=gray] (1,1) rectangle (1.5,2);
 \draw[fill=white] (1.5,1) rectangle (2,2);
 \end{scope}
 \begin{scope}[shift={(1,1)},xscale=0.5]
 \draw[fill=gray] (0,1) rectangle (0.5,2);
 \draw[fill=white] (0.5,1) rectangle (1,2);
 \draw[fill=gray] (1,1) rectangle (1.5,2);
 \draw[fill=white] (1.5,1) rectangle (2,2);
 \end{scope}
 \end{scope}
 \end{scope}
 \end{scope}
 \end{scope}
 \end{scope}
 \end{tikzpicture}}
      \end{tabular}
    \end{tabular}
  \end{center}
  \caption{\nameref{def:overOperation} applied $B=3$ times on the base game
  (\textbf{a}, \textbf{b} and \textbf{c}) followed by transposition (from
  \textbf{c} to \textbf{d}) and repetition of the procedure (\textbf{d},
  \textbf{e} and \textbf{f}) starting from the more complex output game of the
  first iteration}
  \label{fig:iter}
\end{figure}
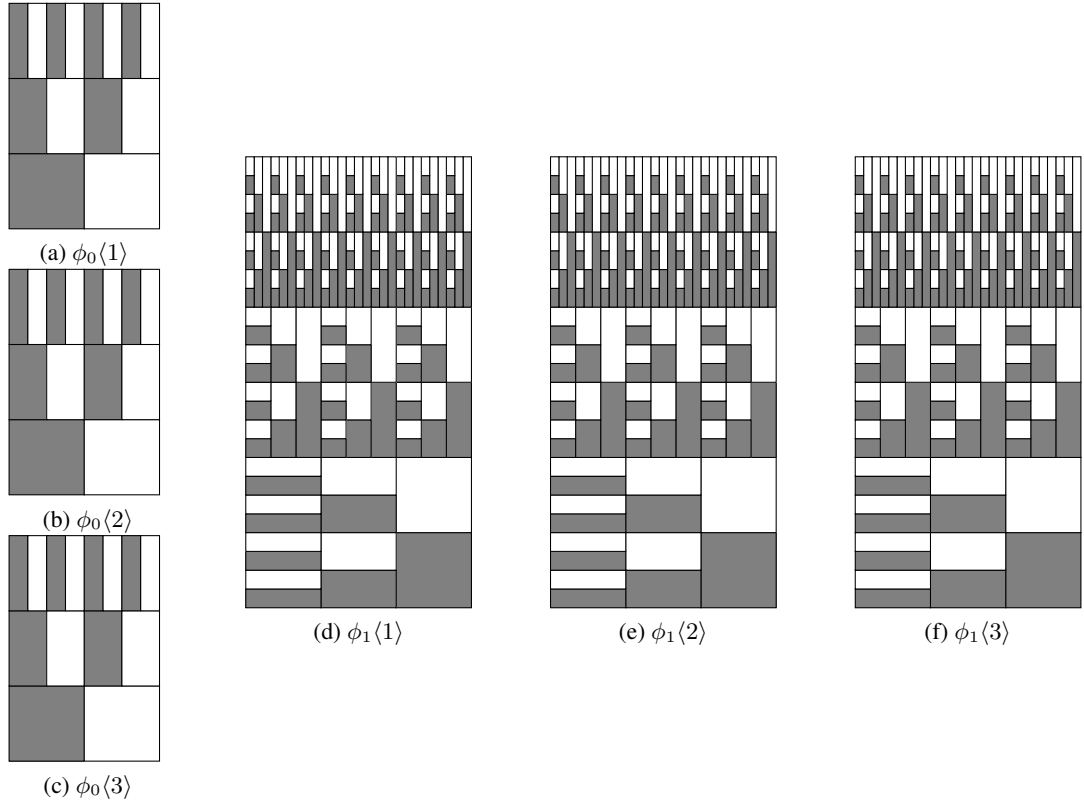

We now define the explicit construction used to generate the counterexample.
We call this family of functions \nameref{definition:AlternatingGame}. As we
shall see later in the proof, the games are built such that the players are
forced to communicate in succession a given number of rounds until the game is
solved. The construction is parameterised by some natural number $B$, which
must be chosen according to specific conditions which will be derived in
Sections \ref{section:LB} and \ref{section:UB}. Note that $B$ is fixed
throughout the paper, aside from Figure \ref{fig:iter} where we set it to $3$
for illustrative purposes, though we only instantiate it once we have derived
the conditions we need it to satisfy.

\begin{definition}[Alternating Communicating Game]\label{definition:AlternatingGame}
  Given $B$ a positive integer, we define
  \begin{gather*}
    \phi_0=\begin{bmatrix}1&0\end{bmatrix}\\
    \phi_{i+1}=\transpose{\left(\overOp{\phi_{i}}{\cstTopMultS}\right)}
  \end{gather*}
\end{definition}

\quad

Figure \ref{fig:iter} illustrates an example of the construction with $B$ set
to $3$ and $i$ going up to $1$. For the actual counterexample the value of $B$
required to prove our lower bound is much larger. Some high-level ideas about
the construction are:
\begin{itemize}
  \item By repeatedly applying the \nameref{def:overOperation} we can force the
  row player to communicate any number of bits.
  \item By transposing the matrix and repeatedly applying the over operation we
  can then force the column player to communicate any number of bits.
  \item The players therefore take turns communicating, and in each round of
  communication must separate the interlaced games from one another.
  \item This can be continued so that we force the players to communicate in
  alternating rounds $i$.
  \item If we set the value of $B$ correctly, the integrality of the bit
  communication means that some amount of communication must be wasted, and
  this amount grows as $i$ gets larger.
\end{itemize}

Though the intuition is simple, proving the lower bound is quite involved. The
value of $B$ must be set so that some amount of communication is wasted but
also such that it is not possible for a player to communicate out of turn. In
order to do so we will have to build the tools necessary to analyse how robust
our construction of interlaced games is to eliminating columns. The main way
we do so is by showing that given a matrix $A$, interlacing it
$\symTotGames$ times with itself and eliminating a certain number of columns
can be directly compared via the subgame relation to a game with fewer copies
of $A$ interlaced but more columns preserved.

Interestingly, the construction is such that standard lower-bound techniques
cannot help in establishing the bound we require. This will motivate our
introduction of a new framework of analysis tailor-designed to constructions
using the \nameref{def:overOperation}. Though the details are specific to the
direct sum problem, once again we believe that the framework, namely the
operation and accompanying specialised analysis, is applicable to communication
complexity more broadly. To show a gap between communicating games in sequence
and communicating them in a batch we will need to establish that the
complexity of our construction is $(i+1)\times k$ for some
$2^{k-1}<\cstTopMultS<2^k$. We give a sketch of proof for why the standard
lower bounds are insufficient to achieve this.

\begin{proposition}
  Given $2^{k-1}<\cstTopMultS<2^{k}$, then rank, fooling set, monochromatic
  rectangles and discrepancy lower bounds cannot prove that
  $\comp{\overOp{\phi_i}{\cstTopMultS}}\geq (i+1) \times k$.
\end{proposition}

\begin{proof}
  Assuming that a matrix $M$ has rank, fooling set, monochromatic rectangles
  or inverse discrepancy $c$, it is the case that for any $l$, $\overOp{M}{l}$
  has at most rank, fooling set, monochromatic rectangles or inverse
  discrepancy $c \times l$.

  Applying this to $\overOp{\phi_i}{\cstTopMultS}$ we can upper bound the rank,
  fooling set, monochromatic rectangles or inverse discrepancy by
  $c \cstTopMultS^{i+1}$, where $c$ is some initialising constant. Taking the
  logarithm of this gives $\log{c}+(i+1)\times \log{B}$ and since $\log{B}<k$,
  when $i$ becomes large enough the gap between the lower bound provable via
  these techniques and the bound we are aiming for becomes arbitrarily large.
\end{proof}

\subsection{Further Definitions and Framework}

In this subsection we introduce the framework which allows us to directly
compare via the subgame relation interlaced games of different size. As
mentioned above, our goal is to show that given an interlacing of some base
game $A$, we can find subgames of the interlacing consisting of a smaller
number of interlaced copies of $A$ but with added conditions on the extracted
rows or columns. The reason this will help us in building our lower bound is
that we can inductively increase the complexity of our construction by
interlacing it with itself whilst provably making it increasingly robust to
extractions of columns.

\subsubsection{Extractions and Projections}

\paragraph{Extracting Rows and Columns.} To start, we define a function to
extract selected rows and columns from a matrix, which will be essential in
studying possible partitions of a matrix by a protocol.

\begin{definition}
  Let $A = (a_{i,j})$ be an $m \times n$ matrix. For
  $R \subseteq \rangeintegers{m}$ a row selection and
  $C \subseteq \rangeintegers{n}$ a column selection,
  $\extractmatrix{A}{R}{C}$ is the $\card{R}\times\card{C}$ matrix
  $B = (b_{i,j})$ defined by $b_{i,j} = a_{r_i,c_j}$ where
  $0 \leq r_0 < r_1 < \dots < r_{\card{R}-1} < m$ are the sorted elements of
  $R$, and $0 \leq c_0 < c_1 < \dots < c_{\card{C}-1} < n$ are the sorted
  elements of $C$.
\end{definition}

\paragraph{Balanced Extraction.} It will be important for us to extract rows
from matrices in a way that is balanced, meaning that the rows are extracted
at regular intervals. This allows us to extract rows from interlaced games
with each interlaced component keeping a given number of its rows. For this we
define the notion of $m,\numberT,\symTotGames$-equipartitioned sets of
integers.

\begin{definition}[Equipartition]\label{def:Equipartition}
  Let $R \subseteq \rangeintegers{m\symTotGames}$ be a set of integers. We say
  that $R$ is \emph{$m,\numberT,\symTotGames$-equipartitioned} if for all
  $\indexOne \in \rangeintegers{\symTotGames}$ we have
  $\card{R \cap \rangeintegers{m\indexOne,m(\indexOne+1)}} = \ceil{\numberT}$.
\end{definition}

\paragraph{Index Encoding.} When working with the indices of matrices, it will
be convenient to represent the integers in a chosen basis so that the digits
encoding them are related to the structure of the interlaced games.

\begin{definition}
  Any integer $c \in \rangeintegers{n^\symTotGames}$ can be written as a sum
  $c = \sum_i b_q n^q$ such that for all $\indexOne$ we have
  $0 \leq b_\indexOne < n$. For any fixed $\symTotGames$, the base-$n$
  representation of numbers in $\rangeintegers{n^\symTotGames}$ exists and is
  unique, so in the rest of this paper we use
  $\basedecomp{b_{\symTotGames-1},\dots,b_0}{n}$ as an alternative way to
  refer to the number $\sum_i b_q n^q$.
\end{definition}
The coefficients are also called \emph{digits}. Traditionally, unicity is
obtained by requiring the highest digit to be non-zero. Here, we obtain
unicity by having a fixed range $\rangeintegers{n^\symTotGames}$ and forcing
the number of digits to be $\symTotGames$. For
$c = \basedecomp{b_{\symTotGames-1},\dots,b_0}{n} \in \rangeintegers{n^\symTotGames}$,
the coefficients satisfy
$b_\indexOne = \remainder{\floor{\frac{c}{n^\indexOne}}}{n}$ for
$\indexOne \in\rangeintegers{\symTotGames}$. The base-$n$ notation allows for
convenient shorthands. For example, notation
$\{ b_3 \suchthat \basedecomp{b_{\symTotGames-1},\dots,b_0}{n} \in C \}$
refers to the same set as notation
$\{ \remainder{\floor{\frac{c}{n^3}}}{n} \suchthat c \in C \}$.

\paragraph{Projections of Games.} Representing integers in this way is
convenient for defining our next concept, the notion of $Q$-projection, where
$Q$ is a set of integers. The notion of $Q$-projection is used to extract
specific subgames which are relevant in our analysis of the complexity of the
construction. More concretely, $Q$-projections are used to map the rows and
columns extracted from a matrix built by applying the
\nameref{def:overOperation} to corresponding rows and columns extracted from a
matrix built with a subset of the games to which the
\nameref{def:overOperation} was applied.

\begin{definition}
  Let $R \subseteq \rangeintegers{m\symTotGames}$,
  $C \subseteq \rangeintegers{n^\symTotGames}$ and
  $Q \subseteq \rangeintegers{\symTotGames}$. Write
  $\symGamePartSize=\card{Q}$. If $\symGamePartSize\geq 1$, let
  $0 \leq q_0 < q_1 < \dots < q_{\symGamePartSize-1} < \symTotGames$ be the
  sorted elements of $Q$.

  The \emph{$Q$-projection} of $R,C$ is the pair $S,D$ defined as follows. If
  $Q=\varnothing$, set
  \[
    S=\varnothing
    \qquad\text{and}\qquad
    D=\{0\}.
  \]
  Otherwise, define
  \[
    S =
    \left\{
      m\indexOne+r
      \suchthat
      \indexOne\in\rangeintegers{\symGamePartSize},\ 
      r\in\rangeintegers{m},\ 
      mq_{\indexOne}+r \in R
    \right\}
    \subseteq \rangeintegers{m\symGamePartSize},
  \]
  and
  \[
    D =
    \left\{
      \basedecomp{b_{q_{\symGamePartSize-1}},\dots,b_{q_0}}{n}
      \suchthat
      \basedecomp{b_{\symTotGames-1},\dots,b_0}{n} \in C
    \right\}
    \subseteq \rangeintegers{n^{\symGamePartSize}}.
  \]
\end{definition}

The next lemma highlights the importance of $Q$-projections. It establishes
that the game obtained by extracting rows and columns $R$ and $C$ from a game
built by applying the \nameref{def:overOperation} $\symTotGames$ times is a
supergame of the game obtained by extracting their $Q$-projections from a game
built by applying the \nameref{def:overOperation} $\card{Q}$ times to $A$.



\begin{lemma}
\label{lem:projection}
For any $m \times n$ matrix $A$ and positive integer $\symTotGames$, let
$R \subseteq \rangeintegers{m\symTotGames}$ and
$C \subseteq \rangeintegers{n^\symTotGames}$ be selections of rows and
columns of $\overOp{A}{\symTotGames}$. If
$Q \subseteq \rangeintegers{\symTotGames}$ is non-empty and $S,D$ is the
$Q$-projection of $R,C$, then
\[
  \extractmatrix{\overOp{A}{\card{Q}}}{S}{D}
  \sqsubseteq
  \extractmatrix{\overOp{A}{\symTotGames}}{R}{C}.
\]
\end{lemma}

\begin{proof}
Let $L=(l_{i,j})=\overOp{A}{\symGamePartSize}$ and
$P=(p_{i,j})=\overOp{A}{\symTotGames}$. Let
$G=(g_{i,j})=\extractmatrix{L}{S}{D}$ and
$H=(h_{i,j})=\extractmatrix{P}{R}{C}$.

Write
\[
  0 \leq q_0 < q_1 < \dots < q_{\symGamePartSize-1} < \symTotGames
\]
for the sorted elements of $Q$, and let
\[
  0 \leq r_0 < r_1 < \dots < r_{\card{R}-1},\quad
  0 \leq s_0 < s_1 < \dots < s_{\card{S}-1},
\]
\[
  0 \leq u_0 < u_1 < \dots < u_{\card{C}-1},\quad
  0 \leq d_0 < d_1 < \dots < d_{\card{D}-1}
\]
be the sorted elements of $R,S,C,D$, respectively.

For each $x=m\ell+t \in S$, define
\[
  \sigma'(x)=mq_\ell+t.
\]
By the definition of the $Q$-projection, $\sigma'(x)\in R$.

For each $d\in D$, choose $\tau'(d)\in C$ such that if
\[
  d=\basedecomp{d_{\symGamePartSize-1},\dots,d_0}{n}
  \qquad\text{and}\qquad
  \tau'(d)=\basedecomp{c_{\symTotGames-1},\dots,c_0}{n},
\]
then
\[
  c_{q_\ell}=d_\ell
  \qquad\text{for every }\ell\in\rangeintegers{\symGamePartSize}.
\]
Such a choice is possible by the definition of $D$.

For each $i\in\rangeintegers{\card{S}}$, let $\sigma(i)$ be the unique index
in $\rangeintegers{\card{R}}$ such that $r_{\sigma(i)}=\sigma'(s_i)$.
Similarly, for each $j\in\rangeintegers{\card{D}}$, let $\tau(j)$ be the
unique index in $\rangeintegers{\card{C}}$ such that
$u_{\tau(j)}=\tau'(d_j)$.

Fix $(i,j)\in \rangeintegers{\card{S}}\times\rangeintegers{\card{D}}$. Write
$s_i=m\ell+t$ with $\ell=\floor{\frac{s_i}{m}}$ and $t=\remainder{s_i}{m}$.
Also write
\[
  d_j=\basedecomp{d_{j,\symGamePartSize-1},\dots,d_{j,0}}{n}
  \qquad\text{and}\qquad
  \tau'(d_j)=\basedecomp{c_{\symTotGames-1},\dots,c_0}{n}.
\]
Then $c_{q_\ell}=d_{j,\ell}$ by construction of $\tau'$.

By the definition of $\overOp{A}{\symGamePartSize}$, the entry
$l_{s_i,d_j}$ depends on component $\ell$, so
\[
  g_{i,j}
  =
  l_{s_i,d_j}
  =
  a_{t,d_{j,\ell}}.
\]
On the other hand, $\sigma'(s_i)=mq_\ell+t$, so
\[
  \floor{\frac{\sigma'(s_i)}{m}}=q_\ell
  \qquad\text{and}\qquad
  \remainder{\sigma'(s_i)}{m}=t.
\]
Applying the definition of $\overOp{A}{\symTotGames}$ therefore gives
\[
  p_{\sigma'(s_i),\tau'(d_j)}
  =
  a_{t,c_{q_\ell}}
  =
  a_{t,d_{j,\ell}}.
\]
Hence
\[
  g_{i,j}
  =
  p_{\sigma'(s_i),\tau'(d_j)}
  =
  h_{\sigma(i),\tau(j)}.
\]
This proves that $G \sqsubseteq H$.
\end{proof}

\subsubsection{Core Lemmas on Extractions and Projections}

We now have the tools to express the core lemmas that will be used in our
lower bound proof. These lemmas concern specific properties of extractions and
projections that will be useful for us to establish the existence of certain
subgames within our construction. We will first prove these on extractions and
projections, and then extend the lemmas from Section \ref{section:CoreLemmas}
to sets of extractions and projections in Section \ref{section:LB}.

\paragraph{Balancing Lemma.} The first lemma is the \nameref{lem:TupleSwap},
which allows us to relate unbalanced row extractions to balanced row
extractions according to the definition of \nameref{def:Equipartition}.

\begin{lemma}[Balancing Lemma]\label{lem:TupleSwap}
  For any $m \times n$ matrix $A$ and positive integer $\symTotGames$, let
  $R \subseteq \rangeintegers{m\symTotGames}$ and
  $C \subseteq \rangeintegers{n^\symTotGames}$ be selections of rows and
  columns of $\overOp{A}{\symTotGames}$. For any $0 \leq \numberT < m$ such
  that $\card{R}\geq \symTotGames \numberT$, there exist
  $S \subseteq \rangeintegers{m\symGamePartSize}$ and
  $D \subseteq \rangeintegers{n^{\symGamePartSize}}$ such that
  \[
    \symGamePartSize
    =
    \ceil{\symTotGames \left(1-\frac{1-\frac{|R|}{\symTotGames m}}{1-\frac{\numberT}{m}} \right)},
  \]
  the set $S$ is $m,\numberT,\symGamePartSize$-equipartitioned whenever
  $\symGamePartSize\geq 1$,
  \[
    \card{D} \geq \frac{\card{C}}{n^{\symTotGames-\symGamePartSize}},
  \]
  and
  \[
    \extractmatrix{\overOp{A}{\symGamePartSize}}{S}{D}
    \sqsubseteq
    \extractmatrix{\overOp{A}{\symTotGames}}{R}{C}.
  \]
\end{lemma}
\begin{proof}
  Let
  \[
    Q' =
    \left\{
      \indexOne \suchthat \indexOne \in \rangeintegers{\symTotGames},
      \card{\rangeintegers{m\indexOne,m(\indexOne+1)} \setminus R} \geq m-\numberT
    \right\}.
  \]
  Since the components in $Q'$ each miss at least $m-\numberT$ rows, we have
  \[
    m\symTotGames-\card{R} \geq \card{Q'}(m-\numberT).
  \]
  Therefore
  \[
    \symGamePartSize
    =
    \ceil{\frac{\card{R}-\symTotGames\numberT}{m-\numberT}}
    \leq
    \symTotGames-\card{Q'}.
  \]
  Choose any subset
  $Q \subseteq \rangeintegers{\symTotGames}\setminus Q'$ of cardinality
  $\symGamePartSize$, and let $S',D$ be the $Q$-projection of $R,C$.

  If $\symGamePartSize=0$, set $S=\varnothing$. Then $D=\{0\}$ by definition,
  and the bound on $\card{D}$ is immediate because $\card{C}\leq n^\symTotGames$.
  The subgame statement is then vacuous. So assume $\symGamePartSize\geq 1$.

  By construction of $Q$, every chosen component contains at least $\numberT$
  rows of $R$. Hence every component of $S'$ contains at least $\numberT$
  rows, and we may choose an $m,\numberT,\symGamePartSize$-equipartitioned
  subset $S \subseteq S'$.

  For the columns, each projected column in $D$ has at most
  $n^{\symTotGames-\symGamePartSize}$ preimages in $C$, since only the digits
  outside $Q$ may vary. Thus
  \[
    \card{D} \geq \frac{\card{C}}{n^{\symTotGames-\symGamePartSize}}.
  \]
  Since $S \subseteq S'$, Lemma~\ref{lem:projection} gives the desired subgame
  relation.
\end{proof}

\paragraph{Projection Lemmas.} Our next two lemmas establish useful properties
of projections. At a high level the \nameref{lem:NewTupleSet} and
\nameref{lem:NewTupleMax} tell us that if we have a game $\overOp{A}{\symTotGames}$
built by interlacing $\symTotGames$ copies of $A$ and we extract a set of
columns $C$ from the game, then it is possible to find subgames meeting
specific conditions which will be essential in our lower bound proof.

In order to show these lemmas, we will require a theorem about projections of
families of subsets:

\begin{theorem}[Product Theorem \citep{Chung1986}]
  \label{thm:product}
  Let $U$ be a finite set and let $A_1, \dots, A_{\symTotGames}$ be subsets of
  $U$ such that every element of $U$ is contained in at least $k$ of
  $A_1, \dots, A_{\symTotGames}$. Let $\mathbf{F}$ be a collection of subsets
  of $U$ and let $\mathbf{F}_i= \{F \cap A_i \suchthat F \in \mathbf{F}\}$ for
  $1 \leq i \leq {\symTotGames}$. Then we have
  $\card{\mathbf{F}}^{k} \leq \prod_{i=1}^{\symTotGames} \card{\mathbf{F}_i}$.
\end{theorem}

We use this theorem to obtain a lower bound on the size of specific families
of subsets. The following corollary will also be necessary for our last core
lemma:

\begin{corollary}\label{cor:product}
  Let $U$ be a finite set such that $U=[n]\times [\symTotGames]$ and let
  $\mathbf{F}$ be a collection of subsets of $U$. For any integer
  $1 \leq \symGamePartSize \leq \symTotGames$, there exists
  $\symTotGames'\subseteq [\symTotGames]$ with
  $\card{\symTotGames'}=\symGamePartSize$ such that
  \[
    |\mathbf{F}|^{\frac{\symGamePartSize}{\symTotGames}}
    \leq
    |\mathbf{F}_{\symTotGames'}|,
  \]
  where
  \[
    A_{\symTotGames'}=\{(\lambda,\gamma)\suchthat \lambda \in [n], \gamma \in \symTotGames'\}
    \qquad\text{and}\qquad
    \mathbf{F}_{\symTotGames'}=\{F\cap A_{\symTotGames'}\suchthat F\in \mathbf{F}\}.
  \]
\end{corollary}

\begin{proof}
  Let $q$ be the minimum positive integer such that $\symGamePartSize$ divides
  $\symTotGames q$, and let $r=\frac{\symTotGames q}{\symGamePartSize}$. For
  $0 \leq i < r$, define
  \[
    A_i=\left\{
      \left(k,\remainder{i\symGamePartSize+j}{\symTotGames}\right)
      \suchthat
      k\in [n],\ j\in [\symGamePartSize]
    \right\}.
  \]
  Every element of $U$ is contained exactly $q$ times. Defining
  $\mathbf{F}_i=\{F\cap A_i\suchthat F \in \mathbf{F}\}$, we can apply
  Theorem~\ref{thm:product} to get
  \[
    |\mathbf{F}|^q\leq \prod_{i=0}^{r-1}|\mathbf{F}_i|.
  \]
  Let $i_{\max}$ maximise $|\mathbf{F}_i|$. Then
  \[
    |\mathbf{F}|^q\leq |\mathbf{F}_{i_{\max}}|^r,
  \]
  and therefore
  \[
    |\mathbf{F}|^{\frac{\symGamePartSize}{\symTotGames}}
    =
    |\mathbf{F}|^{q/r}
    \leq
    |\mathbf{F}_{i_{\max}}|.
  \]
  Taking
  \[
    \symTotGames'=
    \left\{
      \remainder{i_{\max}\symGamePartSize+j}{\symTotGames}
      \suchthat
      j\in [\symGamePartSize]
    \right\}
  \]
  gives $A_{\symTotGames'}=A_{i_{\max}}$ and hence the result.
\end{proof}

The \nameref{thm:product} is used to prove the \nameref{lem:NewTupleSet}, and
Corollary \ref{cor:product} is used to prove the \nameref{lem:NewTupleMax}. We
state and prove these two remaining core lemmas below.

\begin{lemma}[Product of Projections Lemma]\label{lem:NewTupleSet}
  For any $m \times n$ matrix $A$ and positive integer $\symTotGames$, let
  $R \subseteq \rangeintegers{m\symTotGames}$ be an
  $m,\numberT,\symTotGames$-equipartitioned selection of rows of
  $\overOp{A}{\symTotGames}$ and let
  $C \subseteq \rangeintegers{n^\symTotGames}$ be a selection of columns of
  $\overOp{A}{\symTotGames}$. For any partition $R = R_1 \cup R_2$, we can
  write $\symTotGames = \symGamePartSize_1 + \symGamePartSize_2$ such that for
  each $i \in \{1,2\}$ there exist selections
  $S_i \subseteq \rangeintegers{m\symGamePartSize_i}$ and
  $D_i \subseteq \rangeintegers{n^{\symGamePartSize_i}}$ satisfying
  \[
    \card{D_1}\times\card{D_2} \geq \card{C},
  \]
  and, whenever $\symGamePartSize_i\geq 1$,
  \[
    S_i \text{ is $m,\ceil{\numberT/2},\symGamePartSize_i$-equipartitioned}
    \qquad\text{and}\qquad
    \extractmatrix{\overOp{A}{\symGamePartSize_i}}{S_i}{D_i}
    \sqsubseteq
    \extractmatrix{\overOp{A}{\symTotGames}}{R_i}{C}.
  \]
\end{lemma}
\begin{proof}
  Let
  \[
    Q_1 =
    \left\{
      \indexOne \suchthat \indexOne \in \rangeintegers{\symTotGames},
      \card{R_1 \cap \rangeintegers{m\indexOne, m(\indexOne+1)}} \geq \frac{\numberT}{2}
    \right\},
  \]
  and let $Q_2 = \rangeintegers{\symTotGames}\setminus Q_1$. Put
  $\symGamePartSize_i=\card{Q_i}$.

  For each $i \in \{1,2\}$, if $\symGamePartSize_i=0$, set
  $S_i=\varnothing$ and $D_i=\{0\}$. Otherwise let $S_i',D_i$ be the
  $Q_i$-projection of $R_i,C$.

  If $\symGamePartSize_1\geq 1$, then every component of $S_1'$ contains at
  least $\numberT/2$ rows, so we may choose an
  $m,\ceil{\numberT/2},\symGamePartSize_1$-equipartitioned subset
  $S_1 \subseteq S_1'$. Likewise, if $\symGamePartSize_2\geq 1$, then because
  each component of $R$ contains exactly $\numberT$ rows, every component of
  $S_2'$ contains more than $\numberT/2$ rows, so we may choose an
  $m,\ceil{\numberT/2},\symGamePartSize_2$-equipartitioned subset
  $S_2 \subseteq S_2'$.

  Whenever $\symGamePartSize_i\geq 1$, Lemma~\ref{lem:projection} gives
  \[
    \extractmatrix{\overOp{A}{\symGamePartSize_i}}{S_i}{D_i}
    \sqsubseteq
    \extractmatrix{\overOp{A}{\symTotGames}}{R_i}{C}.
  \]

  To bound the columns, define
  \[
    U = \{ (b,\indexOne) \suchthat b\in\rangeintegers{n}, \indexOne\in\rangeintegers{\symTotGames} \},
    \qquad
    A_i = \{ (b,\indexOne) \suchthat b\in\rangeintegers{n}, \indexOne\in Q_i\}.
  \]
  For each $c=\basedecomp{b_{\symTotGames-1},\dots,b_0}{n} \in C$, let
  \[
    f(c)=\{(b_\indexOne,\indexOne)\suchthat \indexOne \in \rangeintegers{\symTotGames}\},
  \]
  and let $\mathbf{F}=\{f(c)\suchthat c\in C\}$.

  For $\symGamePartSize_i\geq 1$, the family
  $\mathbf{F}_i=\{F\cap A_i\suchthat F\in \mathbf{F}\}$ is in bijection with
  $D_i$, so $\card{\mathbf{F}_i}=\card{D_i}$. For $\symGamePartSize_i=0$, we
  have $A_i=\varnothing$, hence $\mathbf{F}_i=\{\varnothing\}$, and also
  $\card{D_i}=1$. Thus in all cases $\card{\mathbf{F}_i}=\card{D_i}$, while
  $\card{\mathbf{F}}=\card{C}$.

  Since $Q_1 \sqcup Q_2 = \rangeintegers{\symTotGames}$, the sets $A_1,A_2$
  partition $U$. Applying Theorem~\ref{thm:product} with $k=1$ yields
  \[
    \card{C} = \card{\mathbf{F}} \leq \card{\mathbf{F}_1}\card{\mathbf{F}_2}
    = \card{D_1}\card{D_2},
  \]
  which is the desired bound.
\end{proof}

\begin{lemma}[Maximum Projection Lemma]\label{lem:NewTupleMax}
  For any $m \times n$ matrix $A$ and positive integer $\symTotGames$, let
  $R \subseteq \rangeintegers{m\symTotGames}$ be an
  $m,\numberT,\symTotGames$-equipartitioned selection of rows of
  $\overOp{A}{\symTotGames}$ and let
  $C \subseteq \rangeintegers{n^\symTotGames}$ be a selection of columns of
  $\overOp{A}{\symTotGames}$. For any integer
  $1 \leq \symGamePartSize < \symTotGames$, there exist
  $S \subseteq \rangeintegers{m\symGamePartSize}$ and
  $D \subseteq \rangeintegers{n^{\symGamePartSize}}$ such that $S$ is
  $m,\numberT,\symGamePartSize$-equipartitioned,
  \[
    \card{D} \geq \card{C}^{\frac{\symGamePartSize}{\symTotGames}},
  \]
  and
  \[
    \extractmatrix{\overOp{A}{\symGamePartSize}}{S}{D}
    \sqsubseteq
    \extractmatrix{\overOp{A}{\symTotGames}}{R}{C}.
  \]
\end{lemma}
\begin{proof}
  For each
  \[
    c=\basedecomp{b_{\symTotGames-1},\dots,b_0}{n}\in C,
  \]
  define
  \[
    f(c)=\{(b_\indexOne,\indexOne)\suchthat \indexOne\in\rangeintegers{\symTotGames}\},
    \qquad
    \mathbf{F}=\{f(c)\suchthat c\in C\}.
  \]
  Applying Corollary~\ref{cor:product} to $\mathbf{F}$ gives a set
  $Q\subseteq [\symTotGames]$ of cardinality $\symGamePartSize$ such that
  \[
    \card{\mathbf{F}}^{\frac{\symGamePartSize}{\symTotGames}}
    \leq
    \card{\mathbf{F}_Q},
  \]
  where
  \[
    A_Q=\{(\lambda,\gamma)\suchthat \lambda \in [n],\ \gamma \in Q\}
    \qquad\text{and}\qquad
    \mathbf{F}_Q=\{F\cap A_Q \suchthat F\in \mathbf{F}\}.
  \]

  Let $S,D$ be the $Q$-projection of $R,C$, and let
  \[
    0 \leq q_0 < q_1 < \dots < q_{\symGamePartSize-1} < \symTotGames
  \]
  be the sorted elements of $Q$.

  By construction,
  \[
    \mathbf{F}_Q
    =
    \left\{
      \{(b_{q_\indexOne},q_\indexOne)\suchthat \indexOne\in\rangeintegers{\symGamePartSize}\}
      \suchthat
      \basedecomp{b_{\symTotGames-1},\dots,b_0}{n}\in C
    \right\}.
  \]
  This set is in bijection with $D$ via the map
  \[
    \{(b_{q_\indexOne},q_\indexOne)\suchthat \indexOne\in\rangeintegers{\symGamePartSize}\}
    \longmapsto
    \basedecomp{b_{q_{\symGamePartSize-1}},\dots,b_{q_0}}{n}.
  \]
  Hence
  \[
    \card{\mathbf{F}_Q}=\card{D}.
  \]
  Likewise, the map $f$ is injective, so
  \[
    \card{\mathbf{F}}=\card{C}.
  \]
  Therefore Corollary~\ref{cor:product} yields
  \[
    \card{D} \geq \card{C}^{\frac{\symGamePartSize}{\symTotGames}}.
  \]

  Since $R$ is $m,\numberT,\symTotGames$-equipartitioned, every component of
  the $Q$-projection keeps exactly $\numberT$ rows, so $S$ is
  $m,\numberT,\symGamePartSize$-equipartitioned. Finally,
  Lemma~\ref{lem:projection} gives
  \[
    \extractmatrix{\overOp{A}{\symGamePartSize}}{S}{D}
    \sqsubseteq
    \extractmatrix{\overOp{A}{\symTotGames}}{R}{C}.
  \]
\end{proof}

We now have the fundamental tools required to relate interlaced games with
missing columns to smaller interlaced games with fewer missing columns. This
will serve as the backbone of our lower bound proof.
\section{Lower Bound Proof}\label{section:LB}

Using the lemmas from Section~\ref{section:CoreLemmas} we can relate games
built from extracting rows and columns from interlaced games to games built
from extracting rows and a smaller ratio of columns from larger interlaced
games. Establishing a lower bound will require an extension of these lemmas to
sets of games.

The first step in doing so is introducing notation to denote special sets of
games.

\begin{definition}[Bracket Notation]\label{def:bracket}
  Given an $m\times n$ matrix $M$, a positive integer $p$, and
  $0<x,y\leq 1$, set $\numberT=\ceil{mx}$ and define
  \[
    \bracket{M}{p}{x}{y}
    =
    \left\{
      \extractmatrix{\overOp{M}{p}}{R}{C}
      \;\middle|\;
      \begin{array}{l}
        R \subseteq \rangeintegers{mp} \text{ is $m,\numberT,p$-equipartitioned, and}\\
        |C|=\ceil{n^p y}
      \end{array}
    \right\}.
  \]
\end{definition}

The sets thus defined are constructed by interlacing a base matrix $A$
$\symTotGames$ times and taking all possible balanced extractions of rows and
columns with cardinality determined by $x$ and $y$ respectively. Part of the
significance of this set of matrices is that a partial protocol
(Definition~\ref{definition:PartialProtocol}) on $\overOp{A}{\symTotGames}$
with the column player communicating $-\ceil{\log y}$ bits will have at least
one of its leaves labelled with rows $X$ and columns $C$ such that
\[
  \extractmatrix{\overOp{A}{\symTotGames}}{X}{C}
\]
is a supergame of at least one matrix in the set
\[
  \bracket{A}{\symTotGames}{x}{y}.
\]
This allows us to bound the progress made by protocols in the best-case
scenario by lower bounding the communication complexity of our set of
matrices.

\begin{example}
  Given
  \[
    \overOp{A}{2}
    =
    \begin{bmatrix}
      1 & 1 & 0 & 0\\
      1 & 0 & 1 & 0
    \end{bmatrix}
  \]
  and setting $y=\frac34$, we have
  \[
  \bracket{A}{2}{1}{\frac{3}{4}} =
  \left\lbrace\left[\begin{array}{ccc}
        1 & 1 & 0 \\
        1 & 0 & 1
      \end{array}\right],
      \left[\begin{array}{ccc}
        1 & 1 & 0 \\
        1 & 0 & 0
      \end{array}\right],
      \left[\begin{array}{ccc}
        1 & 0 & 0 \\
        1 & 1 & 0
      \end{array}\right],
      \left[\begin{array}{ccc}
        1 & 0 & 0 \\
        0 & 1 & 0
      \end{array}\right]\right\rbrace.
  \]
\end{example}

The first argument in the bracket allows us to represent communication which
affected the internal rows of interlaced games instead of separating them.

\begin{example}
  Given
  \[
    I=
    \begin{bmatrix}
      1 & 0\\
      0 & 1
    \end{bmatrix}
  \]
  and $y=\frac12$, we have
  \[
    \overOp{I}{2}
    =
    \begin{bmatrix}
      1 & 1 & 0 & 0\\
      0 & 0 & 1 & 1\\
      1 & 0 & 1 & 0\\
      0 & 1 & 0 & 1
    \end{bmatrix}
  \]
  and
  \[
  \bracket{I}{2}{\frac{1}{2}}{1} =
  \left\lbrace
  \left[\begin{array}{cccc}
    1 & 1 & 0 & 0\\
    1 & 0 & 1 & 0
  \end{array}\right],
  \left[\begin{array}{cccc}
    0 & 0 & 1 & 1\\
    1 & 0 & 1 & 0
  \end{array}\right],
  \left[\begin{array}{cccc}
    1 & 1 & 0 & 0\\
    0 & 1 & 0 & 1
  \end{array}\right],
  \left[\begin{array}{cccc}
    0 & 0 & 1 & 1\\
    0 & 1 & 0 & 1
  \end{array}\right]
  \right\rbrace.
 \]
\end{example}

One can think of the bracket notation as a tool to deal with protocols which
do not act in the manner prescribed by the construction: both players taking
turns separating interlaced games.

In the rest of the paper, when we end up with non-integer values where integer
values are expected, we use the ceiling function by default.

\subsection{Core Lemmas}\label{section:CoreLemmas2}

We start this section with the \nameref{lem:mono}, after which we present the
extensions of the core lemmas from Section~\ref{section:CoreLemmas} to sets of
matrices delimited by the \nameref{def:bracket}.

\begin{lemma}[Monotonicity Lemma]\label{lem:mono}
  Given a $m$ by $n$ matrix $M$, for any integers $1 \leq p' \leq p$,
  $0 < x' \leq x \leq 1$, and $0 < y' \leq y \leq 1$, we have
  \[
    \comp{\bracket{M}{p'}{x'}{y'}}
    \leq
    \comp{\bracket{M}{p}{x}{y}}.
  \]
\end{lemma}
\begin{proof}
  We prove the three monotonicity directions separately and then compose them.

  First, assume $1 \leq p' \leq p$. Let
  \[
    g=\extractmatrix{\overOp{M}{p}}{R}{C} \in \bracket{M}{p}{x'}{y'}.
  \]
  Write
  \[
    \numberT=\ceil{mx'}
    \qquad\text{and}\qquad
    q=p-p'.
  \]
  Let $Q=\rangeintegers{p'}$ be the retained coordinates. Partition the columns
  of $C$ according to their discarded coordinates, namely the base-$n$ digits
  in positions $p',\dots,p-1$. There are exactly $n^q$ such fibres, so one of
  them, call it $C^\star$, satisfies
  \[
    |C^\star|
    \geq
    \left\lceil \frac{|C|}{n^q} \right\rceil
    \geq
    \ceil{n^{p'}y'},
  \]
  because $|C|=\ceil{n^p y'}$.
  Choose a subset
  \[
    C' \subseteq C^\star
    \qquad\text{with}\qquad
    |C'|=\ceil{n^{p'}y'}.
  \]
  Let $S,D$ be the $Q$-projection of $R,C'$. Since $R$ is
  $m,\numberT,p$-equipartitioned, the projected row set $S$ is
  $m,\numberT,p'$-equipartitioned. Because all columns of $C'$ agree on the
  discarded coordinates, the $Q$-projection is injective on $C'$, so
  \[
    |D|=|C'|=\ceil{n^{p'}y'}.
  \]
  Lemma~\ref{lem:projection} therefore gives
  \[
    g'=\extractmatrix{\overOp{M}{p'}}{S}{D}\sqsubseteq
    \extractmatrix{\overOp{M}{p}}{R}{C'}
    \sqsubseteq
    g,
  \]
  and by construction
  \[
    g' \in \bracket{M}{p'}{x'}{y'}.
  \]
  Hence
  \[
    \comp{\bracket{M}{p'}{x'}{y'}}
    \leq
    \comp{\bracket{M}{p}{x'}{y'}}.
  \]

  Next, assume $x' \leq x$. Let
  \[
    g=\extractmatrix{\overOp{M}{p}}{R}{C} \in \bracket{M}{p}{x}{y}
  \]
  and write
  \[
    \numberT=\ceil{mx}
    \qquad\text{and}\qquad
    \numberT'=\ceil{mx'}.
  \]
  Since $\numberT' \leq \numberT$, in each component of the row set $R$ choose
  exactly $\numberT'$ rows and let $R'$ be their union. Then $R' \subseteq R$ is
  $m,\numberT',p$-equipartitioned, so
  \[
    \extractmatrix{\overOp{M}{p}}{R'}{C}
    \in
    \bracket{M}{p}{x'}{y}
    \qquad\text{and}\qquad
    \extractmatrix{\overOp{M}{p}}{R'}{C}\sqsubseteq g.
  \]
  Therefore
  \[
    \comp{\bracket{M}{p}{x'}{y}}
    \leq
    \comp{\bracket{M}{p}{x}{y}}.
  \]

  Finally, assume $y' \leq y$. Let
  \[
    g=\extractmatrix{\overOp{M}{p}}{R}{C} \in \bracket{M}{p}{x}{y}.
  \]
  Since
  \[
    |C|=\ceil{n^p y}\geq \ceil{n^p y'},
  \]
  choose a subset
  \[
    C' \subseteq C
    \qquad\text{with}\qquad
    |C'|=\ceil{n^p y'}.
  \]
  Then
  \[
    \extractmatrix{\overOp{M}{p}}{R}{C'}
    \in
    \bracket{M}{p}{x}{y'}
    \qquad\text{and}\qquad
    \extractmatrix{\overOp{M}{p}}{R}{C'}\sqsubseteq g,
  \]
  so
  \[
    \comp{\bracket{M}{p}{x}{y'}}
    \leq
    \comp{\bracket{M}{p}{x}{y}}.
  \]

  Applying these three monotonicity steps in succession proves the lemma.
\end{proof}

We now extend the \nameref{lem:NewTupleSet} to sets of matrices denoted by the
\nameref{def:bracket}.

\begin{lemma}[Extended Product of Projection Lemma]\label{lem:tuple}
  Given an $m \times n$ matrix
  \[
    g=\extractmatrix{\overOp{M}{\symTotGames}}{R}{C}
    \in
    \bracket{M}{\symTotGames}{x}{y}
  \]
  with $R$ the rows left in $g$ and $C$ the columns, and given a partition
  $R=R_1\cup R_2$ of the rows of $g$, there exist
  $\symGamePartSize_1,\symGamePartSize_2,\yval{1}, \yval{2}$, $S_1,S_2$, and
  $D_1,D_2$ such that
  \[
    \yval{1}\times\yval{2}\geq y
    \qquad\text{and}\qquad
    \symGamePartSize_1+\symGamePartSize_2=\symTotGames.
  \]
  For each $i\in\{1,2\}$, we have
  \[
    \yval{i}=\frac{\card{D_i}}{n^{\symGamePartSize_i}}.
  \]
  In particular, if $\symGamePartSize_i=0$, then $D_i=\{0\}$ and
  $\yval{i}=1$. Moreover, for each $i\in\{1,2\}$ with
  $\symGamePartSize_i\geq 1$, we have
  \[
    \extractmatrix{\overOp{M}{\symGamePartSize_i}}{S_i}{D_i}\sqsubseteq
    \extractmatrix{\overOp{M}{\symTotGames}}{R_i}{C}
  \]
  and
  \[
    \extractmatrix{\overOp{M}{\symGamePartSize_i}}{S_i}{D_i}
    \in
    \bracket{M}{\symGamePartSize_i}{\frac{x}{2}}{\yval{i}}.
  \]
\end{lemma}

\begin{proof}
  The projected child data follow from Lemma~\ref{lem:NewTupleSet}. Since
  $g \in \bracket{M}{\symTotGames}{x}{y}$, the definition of the
  \nameref{def:bracket} gives that $R$ is
  $m,\numberT,\symTotGames$-equipartitioned with $\numberT=\ceil{xm}$.

  For each $i\in\{1,2\}$ with $\symGamePartSize_i\geq 1$,
  Lemma~\ref{lem:NewTupleSet} gives that $S_i$ is
  $m,\ceil{\numberT/2},\symGamePartSize_i$-equipartitioned. We also have that
  $D_1$ and $D_2$ satisfy $\card{D_1}\times \card{D_2} \geq |C|$. By
  construction,
  \[
    \yval{1}=\frac{\card{D_1}}{n^{\symGamePartSize_1}},
    \qquad
    \yval{2}=\frac{\card{D_2}}{n^{\symGamePartSize_2}},
    \qquad\text{and}\qquad
    \frac{|C|}{n^{\symTotGames}}\geq y,
  \]
  therefore
  \[
    \yval{1}\times \yval{2}
    =
    \frac{\card{D_1}\times\card{D_2}}{n^{\symTotGames}}
    \geq
    \frac{|C|}{n^{\symTotGames}}
    \geq
    y.
  \]
  Since
  \[
    \ceil{\frac{\numberT}{2}}=\ceil{\frac{\ceil{mx}}{2}}=\ceil{\frac{mx}{2}},
  \]
  the positive child row selections are exactly the row sets required by the
  bracket notation at parameter $\frac{x}{2}$. Therefore, for each
  $i\in\{1,2\}$ with $\symGamePartSize_i\geq 1$,
  \[
    \extractmatrix{\overOp{M}{\symGamePartSize_i}}{S_i}{D_i}
    \in
    \bracket{M}{\symGamePartSize_i}{\frac{x}{2}}{\yval{i}}.
  \]
  This proves the lemma.
\end{proof}

Likewise with the \nameref{lem:NewTupleMax}.

\begin{lemma}[Extended Maximum Projection Lemma]\label{lem:TupleCorollary}
  For all $1\leq \symGamePartSize\leq \symTotGames$, we have
  \[
    \comp{\bracket{M}{\symTotGames}{x}{y}}
    \geq
    \comp{\bracket{M}{\symGamePartSize}{x}{y^{\frac{\symGamePartSize}{\symTotGames}}}}.
  \]
\end{lemma}
\begin{proof}
  If $\symGamePartSize=\symTotGames$, then
  \[
    \bracket{M}{\symGamePartSize}{x}{y^{\frac{\symGamePartSize}{\symTotGames}}}
    =
    \bracket{M}{\symTotGames}{x}{y},
  \]
  so the claim is immediate. Assume henceforth that
  \[
    1\leq \symGamePartSize<\symTotGames.
  \]

  Given an $m$ by $n$ matrix
  \[
    g=\extractmatrix{\overOp{M}{\symTotGames}}{R}{C}
    \in
    \bracket{M}{\symTotGames}{x}{y}
  \]
  with $R$ being $m,\numberT,\symTotGames$-equipartitioned,
  Lemma~\ref{lem:NewTupleMax} gives us for any $\symGamePartSize$ a matrix
  \[
    g'=\extractmatrix{\overOp{M}{\symGamePartSize}}{S}{D}
  \]
  with $S$ being $m,\numberT,\symGamePartSize$-equipartitioned and
  \[
    |D|\geq|C|^{\frac{\symGamePartSize}{\symTotGames}}
  \]
  such that $g' \sqsubseteq g$.

  Therefore
  \[
    \frac{|D|}{n^{\symGamePartSize}}
    \geq
    \left(\frac{|C|}{n^{\symTotGames}}\right)^{\frac{\symGamePartSize}{\symTotGames}}
    \geq
    y^{\frac{\symGamePartSize}{\symTotGames}},
  \]
  where the last inequality uses the bracket definition
  $|C|=\ceil{n^{\symTotGames}y}$.

  Since $|D|$ is an integer, we can choose a subset
  \[
    D' \subseteq D
    \qquad\text{with}\qquad
    |D'|=\ceil{n^{\symGamePartSize}y^{\frac{\symGamePartSize}{\symTotGames}}}.
  \]
  Then
  \[
    g''=\extractmatrix{\overOp{M}{\symGamePartSize}}{S}{D'}
    \in
    \bracket{M}{\symGamePartSize}{x}{y^{\frac{\symGamePartSize}{\symTotGames}}},
  \]
  and $g''\sqsubseteq g'\sqsubseteq g$. Therefore every matrix in
  $\bracket{M}{\symTotGames}{x}{y}$ has a subgame in
  $\bracket{M}{\symGamePartSize}{x}{y^{\frac{\symGamePartSize}{\symTotGames}}}$,
  giving the desired complexity bound.
\end{proof}

And finally we extend the \nameref{lem:TupleSwap}.

\begin{lemma}[Extended Balancing Lemma]\label{lem:swap}
  For any $m \times n$ matrix $M$, integer $\symTotGames$, $1<\alpha$,
  $0 < x < 1$, $0 < \alpha x \leq 1$, and $0 < y \leq 1$, if
  $mx \in \mathbb{N}$ and
  \[
    p^\star
    =
    \left\lfloor \frac{\symTotGames(\alpha-1)x}{1-x} \right\rfloor
    \geq 1,
  \]
  then
  \[
    \comp{\bracket{\overOp{M}{\symTotGames}}{1}{\alpha x}{y}}
    \geq
    \comp{\bracket{M}{p^\star}{x}{y}}.
  \]
\end{lemma}
\begin{proof}
  Let
  \[
    \numberT = mx.
  \]
  Since $x<1$, we have $\numberT < m$, so Lemma~\ref{lem:TupleSwap} applies.

  Given $g \in \bracket{\overOp{M}{\symTotGames}}{1}{\alpha x}{y}$, write
  \[
    g=\extractmatrix{\overOp{M}{\symTotGames}}{R}{C}.
  \]
  Since
  \[
    |R|=\ceil{m\symTotGames \alpha x}
    \qquad\text{and}\qquad
    \alpha > 1,
  \]
  we have
  \[
    |R| \geq m\symTotGames x = \symTotGames\numberT.
  \]
  Applying the \nameref{lem:TupleSwap} with $T=\numberT$ gives sets $S,D$ and
  an integer
  \[
    \symGamePartSize
    =
    \left\lceil
      \symTotGames \left(1-\frac{1-\frac{|R|}{\symTotGames m}}{1-\frac{\numberT}{m}} \right)
    \right\rceil
  \]
  such that
  \[
    |D|\geq\frac{|C|}{n^{\symTotGames-\symGamePartSize}}.
  \]
  By the definition of the bracket notation,
  \[
    \frac{|R|}{m\symTotGames}
    \geq
    \alpha x.
  \]
  Therefore
  \[
    \symGamePartSize
    \geq
    \left\lceil
      \symTotGames \left(\frac{\alpha x-x}{1-x} \right)
    \right\rceil
    =
    \left\lceil
      \frac{\symTotGames(\alpha-1)x}{1-x}
    \right\rceil.
  \]
  Hence $\symGamePartSize\geq p^\star\geq 1$, so the positive-size clause of
  Lemma~\ref{lem:TupleSwap} gives
  \[
    g'=\extractmatrix{\overOp{M}{\symGamePartSize}}{S}{D}
    \sqsubseteq
    \extractmatrix{\overOp{M}{\symTotGames}}{R}{C},
  \]
  where $S$ is $m,\numberT,\symGamePartSize$-equipartitioned. Since
  $\numberT = mx$, this satisfies the row condition in
  $\bracket{M}{\symGamePartSize}{x}{y}$. Also,
  \[
    \frac{|D|}{n^{\symGamePartSize}}
    \geq
    \frac{|C|}{n^{\symTotGames}}
    \geq
    y.
  \]
  Since $|D|$ is an integer and $\frac{|D|}{n^{\symGamePartSize}}\geq y$, we can
  choose a subset
  \[
    D' \subseteq D
    \qquad\text{with}\qquad
    |D'|=\ceil{n^{\symGamePartSize}y}.
  \]
  Then
  \[
    g''=\extractmatrix{\overOp{M}{\symGamePartSize}}{S}{D'}
    \in
    \bracket{M}{\symGamePartSize}{x}{y},
  \]
  and $g''\sqsubseteq g'\sqsubseteq g$.
  Since $p^\star \leq \symGamePartSize$, the \nameref{lem:mono} implies
  \[
    \comp{\bracket{M}{\symGamePartSize}{x}{y}}
    \geq
    \comp{\bracket{M}{p^\star}{x}{y}}.
  \]
  Combining this with the previous subgame extraction from $g''$ yields the claim.
\end{proof}

Since our construction transposes the obtained matrix at each round before
continuing to iterate the \nameref{def:overOperation}, we will require the
following.

\begin{lemma}\label{lem:transposeComp}
  \[
    \comp{\bracket{M}{1}{x}{y}}
    =
    \comp{\bracket{\transpose{M}}{1}{y}{x}}.
  \]
\end{lemma}
\begin{proof}
  Since $\overOp{M}{1}=M$, Definition~\ref{def:bracket} shows that
  $\bracket{M}{1}{x}{y}$ is exactly the set of submatrices
  \[
    \extractmatrix{M}{R}{C}
  \]
  with
  \[
    |R|=\ceil{mx}
    \qquad\text{and}\qquad
    |C|=\ceil{ny}.
  \]
  Likewise, $\bracket{\transpose{M}}{1}{y}{x}$ is exactly the set of
  submatrices
  \[
    \extractmatrix{\transpose{M}}{C}{R}
  \]
  with the same cardinality conditions. Transposition therefore gives a
  bijection
  \[
    \extractmatrix{M}{R}{C}
    \longleftrightarrow
    \extractmatrix{\transpose{M}}{C}{R}
  \]
  between the two bracket families. Deterministic communication complexity is
  invariant under transposition, since any protocol for one matrix becomes a
  protocol for the transpose after swapping the two players. Taking the
  minimum complexity over the two corresponding bracket families proves the
  claim.
\end{proof}

Our final lemma in this section is Lemma~\ref{lem:old-partition}. This is the
workhorse of our analysis in that it is the central step in the induction
which follows. Indeed, it lower bounds the complexity of a game in terms of
the complexity of the worst-case game obtained after one step of the best-case
protocol.

\begin{lemma}\label{lem:old-partition}
  For any matrix $M$, any $0 < x \leq \frac{1}{2}$, any $0 < y \leq 1$, and
  any $\delta \in \{0,1\}$, if
  \[
    \comp{\bracket{M}{2\symTotGames+\delta}{2x}{y}} \geq 1,
  \]
  then
  \begin{equation}
    \begin{aligned}
      \comp{\bracket{M}{2\symTotGames+\delta}{2x}{y}}
      \geq
      1 + \min\Biggl(
        &\comp{\bracket{M}{2\symTotGames+\delta}{x}{y}},\\
        &\min_{\substack{\symGamePartSize\in \rangeintegers{\symTotGames}\\a\in [0,1]}}
        \max\Bigl(
          \comp{\bracket{M}{\symTotGames+\delta+\symGamePartSize}{x}{y^a}},
          \comp{\bracket{M}{\symTotGames-\symGamePartSize}{x}{y^{1-a}}}
        \Bigr),\\
        &\comp{\bracket{M}{2\symTotGames+\delta}{2x}{\frac{y}{2}}}
      \Biggr).
    \end{aligned}
    \label{eq:old-partition}
  \end{equation}
\end{lemma}
\begin{proof}
  Fix
  \[
    g=\extractmatrix{\overOp{M}{2\symTotGames+\delta}}{R}{C}
    \in
    \bracket{M}{2\symTotGames+\delta}{2x}{y}.
  \]
  Any protocol for $g$ of positive depth begins with either a row-player bit or a
  column-player bit.

  If the row player emits the first bit, then the root partitions the rows as
  $R=R_1 \sqcup R_2$. Apply Lemma~\ref{lem:tuple} to
  \[
    g \in \bracket{M}{2\symTotGames+\delta}{2x}{y}
  \]
  with that row partition. We obtain integers $p_1,p_2$, densities $y_1,y_2$,
  and sets $S_1,S_2,D_1,D_2$ with
  \[
    p_1+p_2=2\symTotGames+\delta
    \qquad\text{and}\qquad
    y_1y_2 \ge y.
  \]

  Relabel if needed so that $p_1 \ge p_2$. Then $p_1 \ge 1$.
  Lemma~\ref{lem:tuple} therefore gives a subgame
  \[
    g_1=\extractmatrix{\overOp{M}{p_1}}{S_1}{D_1}
    \in
    \bracket{M}{p_1}{x}{y_1}
  \]
  with
  \[
    g_1 \sqsubseteq \extractmatrix{\overOp{M}{2\symTotGames+\delta}}{R_1}{C}
    \sqsubseteq g.
  \]

  If $p_2=0$, then $p_1=2\symTotGames+\delta$, and Lemma~\ref{lem:tuple} gives
  \[
    y_2=1.
  \]
  Hence $y_1 \ge y$, and monotonicity gives
  \[
    \comp{g}
    \ge
    1+\comp{\bracket{M}{2\symTotGames+\delta}{x}{y}}.
  \]

  Assume henceforth that $p_2 \ge 1$. Lemma~\ref{lem:tuple} then also gives
  \[
    g_2=\extractmatrix{\overOp{M}{p_2}}{S_2}{D_2}
    \in
    \bracket{M}{p_2}{x}{y_2}
  \]
  with
  \[
    g_2 \sqsubseteq \extractmatrix{\overOp{M}{2\symTotGames+\delta}}{R_2}{C}
    \sqsubseteq g.
  \]

  Then
  \[
    p_1=\symTotGames+\delta+\symGamePartSize,
    \qquad
    p_2=\symTotGames-\symGamePartSize
  \]
  for some $\symGamePartSize\in\rangeintegers{\symTotGames}$.

  If $y=1$, then necessarily $y_1=y_2=1$, so we may choose any $a\in[0,1]$.
  Otherwise $0<y<1$, and since Lemma~\ref{lem:tuple} gives
  \[
    y_2=\frac{|D_2|}{n^{p_2}}\le 1,
  \]
  we have $y_1 \ge y$ from $y_1y_2 \ge y$. Define
  \[
    a=\frac{\log y_1}{\log y}\in[0,1].
  \]
  Then $y_1=y^a$, and from $y_1y_2 \ge y$ we get
  \[
    y_2 \ge \frac{y}{y_1} = y^{1-a}.
  \]
  By monotonicity,
  \[
    \comp{\bracket{M}{p_1}{x}{y_1}}
    \ge
    \comp{\bracket{M}{p_1}{x}{y^a}},
    \qquad
    \comp{\bracket{M}{p_2}{x}{y_2}}
    \ge
    \comp{\bracket{M}{p_2}{x}{y^{1-a}}}.
  \]
  Any protocol for $g$ whose first bit is sent by the row player induces
  protocols of depth one less on the two child subgames. Therefore
  \[
    \comp{g}
    \ge
    1+\max\left(
      \comp{\bracket{M}{p_1}{x}{y^a}},
      \comp{\bracket{M}{p_2}{x}{y^{1-a}}}
    \right).
  \]

  If the column player emits a bit first, then the columns are partitioned into
  two subgames. One of the two children keeps at least half of the columns, and
  therefore contains a matrix from
  \[
    \bracket{M}{2\symTotGames+\delta}{2x}{\frac{y}{2}}
  \]
  as a subgame. By monotonicity, this gives
  \[
    \comp{g}
    \ge
    1+\comp{\bracket{M}{2\symTotGames+\delta}{2x}{\frac{y}{2}}}.
  \]

  Since the choice of $g$ was arbitrary, taking the minimum over $g$ and over the
  possible first moves yields the claimed bound.
\end{proof}

\subsection{Inductive Proof of the Lower Bound}\label{sec:inductiveLB}

We now prove the lower bound. The argument isolates general conditions under
which the \nameref{def:overOperation} increases the complexity of a game. If
one starts with a game in which communication from the column player has
little effect until the row player has partitioned the game appropriately,
then after interlacing and transposing the same phenomenon reappears with the
roles of the two players reversed.

As in Sections~\ref{section:CoreLemmas} and \ref{section:CoreLemmas2}, most of
the statements are formulated for general matrices and therefore describe the
behaviour of the \nameref{def:overOperation} rather than of the specific
\nameref{definition:AlternatingGame}. Only at the end do we specialise to the
alternating family and derive the lower bound for $\phi_i$.

\paragraph{Interlacing Once.}
First, we analyse the effect of iterating the \nameref{def:overOperation}. The
local partition argument links the complexity of sets obtained by extracting
rows and columns from $\overOp{M}{p}$ to those obtained from $\overOp{M}{2p}$.
When this is iterated, a single column-density parameter is no longer enough:
the argument naturally tracks the three neighbouring densities
$y$, $\frac{y}{2}$, and $\frac{y}{4}$. We therefore package these three lower
bounds into the quantity
\[
  \Lambda_M(p,x,y)
  :=
  \min_{0 \leq j < 3}
  \left(
    j + \comp{\bracket{M}{p}{x}{\frac{y}{2^j}}}
  \right).
\]

The next lemma is the bridge that governs repeated interlacing.

\begin{lemma}[Iterated Partition Lemma]\label{lem:new-partition}
  Let $2<\rho$, and set
  \[
    \beta:=\frac{\rho-1}{\rho-2}.
  \]
  For any matrix $M$, non-negative integers $0\le s\le k$, integer $p\ge 1$,
  and reals $0<x\le 2^{-k}$ and $0<y\le 1$, if
  \[
    \comp{\bracket{M}{p}{x}{y/4}}\ge 1
    \qquad\text{and}\qquad
    (\rho-1)^k\le \rho^{k-s},
  \]
  then
  \begin{align}
    \Lambda_M\!\left(
      \left\lfloor 2^k\beta^s p\right\rfloor,
      2^k x,
      y^{\rho^s}
    \right)
    &\ge
    k+\Lambda_M(p,x,y), \label{eq:iterated-partition-bundled}\\[1ex]
    \comp{\bracket{M}{\left\lfloor 2^k\beta^s p\right\rfloor}{2^k x}{y^{\rho^s}}}
    &\ge
    k+\Lambda_M(p,x,y). \label{eq:iterated-partition-scalar}
  \end{align}
\end{lemma}

The first inequality is the substantive one, and the second is the form used
later. The detailed proof is deferred to
Appendix~\ref{appendix:lower-bound-details}. The point is that the local
partition argument can be iterated while keeping track of the three
neighbouring densities above.

\begin{corollary}\label{cor:iterated-partition-seed}
  Under the hypotheses of Lemma~\ref{lem:new-partition}, suppose in addition
  that for some real $H$,
  \[
    \comp{\bracket{M}{p}{x}{y}} \ge H,
    \qquad
    \comp{\bracket{M}{p}{x}{y/2}} \ge H-1,
    \qquad
    \comp{\bracket{M}{p}{x}{y/4}} \ge H-2.
  \]
  Then
  \[
    \comp{\bracket{M}{\left\lfloor 2^k\beta^s p\right\rfloor}{2^k x}{y^{\rho^s}}}
    \ge
    k+H.
  \]
\end{corollary}
\begin{proof}
  The three displayed hypotheses are exactly the statement
  \[
    \Lambda_M(p,x,y)\ge H.
  \]
  Apply \eqref{eq:iterated-partition-scalar}.
\end{proof}

To use the iterated partition lemma in the alternating construction, it
remains to pass from the seed at density $2^{-1}$ to the density $2^{-3/8}$
that appears after one round of alternation.

\begin{lemma}\label{lem:seed-collapse}
  Let $M$ be a matrix, let $\frac{3}{8}<a$, and let $k\ge 5$ be an integer.
  If
  \[
    \comp{\bracket{M}{1}{2^{-k-a}}{2^{-3}}}\ge 1,
  \]
  then
  \[
    \Lambda_M(6,2^{3-k-a},2^{-1})
    \geq
    3 + \Lambda_M(1,2^{-k-a},2^{-3/8}).
  \]
\end{lemma}
\begin{proof}
  Monotonicity gives
  \[
    \comp{\bracket{M}{6}{2^{3-k-a}}{2^{-3}}}\ge 1,
    \qquad
    \comp{\bracket{M}{2}{2^{2-k-a}}{2^{-11/4}}}\ge 1,
  \]
  \[
    \comp{\bracket{M}{2}{2^{1-k-a}}{2^{-11/4}}}\ge 1,
    \qquad
    \comp{\bracket{M}{3}{2^{2-k-a}}{2^{-11/4}}}\ge 1.
  \]
  Therefore
  \[
    \begin{aligned}
      \Lambda_M(6,2^{3-k-a},2^{-1})
      &\geq 1 + \min\Bigl(
        \Lambda_M(3,2^{2-k-a},2^{-3/4}),
        \Lambda_M(3,2^{2-k-a},2^{-1/4})
      \Bigr) \\
      &\geq 1 + \Lambda_M(3,2^{2-k-a},2^{-3/4}) \\
      &\geq 2 + \Lambda_M(2,2^{1-k-a},2^{-3/4}) \\
      &\geq 3 + \Lambda_M(1,2^{-k-a},2^{-3/8}).
    \end{aligned}
  \]
  Here the first and last steps use Lemma~\ref{lem:lambda-col-step} with
  $\tau=\frac13$ and $\tau=1$ respectively, the second is monotonicity, and
  the third uses Lemma~\ref{lem:lambda-row-step}.
\end{proof}

\paragraph{One Round of Alternation.}
For each $i$ in the \nameref{definition:AlternatingGame} definition, we first
interlace $\Phi_i$ with itself a given number of times and then transpose the
result to obtain $\Phi_{i+1}$. The work done so far tracks the conditions on
row and column extractions when repeating the \nameref{def:overOperation}. The
\nameref{def:bracket} treats rows and columns differently, since only the row
extraction is required to satisfy the \nameref{def:Equipartition} condition.
Lemma~\ref{lem:swap} is what allows us to pass from the iterated partition
lemma to a statement that remains useful after taking the transpose.

\begin{theorem}\label{thm:induction-k}
  For any $\frac{3}{8} < a$ and integer $5 \leq k$ such that
  \[
    \sqrt{k+a} - 1 \leq (k+a)^{\frac{1}{2}-\frac{1}{k-3}},
  \]
  set
  \[
    Q_k
    :=
    \left\lceil
      2^k\times\frac{3\times 2^a-3}{2^{\frac{13}{8}}\times 2^{a}-4}
      \times
      \left(\frac{\sqrt{k+a}-1}{\sqrt{k+a}-2}\right)^2
    \right\rceil.
  \]
  For any matrix $M$ with $m2^{-a}\in\mathbb{N}$, where $m$ is the number of
  rows of $M$, if
  \[
    \comp{\bracket{M}{1}{2^{-k-a}}{2^{-3}}} \geq 1,
  \]
  then
  \[
    \comp{\bracket{\overOp{M}{Q_k}}{1}{2^{-3/8}}{2^{-k-a}}}
    \geq
    k + \Lambda_M(1,2^{-k-a},2^{-3/8}).
  \]
\end{theorem}
\begin{proof}
  Let
  \[
    \rho = \sqrt{k+a},
    \qquad
    \alpha = 2^{a-\frac{3}{8}},
    \qquad
    x_0 = 2^{-k-a},
  \]
  and set
  \[
    P_k
    :=
    \left\lfloor
      Q_k\times\frac{(\alpha-1)2^{-a}}{1-2^{-a}}
    \right\rfloor.
  \]
  By the definition of $Q_k$,
  \[
    P_k
    \geq
    \left\lfloor
      6 \times 2^{k-3}\left(\frac{\rho-1}{\rho-2}\right)^2
    \right\rfloor.
  \]
  Hence
  \begin{align*}
    \comp{\bracket{\overOp{M}{Q_k}}{1}{2^{-3/8}}{x_0}}
    &\geq \comp{\bracket{M}{P_k}{2^{-a}}{x_0}}
    & \text{Lemma~\ref{lem:swap}} \\
    &\geq \comp{\bracket{M}{
      \left\lfloor 6 \times 2^{k-3}\left(\frac{\rho-1}{\rho-2}\right)^2 \right\rfloor
    }{2^{-a}}{x_0}}
    & \text{Lemma~\ref{lem:mono}}.
  \end{align*}

  Apply Corollary~\ref{cor:iterated-partition-seed} with
  \[
    \tilde{k}=k-3,
    \qquad
    s=2,
    \qquad
    p=6,
    \qquad
    x=2^{3-k-a},
    \qquad
    y=2^{-1}.
  \]
  Its numerical side condition is exactly
  \[
    (\rho-1)^{k-3}\le \rho^{k-5},
  \]
  which is the hypothesis of the theorem, and its weak side condition
  \[
    \comp{\bracket{M}{6}{2^{3-k-a}}{2^{-3}}}\ge 1
  \]
  follows from
  \[
    \comp{\bracket{M}{1}{2^{-k-a}}{2^{-3}}}\ge 1
  \]
  by monotonicity. Therefore
  \[
    \comp{\bracket{M}{
      \left\lfloor 6 \times 2^{k-3}\left(\frac{\rho-1}{\rho-2}\right)^2 \right\rfloor
    }{2^{-a}}{x_0}}
    \geq
    k-3 + \Lambda_M(6,2^{3-k-a},2^{-1}).
  \]
  Lemma~\ref{lem:seed-collapse} now gives
  \[
    \Lambda_M(6,2^{3-k-a},2^{-1})
    \geq
    3 + \Lambda_M(1,2^{-k-a},2^{-3/8}),
  \]
  so the theorem follows.
\end{proof}

From this point onward, we specialise
Definition~\ref{definition:AlternatingGame} by fixing
\[
  B := Q := 255\cdot 2^{k-8},
\]
with $k=10000$ and $a=10$. Thus, for the remainder of the paper,
\[
  \phi_{i+1}=\transpose{\overOp{\phi_i}{Q}}
  \qquad\text{for all } i\ge 0.
\]

\begin{lemma}\label{lem:phi-row-divisibility}
  For $k=10000$, $a=10$, and any $i \geq 1$, both the number of rows and the
  number of columns of $\phi_i$ are divisible by $2^{a+2}$.
\end{lemma}
\begin{proof}
  Write
  \[
    Q = 255\cdot 2^{k-8}.
  \]
  Since $k=10000$, the number $Q$ is divisible by $2^{a+2}=2^{12}$.

  For $i=1$, the matrix
  \[
    \phi_1 = \transpose{\overOp{\phi_0}{Q}}
  \]
  has $2^Q$ rows and $Q$ columns, so both dimensions are divisible by
  $2^{a+2}$.

  For the induction step, suppose $\phi_i$ has both dimensions divisible by
  $2^{a+2}$. If $\phi_i$ has $r_i$ rows and $c_i$ columns, then
  \[
    \phi_{i+1} = \transpose{\overOp{\phi_i}{Q}}
  \]
  has
  \[
    c_i^Q \text{ rows}
    \qquad\text{and}\qquad
    Qr_i \text{ columns.}
  \]
  Since $c_i$ and $Q$ are both divisible by $2^{a+2}$, so are $c_i^Q$ and
  $Qr_i$. This completes the induction.
\end{proof}

To specialise the argument to the alternating family, we also record the
numerical estimates for the fixed values $k=10000$ and $a=10$.

\begin{lemma}\label{lem:shifted-rung-arithmetic}
  For $k=10000$ and $a=10$, write
  \[
    \rho = \sqrt{k+a},
    \qquad
    \beta = \frac{\rho-1}{\rho-2},
    \qquad
    Q = 255\cdot 2^{k-8},
  \]
  and for each $j\in\{0,1,2\}$ set
  \[
    k_j = k-j,
    \qquad
    a_j = a+j,
    \qquad
    P_j = \left\lfloor Q\times\frac{2^{a-\frac{3}{8}}-1}{2^{a_j}-1} \right\rfloor.
  \]
  Then, for each $j\in\{0,1,2\}$,
  \[
    (\rho-1)^{k_j-3}\le \rho^{k_j-5}
    \qquad\text{and}\qquad
    P_j \geq \left\lfloor 6\times 2^{k_j-3}\beta^2 \right\rfloor.
  \]
\end{lemma}
\begin{proof}
  Since
  \[
    100^2<k+a=10010<101^2,
  \]
  we have
  \[
    100<\rho<101.
  \]
  Therefore
  \[
    (\rho-1)^{k_j-3}
    =
    \rho^{k_j-3}\left(1-\frac{1}{\rho}\right)^{k_j-3}
    \leq
    \rho^{k_j-3}e^{-(k_j-3)/\rho}
    \leq
    \rho^{k_j-3}e^{-9995/101}.
  \]
  As $e^{-9995/101}<10^{-40}<\rho^{-2}$, this gives
  \[
    (\rho-1)^{k_j-3}\le \rho^{k_j-5}
  \]
  for each $j\in\{0,1,2\}$.

  Also $\beta<\frac{99}{98}$, and since
  \[
    \left(\frac{13}{10}\right)^8 > 8,
  \]
  we have
  \[
    2^{a-\frac{3}{8}}-1
    =
    \frac{2^{10}}{2^{3/8}}-1
    >
    \frac{10240}{13}-1
    =
    \frac{10227}{13}.
  \]
  Thus, for $j=0,1,2$,
  \[
    2^{-j}\times
    \frac{3(2^{a_j}-1)}{4(2^{a-\frac{3}{8}}-1)}
    \times
    \left(\frac{\sqrt{k+a}-1}{\sqrt{k+a}-2}\right)^2
    <
    2^{-j}\times
    \frac{3(2^{10+j}-1)}{4(10227/13)}
    \times
    \left(\frac{99}{98}\right)^2.
  \]
  The three resulting rational bounds are
  \[
    \frac{130343499}{130960144},
    \qquad
    \frac{260814411}{261920288},
    \qquad
    \frac{74536605}{74834368},
  \]
  and each is smaller than $\frac{255}{256}$. Hence
  \[
    Q\times\frac{2^{a-\frac{3}{8}}-1}{2^{a_j}-1}
    \geq
    6\times 2^{k_j-3}\beta^2
  \]
  for each $j$, and flooring gives the second claim.
\end{proof}

\begin{proposition}\label{prop:phi-growth-step}
  For $k=10000$, $a=10$, define
  \[
    \eta_j = 2^{-3/8-j}
    \qquad\text{for } j\in\{0,1,2\}.
  \]
  For any $i \geq 1$, if
  \[
    \comp{\bracket{\phi_i}{1}{2^{-k-a}}{2^{-3}}} \geq 1,
  \]
  then
  \[
    \comp{\bracket{\phi_{i+1}}{1}{2^{-k-a}}{\eta_j}}
    \geq
    k-j+\Lambda_{\phi_i}(1,2^{-k-a},\eta_0)
    \qquad\text{for each } j\in\{0,1,2\}.
  \]
  Consequently,
  \[
    \Lambda_{\phi_{i+1}}(1,2^{-k-a},\eta_0)
    \geq
    k+\Lambda_{\phi_i}(1,2^{-k-a},\eta_0).
  \]
\end{proposition}
\begin{proof}
  Write
  \[
    x_0 = 2^{-k-a},
    \qquad
    \rho = \sqrt{k+a},
    \qquad
    Q = 255\cdot 2^{k-8}.
  \]
  By Lemma~\ref{lem:phi-row-divisibility}, the number of rows of $\phi_i$ is
  divisible by $2^{a+2}$, hence by $2^{a+j}$ for each $j\in\{0,1,2\}$.
  Fix $j\in\{0,1,2\}$ and set
  \[
    k_j = k-j,
    \qquad
    a_j = a+j,
    \qquad
    x_j = 2^{-a_j},
    \qquad
    P_j = \left\lfloor Q\times\frac{2^{a-\frac{3}{8}}-1}{2^{a+j}-1} \right\rfloor.
  \]
  Then $k_j+a_j=k+a$, so the same $\rho$ applies to each shifted pair, and
  \[
    \eta_j = 2^{a-\frac{3}{8}}x_j.
  \]
  Since the number of rows of $\phi_i$ is divisible by $2^{a+j}$,
  Lemma~\ref{lem:swap} applies with row parameter $x_j$, giving
  \begin{align*}
    \comp{\bracket{\phi_{i+1}}{1}{x_0}{\eta_j}}
    &= \comp{\bracket{\overOp{\phi_i}{Q}}{1}{\eta_j}{x_0}}
    & \text{Definition of $\phi_{i+1}$ and Lemma~\ref{lem:transposeComp}} \\
    &\geq \comp{\bracket{\phi_i}{P_j}{2^{-a-j}}{x_0}}
    & \text{Lemma~\ref{lem:swap}}.
  \end{align*}
  Lemma~\ref{lem:shifted-rung-arithmetic} gives
  \[
    P_j \geq \left\lfloor 6\times 2^{k_j-3}\beta^2 \right\rfloor
  \]
  and
  \[
    (\rho-1)^{k_j-3}\le \rho^{k_j-5}
  \]
  for each $j\in\{0,1,2\}$.
  Therefore
  \[
    \comp{\bracket{\phi_{i+1}}{1}{x_0}{\eta_j}}
    \geq
    \comp{\bracket{\phi_i}{
      \left\lfloor 6\times 2^{k_j-3}\beta^2 \right\rfloor
    }{2^{-a-j}}{x_0}}
  \]
  by monotonicity.

  Apply Corollary~\ref{cor:iterated-partition-seed} to $\phi_i$ with
  \[
    \tilde{k}=k_j-3,
    \qquad
    s=2,
    \qquad
    p=6,
    \qquad
    x=2^{3-k-a},
    \qquad
    y=2^{-1}.
  \]
  Here $2^{\tilde{k}}x=2^{-a_j}$ and $y^{\rho^2}=x_0$, while the weak side
  condition follows from the hypothesis by monotonicity. Thus
  \[
    \comp{\bracket{\phi_i}{
      \left\lfloor 6\times 2^{k_j-3}\beta^2 \right\rfloor
    }{2^{-a-j}}{x_0}}
    \geq
    k_j-3 + \Lambda_{\phi_i}(6,2^{3-k-a},2^{-1}).
  \]
  Lemma~\ref{lem:seed-collapse} gives
  \[
    \Lambda_{\phi_i}(6,2^{3-k-a},2^{-1})
    \geq
    3 + \Lambda_{\phi_i}(1,x_0,2^{-3/8}).
  \]
  Hence
  \[
    \comp{\bracket{\phi_{i+1}}{1}{x_0}{\eta_j}}
    \geq
    k-j+\Lambda_{\phi_i}(1,x_0,2^{-3/8})
    \qquad\text{for each } j\in\{0,1,2\}.
  \]
  Taking the defining minimum over $j$ yields
  \[
    \Lambda_{\phi_{i+1}}(1,x_0,2^{-3/8})
    \geq
    k+\Lambda_{\phi_i}(1,x_0,2^{-3/8}),
  \]
  as claimed.
\end{proof}

Finally we can derive our bound on $\phi_i$. We bootstrap the argument with
Lemma~\ref{lem:rankclaim}, which will be used both for the weak side condition
and for the three-rung base case at $\phi_1$.

\begin{lemma}\label{lem:rankclaim}
  For $0 < x,y \leq 1$ with $\symTotGames + \log y > 0$, we have
  \[
    \comp{\bracket{\phi_0}{\symTotGames}{x}{y}}
    \geq
    \ceil{\log (\symTotGames + \log y)}.
  \]
\end{lemma}
\begin{proof}
  Remember that
  \[
    \overOp{\phi_0}{1}= \begin{bmatrix} 1 & 0 \end{bmatrix}.
  \]
  The columns of $\overOp{\phi_0}{\symTotGames}$ consist of every bit string of
  length $\symTotGames$. So every matrix in
  $\bracket{\phi_0}{\symTotGames}{x}{y}$ is an
  $\symTotGames \times N$ matrix whose
  \[
    N=\ceil{2^{\symTotGames}y}
  \]
  columns are distinct Boolean vectors.

  Let $d$ be the rank of such a matrix. Its columns lie in a real subspace
  $V \leq \mathbb{R}^{\symTotGames}$ of dimension $d$. Choose $d$ coordinates
  on which the coordinate projection $\pi : V \to \mathbb{R}^d$ is injective.
  Then $\pi$ is also injective on the Boolean columns, and each projected
  column lies in $\{0,1\}^d$. Hence there are at most $2^d$ distinct Boolean
  columns in $V$, so
  \[
    2^{\symTotGames}y \leq N \leq 2^d.
  \]
  Therefore
  \[
    d \geq \log(2^{\symTotGames}y) = \symTotGames + \log y.
  \]
  The standard log-rank lower bound for deterministic communication complexity
  (see \citet{Kushilevitz1997} or \citet{rao2020communication}) now gives
  \[
    \comp{\bracket{\phi_0}{\symTotGames}{x}{y}}
    \geq
    \ceil{\log(\symTotGames + \log y)},
  \]
  which is the claim.
\end{proof}

\begin{lemma}\label{lem:phi-zero-seed}
  Let $Q$ be a positive integer, let $0 < \theta \leq 1$, let $0 < y \leq 1$,
  and let $0 < x \leq 1$. Write
  \[
    q = \left\lfloor Q\theta \right\rfloor.
  \]
  If $q \geq 1$, then
  \[
    \comp{\bracket{\overOp{\phi_0}{Q}}{1}{\theta}{y}}
    \geq
    \comp{\bracket{\phi_0}{q}{x}{y}}.
  \]
\end{lemma}
\begin{proof}
  Let
  \[
    g = \extractmatrix{\overOp{\phi_0}{Q}}{R}{C}
    \in
    \bracket{\overOp{\phi_0}{Q}}{1}{\theta}{y}.
  \]
  Since $\phi_0$ has one row and two columns,
  \[
    \card{R} = \ceil{Q\theta}
    \qquad\text{and}\qquad
    \card{C} = \ceil{2^Q y}.
  \]
  Choose any subset $Q' \subseteq R$ with $\card{Q'} = q$, and let $D$ be the
  $Q'$-projection of $C$. Each projected column has at most $2^{Q-q}$ lifts,
  so
  \[
    \card{D}
    \geq
    \frac{\card{C}}{2^{Q-q}}
    \geq
    2^q y.
  \]
  Hence $\card{D}\ge \ceil{2^q y}$, and we may choose
  \[
    D' \subseteq D
    \qquad\text{with}\qquad
    \card{D'} = \ceil{2^q y}.
  \]
  By Lemma~\ref{lem:projection},
  \[
    \extractmatrix{\overOp{\phi_0}{q}}{\rangeintegers{q}}{D'}
    \sqsubseteq
    \extractmatrix{\overOp{\phi_0}{q}}{\rangeintegers{q}}{D}
    \sqsubseteq
    g.
  \]
  Because $\phi_0$ has one row, every positive row parameter $x$ satisfies
  $\ceil{x}=1$, so the left-hand matrix belongs to
  $\bracket{\phi_0}{q}{x}{y}$. Thus each
  $g\in \bracket{\overOp{\phi_0}{Q}}{1}{\theta}{y}$ contains a member of
  $\bracket{\phi_0}{q}{x}{y}$ as a subgame, and the claimed lower bound
  follows.
\end{proof}

\begin{lemma}\label{lem:phi-one-weak-side-condition}
  For $k=10000$ and $a=10$, we have
  \[
    \comp{\bracket{\phi_{1}}{1}{2^{-k-a}}{2^{-3}}} \geq 1.
  \]
\end{lemma}
\begin{proof}
  Write
  \[
    Q = 255\cdot 2^{k-8}
    \qquad\text{and}\qquad
    q = \left\lfloor \frac{Q}{8} \right\rfloor.
  \]
  Since
  \[
    q
    =
    \left\lfloor 255\cdot 2^{k-11} \right\rfloor
    >
    2^{k-4},
  \]
  we have
  \[
    q-(k+a) > 2^{k-4}-(k+a) > 2
  \]
  for $k=10000$ and $a=10$. Therefore
  \begin{align*}
    \comp{\bracket{\phi_{1}}{1}{2^{-k-a}}{2^{-3}}}
    &= \comp{\bracket{\transpose{\overOp{\phi_{0}}{Q}}}{1}{2^{-k-a}}{2^{-3}}}
    & \text{Definition of $\phi_1$} \\
    &= \comp{\bracket{\overOp{\phi_{0}}{Q}}{1}{2^{-3}}{2^{-k-a}}}
    & \text{Lemma~\ref{lem:transposeComp}} \\
    &\geq \comp{\bracket{\phi_0}{q}{2^{-a}}{2^{-k-a}}}
    & \text{Lemma~\ref{lem:phi-zero-seed}} \\
    &\geq \ceil{\log \left(q-(k+a)\right)}
    & \text{Lemma~\ref{lem:rankclaim}} \\
    &\geq 1.
  \end{align*}
\end{proof}

\begin{lemma}\label{lem:phi-weak-side-propagates}
  For $k=10000$, $a=10$, and any $i \geq 1$, if
  \[
    \comp{\bracket{\phi_i}{1}{2^{-k-a}}{2^{-3}}} \geq 1,
  \]
  then
  \[
    \comp{\bracket{\phi_{i+1}}{1}{2^{-k-a}}{2^{-3}}} \geq 1.
  \]
\end{lemma}
\begin{proof}
  Write
  \[
    Q = 255\cdot 2^{k-8}
    \qquad\text{and}\qquad
    P = \left\lfloor Q\times\frac{2^{a-3}-1}{2^a-1} \right\rfloor.
  \]
  By Lemma~\ref{lem:phi-row-divisibility}, the number of rows of $\phi_i$ is
  divisible by $2^a$, so Lemma~\ref{lem:swap} applies with row parameter
  $2^{-a}$. Hence
  \begin{align*}
    \comp{\bracket{\phi_{i+1}}{1}{2^{-k-a}}{2^{-3}}}
    &= \comp{\bracket{\transpose{\overOp{\phi_i}{Q}}}{1}{2^{-k-a}}{2^{-3}}}
    & \text{Definition of $\phi_{i+1}$} \\
    &= \comp{\bracket{\overOp{\phi_i}{Q}}{1}{2^{-3}}{2^{-k-a}}}
    & \text{Lemma~\ref{lem:transposeComp}} \\
    &\geq \comp{\bracket{\phi_i}{P}{2^{-a}}{2^{-k-a}}}
    & \text{Lemma~\ref{lem:swap}} \\
    &\geq \comp{\bracket{\phi_i}{1}{2^{-a}}{2^{-\frac{k+a}{P}}}}
    & \text{Lemma~\ref{lem:TupleCorollary}}.
  \end{align*}
  Since
  \[
    P
    =
    \left\lfloor 255\cdot 2^{k-8}\times\frac{127}{1023} \right\rfloor
    >
    2^{k-4},
  \]
  we have $\frac{k+a}{P}<3$, so $2^{-(k+a)/P}>2^{-3}$. Also
  $2^{-k-a}\le 2^{-a}$. Thus monotonicity gives
  \[
    \comp{\bracket{\phi_i}{1}{2^{-a}}{2^{-\frac{k+a}{P}}}}
    \geq
    \comp{\bracket{\phi_i}{1}{2^{-k-a}}{2^{-3}}}
    \geq
    1,
  \]
  which proves the claim.
\end{proof}

\begin{lemma}\label{lem:phi-one-lambda}
  For $k=10000$ and $a=10$, we have
  \[
    \Lambda_{\phi_1}(1,2^{-k-a},2^{-3/8}) \geq k.
  \]
\end{lemma}
\begin{proof}
  Write
  \[
    Q = 255\cdot 2^{k-8},
    \qquad
    \eta_0 = 2^{-3/8},
    \qquad
    \eta_1 = 2^{-11/8},
    \qquad
    \eta_2 = 2^{-19/8}.
  \]
  Since
  \[
    \left(\frac{13}{10}\right)^8 > 8,
  \]
  we have $2^{-3/8} > \frac{10}{13}$. Therefore
  \[
    \frac{255}{256}\eta_0 > \frac34,
    \qquad
    \frac{255}{256}\eta_1 > \frac38,
    \qquad
    \frac{255}{256}\eta_2 > \frac3{16}.
  \]
  Hence
  \[
    \left\lfloor Q\eta_0 \right\rfloor-(k+a) > 2^{k-1},
    \qquad
    \left\lfloor Q\eta_1 \right\rfloor-(k+a) > 2^{k-2},
    \qquad
    \left\lfloor Q\eta_2 \right\rfloor-(k+a) > 2^{k-3}.
  \]
  For each $0 \leq j < 3$, Lemma~\ref{lem:transposeComp} and then
  Lemma~\ref{lem:phi-zero-seed} give
  \[
    \comp{\bracket{\phi_1}{1}{2^{-k-a}}{\eta_j}}
    \geq
    \comp{\bracket{\phi_0}{\left\lfloor Q\eta_j \right\rfloor}{2^{-a}}{2^{-k-a}}}.
  \]
  The displayed lower bounds show that Lemma~\ref{lem:rankclaim} applies in
  each case, so
  \[
    \comp{\bracket{\phi_1}{1}{2^{-k-a}}{\eta_0}} \geq k,
    \qquad
    \comp{\bracket{\phi_1}{1}{2^{-k-a}}{\eta_1}} \geq k-1,
    \qquad
    \comp{\bracket{\phi_1}{1}{2^{-k-a}}{\eta_2}} \geq k-2.
  \]
  Taking the minimum of
  \[
    \comp{\bracket{\phi_1}{1}{2^{-k-a}}{\eta_0}},
    \qquad
    1 + \comp{\bracket{\phi_1}{1}{2^{-k-a}}{\eta_1}},
    \qquad
    2 + \comp{\bracket{\phi_1}{1}{2^{-k-a}}{\eta_2}}
  \]
  proves the claim.
\end{proof}

\begin{corollary}\label{cor:phi-bundled}
  For $k=10000$, $a=10$, and any $i \geq 1$, we have
  \[
    \comp{\bracket{\phi_i}{1}{2^{-k-a}}{2^{-3}}} \geq 1
    \qquad\text{and}\qquad
    \Lambda_{\phi_i}(1,2^{-k-a},2^{-3/8}) \geq ki.
  \]
\end{corollary}
\begin{proof}
  We argue by induction on $i$.

  For $i=1$, the weak side condition is
  Lemma~\ref{lem:phi-one-weak-side-condition}, and the three-rung lower bound
  is Lemma~\ref{lem:phi-one-lambda}.

  For the induction step, assume
  \[
    \comp{\bracket{\phi_i}{1}{2^{-k-a}}{2^{-3}}} \geq 1
    \qquad\text{and}\qquad
    \Lambda_{\phi_i}(1,2^{-k-a},2^{-3/8}) \geq ki.
  \]
  Lemma~\ref{lem:phi-weak-side-propagates} gives
  \[
    \comp{\bracket{\phi_{i+1}}{1}{2^{-k-a}}{2^{-3}}} \geq 1,
  \]
  and Proposition~\ref{prop:phi-growth-step} gives
  \[
    \Lambda_{\phi_{i+1}}(1,2^{-k-a},2^{-3/8})
    \geq
    k+\Lambda_{\phi_i}(1,2^{-k-a},2^{-3/8})
    \geq
    k(i+1).
  \]
  This completes the induction.
\end{proof}

\begin{corollary}\label{cor:lower-bound-phi}
  For $k=10000$, $a=10$, and any $i \geq 1$, we have
  \[
    \comp{\phi_i} \geq ki.
  \]
\end{corollary}
\begin{proof}
  Corollary~\ref{cor:phi-bundled} gives
  \[
    \Lambda_{\phi_i}(1,2^{-k-a},2^{-3/8}) \geq ki.
  \]
  Since
  \[
    \bracket{\phi_i}{1}{1}{1} = \{\phi_i\},
  \]
  we have
  \[
    \comp{\phi_i}
    =
    \comp{\bracket{\phi_i}{1}{1}{1}}
    \geq
    \comp{\bracket{\phi_i}{1}{2^{-k-a}}{2^{-3/8}}}
    \geq
    \Lambda_{\phi_i}(1,2^{-k-a},2^{-3/8}),
  \]
  where the first inequality is monotonicity and the second follows from the
  definition of $\Lambda_{\phi_i}$ as a minimum whose first term is
  $\comp{\bracket{\phi_i}{1}{2^{-k-a}}{2^{-3/8}}}$. This proves the claim.
\end{proof}
\section{Upper Bound Proof and Refutation of the Conjecture}\label{section:UB}

\subsection{Upper Bound for the Direct Sum of the Counterexample}

Having shown the lower bound, we now require an upper bound on the
communication required when solving instances in batch. In this section, we
first prove such an upper bound and then conclude the paper by establishing
that our family of functions refutes the direct sum conjecture. We finally
record a further quantitative consequence comparing
$\comp{\phi_i^{178}}$ and $\comp{\phi_i}$.

Throughout this section, we view our communication games as Boolean functions
with the inputs partitioned into two. We also extend the previous definitions
in the evident way to functions whose outputs are natural numbers rather than
Booleans.

\begin{definition}[Direct Sum]\label{def:directSum}
  Let $f:X\times Y\to\{0,1\}$ be a function. Set
  \[
    X' = X^l
    \qquad\text{and}\qquad
    Y' = Y^l.
  \]
  We define
  \[
    f^l:X'\times Y' \to \{0,1\}^l
  \]
  by
  \[
    f^l((x_1,\ldots,x_l),(y_1,\ldots,y_l))
    =
    (f(x_1,y_1),\ldots,f(x_l,y_l)).
  \]
\end{definition}

The following observation will be useful.
\begin{observation}\label{obs:transds}
  \[
    \comp{f^l} = \comp{(\transpose{f})^l}.
  \]
\end{observation}

\begin{definition}
  Given a set $X$, write
  \[
    X'=\left(\sqcup_{i\in [\kappa]}X\right)^l.
  \]
  For each $u=(u_1,\ldots,u_l)\in [\kappa]^l$, define
  \[
    \gamma_{X'}(u)
    :=
    \left\{
      ((u_1,v_1),\ldots,(u_l,v_l)) \in X'
      \suchthat
      v_1,\ldots,v_l\in X
    \right\}.
  \]
\end{definition}
\begin{definition}
  Given a set $X$, write
  \[
    X'=\left(\sqcup_{i\in [\kappa]}X\right)^l.
  \]
  We define the partition
  \[
    \Gamma_{X'}
    :=
    \{\gamma_{X'}(u)\suchthat u\in [\kappa]^l\}.
  \]
\end{definition}
\begin{definition}\label{def:sigmaTauUB}
  Given a set $X$, let
  \[
    X'=\left(\sqcup_{i\in [\kappa]}X\right)^l,
    \qquad
    Y'=(Y^\kappa)^l,
  \]
  and fix $u=(u_1,\ldots,u_l)\in [\kappa]^l$. We define maps
  \[
    \sigma:X'\to X^l
    \qquad\text{and}\qquad
    \tau_u:Y'\to Y^l
  \]
  by
  \[
    \sigma(((u_1,x_1),\ldots,(u_l,x_l)))
    =
    (x_1,\ldots,x_l)
  \]
  and
  \[
    \tau_u(((y_{1,1},\ldots,y_{1,\kappa}),\ldots,(y_{l,1},\ldots,y_{l,\kappa})))
    =
    (y_{1,u_1},\ldots,y_{l,u_l}).
  \]
\end{definition}
\begin{lemma}\label{lem:subgameLemma}
  Let $f$ be a Boolean function on sets $X$ and $Y$. Write
  \[
    X'=\left(\sqcup_{i\in [\kappa]}X\right)^l
    \qquad\text{and}\qquad
    Y'=(Y^\kappa)^l.
  \]
  For any $u\in [\kappa]^l$, $x\in\gamma_{X'}(u)$, and $y\in Y'$,
  we have
  \[
    \overOp{f}{\kappa}^l(x,y)=f^l(\sigma(x),\tau_u(y)).
  \]
\end{lemma}
\begin{proof}
  Write
  \[
    x=((u_1,x_1),\ldots,(u_l,x_l))
    \qquad\text{and}\qquad
    y=(y_1,\ldots,y_l),
  \]
  where each $y_i=(y_{i,1},\ldots,y_{i,\kappa})$. Then
  \[
    \begin{aligned}
      \overOp{f}{\kappa}^l(x,y)
      &=
      \bigl(
        \overOp{f}{\kappa}((u_1,x_1),y_1),
        \ldots,
        \overOp{f}{\kappa}((u_l,x_l),y_l)
      \bigr) \\
      &=
      (f(x_1,y_{1,u_1}),\ldots,f(x_l,y_{l,u_l})) \\
      &=
      f^l(\sigma(x),\tau_u(y)).
    \end{aligned}
  \]
\end{proof}
\begin{theorem}\label{thm:OneRoundUpper}
  \[
    \comp{\overOp{f}{\kappa}^l}
    \leq
    \comp{f^l}+\ceil{l\log \kappa}.
  \]
\end{theorem}
\begin{proof}
  Fix $u\in [\kappa]^l$. After Alice communicates $u$, both
  players know that the input lies in $\gamma_{X'}(u)\times Y'$. Alice can
  then locally compute $\sigma(x)\in X^l$, and Bob can locally compute
  $\tau_u(y)\in Y^l$. By Lemma~\ref{lem:subgameLemma}, for every
  $(x,y)\in\gamma_{X'}(u)\times Y'$ we have
  \[
    \overOp{f}{\kappa}^l(x,y)=f^l(\sigma(x),\tau_u(y)).
  \]
  Hence, after the initial message specifying $u$, the players can simulate
  an optimal protocol for $f^l$ on the input
  $(\sigma(x),\tau_u(y))$. Since there are $\kappa^l$ possible values of $u$,
  the initial communication costs at most $\ceil{l\log\kappa}$ bits, and
  therefore
  \[
    \comp{\overOp{f}{\kappa}^l}
    \leq
    \comp{f^l}+\ceil{l\log\kappa}.
  \]
\end{proof}

\subsection{The Direct Sum Conjecture is False}

\begin{corollary}\label{cor:upper-bound-phi}
  For $k=10000$, $i \geq 0$, and $h \geq 1$, we have
  \[
    \comp{\phi_i^{178h}} \leq 178h+i(178kh-h).
  \]
\end{corollary}
\begin{proof}
  We prove the result by induction on $i$. For $i=0$, the function
  $\phi_0^{178h}$ has one row and $2^{178h}$ distinct outputs, so any
  depth-$d$ protocol has at most $2^d$ leaves and hence must satisfy
  $d\ge 178h$. Conversely, Bob can send the $178h$ output bits, so
  \[
    \comp{\phi_0^{178h}}=178h.
  \]
  For the induction step,
  \begin{align*}
    \comp{\phi_{i+1}^{178h}}
    &= \comp{\left(\transpose{\overOp{\phi_i}{2^k\times\frac{255}{256}}}\right)^{178h}}
    & \text{Definition of $\phi_{i+1}$} \\
    &= \comp{\overOp{\phi_i}{2^k\times\frac{255}{256}}^{178h}}
    & \text{Observation~\ref{obs:transds}} \\
    &\leq \comp{\phi_i^{178h}}+\ceil{178h\log\left(2^k\times\frac{255}{256}\right)}
    & \text{Theorem~\ref{thm:OneRoundUpper}} \\
    &\leq \comp{\phi_i^{178h}}+178hk+\ceil{178h\log 255}-178h\times 8
    \\
    &\leq \comp{\phi_i^{178h}}+178hk+h\ceil{178\log 255}-178h\times 8
    \\
    &\leq \comp{\phi_i^{178h}}+178hk-h
    & \text{Numerical evaluation} \\
    &\leq 178h+(i+1)(178kh-h)
    & \text{Induction hypothesis.}
  \end{align*}
\end{proof}

We now prove Theorem~\ref{th:main}.
\begin{proof}
  For any natural number $i$, the dimensions $r_i\times c_i$ of $\phi_i$ are
  given by
  \begin{align*}
    r_{2i} &= B^{\frac{B^{i+1}-B}{B-1}},
    &
    c_{2i} &= 2^{B^i}B^{\frac{B^i-1}{B-1}},
    \\
    r_{2i+1} &= 2^{B^{i+1}}B^{\frac{B^{i+1}-B}{B-1}},
    &
    c_{2i+1} &= B^{\frac{B^{i+1}-1}{B-1}}.
  \end{align*}
  For any $n$, let $i_n$ be the largest $i$ such that
  \[
    2^n \geq r_i
    \qquad\text{and}\qquad
    2^n \geq c_i.
  \]
  Since $2^n\ge r_{i_n}$ and $2^n\ge c_{i_n}$, choose surjections
  \[
    \sigma_n:\{0,1\}^n\twoheadrightarrow [r_{i_n}]
    \qquad\text{and}\qquad
    \tau_n:\{0,1\}^n\twoheadrightarrow [c_{i_n}],
  \]
  and define
  \[
    f_n(u,v)=\phi_{i_n}(\sigma_n(u),\tau_n(v)).
  \]
  Thus $f_n$ is obtained from $\phi_{i_n}$ by duplicating rows and columns.
  Choose subsets
  \[
    X_n\subseteq \{0,1\}^n
    \qquad\text{and}\qquad
    Y_n\subseteq \{0,1\}^n
  \]
  such that $\sigma_n|_{X_n}$ and $\tau_n|_{Y_n}$ are bijections. Restricting
  $f_n$ to $X_n\times Y_n$ recovers $\phi_{i_n}$, so
  \[
    \comp{f_n}\ge \comp{\phi_{i_n}}.
  \]
  Conversely, Alice and Bob can locally apply $\sigma_n$ and $\tau_n$ to
  their inputs and then run an optimal protocol for $\phi_{i_n}$, giving
  \[
    \comp{f_n}\le \comp{\phi_{i_n}}.
  \]
  Hence
  \[
    \comp{f_n}=\comp{\phi_{i_n}}.
  \]
  Applying the same argument independently on each coordinate gives
  \[
    \comp{f_n^\ell}=\comp{\phi_{i_n}^\ell}
    \qquad\text{for every }\ell\ge 1.
  \]

  Let us now prove the displayed claim. Given $N$, $L$, and $C$, choose
  \[
    \ell = 178(L+1)
    \qquad\text{and}\qquad
    n=\max(N,\ceil{\log r_i},\ceil{\log c_i}),
  \]
  where $i=178(C+2)$. Then
  \begin{align*}
    \comp{f_n}
    &= \comp{\phi_{i_n}}
    \\
    &\geq ki_n
    & \text{Corollary~\ref{cor:lower-bound-phi}} \\
    &\geq
    \frac{\comp{\phi_{i_n}^{178(L+1)}}-178(L+1)+(L+1)i_n}{178(L+1)}
    & \text{Corollary~\ref{cor:upper-bound-phi} with $h=L+1$} \\
    &= \frac{\comp{f_n^\ell}}{\ell}-1+\frac{i_n}{178}
    \\
    &>
    \frac{\comp{f_n^\ell}}{\ell}+C,
  \end{align*}
  since $i_n\ge i=178(C+2)$.
\end{proof}

The same family also yields the following additive consequence.

\begin{theorem}\label{theorem:multForm}
  \[
    \comp{\phi_i^{178}}
    \leq
    178\comp{\phi_i}+177.
  \]
\end{theorem}
\begin{proof}
  If $i=0$, then $\comp{\phi_0}=1$ and
  \[
    \comp{\phi_0^{178}}=178\le 178\comp{\phi_0}+177,
  \]
  so the claim is immediate. Assume henceforth that $i \geq 1$.
  \begin{align*}
    \comp{\phi_i^{178}}
    &\leq 178+(1780000-1)i
    & \text{Corollary~\ref{cor:upper-bound-phi} with $h=1$} \\
    &\leq 177+1780000i
    \\
    &\leq 177+178\comp{\phi_i}
    & \text{Corollary~\ref{cor:lower-bound-phi}} \\
    &= 178\comp{\phi_i}+177.
  \end{align*}
\end{proof}

\appendix
\section{Details of the Lower-Bound Argument}\label{appendix:lower-bound-details}

This appendix supplies the detailed lower-bound derivations deferred from
Section~\ref{section:LB}. We first isolate the one-step partition recurrence,
then derive the three-rung partition lemma and its bundled ladder
consequences, and then prove the full iterated partition lemma.

\subsection{Local Partition Bookkeeping}\label{app:partition-bookkeeping}

We start by extracting from Lemma~\ref{lem:old-partition} the one-step
recurrence that will feed the three-rung argument.

\begin{lemma}[Partition recurrence]
  \label{lem:partition-recurrence}
  For a matrix $M$, integer $\symTotGames \geq 1$, real $0 < \tau \leq 1$,
  $\delta \in \{0,1\}$, and reals $0 < x \leq \frac{1}{2}$ and $0 < y \leq 1$,
  if
  \[
    \comp{\bracket{M}{2\symTotGames+\delta}{2x}{y}} \geq 1,
  \]
  then
  \begin{align}
    \comp{\bracket{M}{2\symTotGames+\delta}{2x}{y}}
    &\geq 1 + \min
      \begin{cases}
      \comp{\bracket{M}{\symTotGames+\delta}{x}{y^{\frac{\symTotGames+\delta}{\symTotGames(1+\tau)+\delta}}}} \\
      \comp{\bracket{M}{\lfloor \symTotGames(1-\tau)\rfloor+1}{x}{y^{\frac{\tau}{1+\tau}}}} \\
      \comp{\bracket{M}{2\symTotGames+\delta}{2x}{\frac{y}{2}}}
      \end{cases}. \label{eq:partition-recurrence}
  \end{align}
\end{lemma}
\begin{proof}
  Let
  \[
    E_0=\comp{\bracket{M}{2\symTotGames+\delta}{x}{y}}
    \qquad\text{and}\qquad
    R=\comp{\bracket{M}{2\symTotGames+\delta}{2x}{\frac{y}{2}}}.
  \]
  Lemma~\ref{lem:old-partition} gives
  \[
    \comp{\bracket{M}{2\symTotGames+\delta}{2x}{y}}
    \geq
    1 + \min(E_0,E_{\mathrm{mid}},R),
  \]
  where
  \[
    E_{\mathrm{mid}}=
    \min_{\substack{\symGamePartSize\in\rangeintegers{\symTotGames}\\a\in [0,1]}}
    \max\left(
      \comp{\bracket{M}{\symTotGames+\delta+\symGamePartSize}{x}{y^a}},
      \comp{\bracket{M}{\symTotGames-\symGamePartSize}{x}{y^{1-a}}}
    \right).
  \]

  Lemma~\ref{lem:TupleCorollary} sends the endpoint term to
  \[
    \bracket{M}{\symTotGames+\delta}{x}{
      y^{\frac{\symTotGames+\delta}{2\symTotGames+\delta}}
    }.
  \]
  Since $\tau\leq 1$, we have
  \[
    \frac{\symTotGames+\delta}{2\symTotGames+\delta}
    \leq
    \frac{\symTotGames+\delta}{\symTotGames(1+\tau)+\delta},
  \]
  so monotonicity lowers that endpoint term to
  \[
    \comp{\bracket{M}{\symTotGames+\delta}{x}{y^{\frac{\symTotGames+\delta}{\symTotGames(1+\tau)+\delta}}}}.
  \]

  Fix $\symGamePartSize\in\rangeintegers{\symTotGames}$ and $a\in[0,1]$. If
  $\symGamePartSize\geq \symTotGames\tau$, then Lemma~\ref{lem:TupleCorollary}
  sends the first child term to
  \[
    \bracket{M}{\symTotGames+\delta}{x}{
      y^{a\frac{\symTotGames+\delta}{\symTotGames+\delta+\symGamePartSize}}
    }.
  \]
  Since $a\leq 1$ and $\symGamePartSize\geq \symTotGames\tau$, we have
  \[
    a\frac{\symTotGames+\delta}{\symTotGames+\delta+\symGamePartSize}
    \leq
    \frac{\symTotGames+\delta}{\symTotGames(1+\tau)+\delta},
  \]
  so the first child is at least the first displayed branch by monotonicity.

  If $\symGamePartSize<\symTotGames\tau$ and
  $a\leq \frac{1}{1+\tau}$, then the same application of
  Lemma~\ref{lem:TupleCorollary} yields
  \[
    \bracket{M}{\symTotGames+\delta}{x}{
      y^{a\frac{\symTotGames+\delta}{\symTotGames+\delta+\symGamePartSize}}
    },
  \]
  and now
  \[
    a\frac{\symTotGames+\delta}{\symTotGames+\delta+\symGamePartSize}
    \leq
    \frac{1}{1+\tau}
    \leq
    \frac{\symTotGames+\delta}{\symTotGames(1+\tau)+\delta}.
  \]
  So the first child again dominates the first displayed branch.

  If $\symGamePartSize<\symTotGames\tau$ and
  $a\geq \frac{1}{1+\tau}$, then
  \[
    \symTotGames-\symGamePartSize\geq \lfloor \symTotGames(1-\tau)\rfloor+1,
    \qquad
    1-a\leq \frac{\tau}{1+\tau},
  \]
  so monotonicity gives
  \[
    \comp{\bracket{M}{\symTotGames-\symGamePartSize}{x}{y^{1-a}}}
    \geq
    \comp{\bracket{M}{\lfloor \symTotGames(1-\tau)\rfloor+1}{x}{y^{\frac{\tau}{1+\tau}}}}.
  \]
  Therefore the quantity inside the minimum defining $E_{\mathrm{mid}}$ is
  always at least one of the first two displayed branches, and hence
  \[
    E_{\mathrm{mid}}
    \geq
    \min \begin{cases}
      \comp{\bracket{M}{\symTotGames+\delta}{x}{y^{\frac{\symTotGames+\delta}{\symTotGames(1+\tau)+\delta}}}} \\
      \comp{\bracket{M}{\lfloor \symTotGames(1-\tau)\rfloor+1}{x}{y^{\frac{\tau}{1+\tau}}}}
    \end{cases}.
  \]

  Combining this with the endpoint bound gives exactly
  \eqref{eq:partition-recurrence}.
\end{proof}

\subsection{Three-Rung Partition Lemma}\label{app:three-rung-partition}

We now organise the local partition argument so that the three densities
$y$, $\frac{y}{2}$, and $\frac{y}{4}$ are carried simultaneously.

\begin{lemma}[Partition Lemma]\label{lem:partition}\label{lem:partition-three-rung}
  Let $M$ be a matrix, let $\symTotGames \in \mathbb{N}$ with
  $\symTotGames \geq 1$, let $\delta \in \{0,1\}$, and let
  $0 < x \leq \frac12$ and $0 < y \leq 1$. Assume that
  \[
    \comp{\bracket{M}{2\symTotGames}{2x}{\frac{y}{4}}} \geq 1.
  \]
  Then we have
  \begin{align}
    \comp{\bracket{M}{2\symTotGames+\delta}{2x}{y}}
    &\geq 1 + \min_{0 \leq j < 3}\left(
      j + \comp{\bracket{M}{\symTotGames+\delta}{x}{\frac{y}{2^j}}}
    \right), \label{eq:partition-three-rung-row}
  \end{align}
  and for every $0 < \tau \leq 1$, writing
  \[
    u = y^{\frac{1}{1+\tau}}
    \qquad\text{and}\qquad
    v = y^{\frac{\tau}{1+\tau}},
  \]
  we have
  \begin{align}
    \comp{\bracket{M}{2\symTotGames}{2x}{y}}
    \geq 1 + \min \Bigg(
      &\min_{0 \leq j < 3}\left(
        j + \comp{\bracket{M}{\symTotGames}{x}{\frac{u}{2^j}}}
      \right), \notag\\
      &\min_{0 \leq j < 3}\left(
        j + \comp{\bracket{M}{\lfloor \symTotGames(1-\tau)\rfloor+1}{x}{\frac{v}{2^j}}}
      \right)
    \Bigg). \label{eq:partition-three-rung-col}
  \end{align}
\end{lemma}
\begin{proof}
  For the row-partition cascade, define
  \[
    R_j = \comp{\bracket{M}{2\symTotGames+\delta}{2x}{\frac{y}{2^j}}}
    \qquad\text{and}\qquad
    S_j = \comp{\bracket{M}{\symTotGames+\delta}{x}{\frac{y}{2^j}}}
    \qquad\text{for } 0 \leq j \leq 2.
  \]
  The assumption is exactly the bottom-rung nonemptiness input. By
  Lemma~\ref{lem:mono}, it implies
  \[
    R_0 \geq 1,
    \qquad
    R_1 \geq 1,
    \qquad\text{and}\qquad
    R_2 \geq 1.
  \]
  Applying Lemma~\ref{lem:old-partition} with the density parameter set to
  $\frac{y}{2^j}$ gives
  \[
    \begin{aligned}
      R_j \geq 1 + \min\Bigl(
        &\comp{\bracket{M}{2\symTotGames+\delta}{x}{\frac{y}{2^j}}},\\
        &\min_{\substack{\symGamePartSize\in \rangeintegers{\symTotGames}\\a\in [0,1]}}
        \max\bigl(
          \comp{\bracket{M}{\symTotGames+\delta+\symGamePartSize}{x}{\left(\frac{y}{2^j}\right)^a}},
          \comp{\bracket{M}{\symTotGames-\symGamePartSize}{x}{\left(\frac{y}{2^j}\right)^{1-a}}}
        \bigr),\\
        &\comp{\bracket{M}{2\symTotGames+\delta}{2x}{\frac{y}{2^{j+1}}}}
      \Bigr)
    \end{aligned}
  \]
  for $0 \leq j \leq 2$.
  By monotonicity in the first argument,
  \[
    \comp{\bracket{M}{2\symTotGames+\delta}{x}{\frac{y}{2^j}}}
    \geq
    S_j.
  \]
  For every admissible $\symGamePartSize$ and $a$,
  Lemma~\ref{lem:TupleCorollary} sends the first child term to
  \[
    \bracket{M}{\symTotGames+\delta}{x}{
      \left(\frac{y}{2^j}\right)^{
        a\frac{\symTotGames+\delta}{\symTotGames+\delta+\symGamePartSize}
      }
    }.
  \]
  Since $a\leq 1$ and $\symGamePartSize\geq 0$, we have
  \[
    a\frac{\symTotGames+\delta}{\symTotGames+\delta+\symGamePartSize}\leq 1,
  \]
  so another application of Lemma~\ref{lem:mono} gives
  \[
    \comp{\bracket{M}{\symTotGames+\delta+\symGamePartSize}{x}{\left(\frac{y}{2^j}\right)^a}}
    \geq
    S_j.
  \]
  Therefore
  \[
    R_j
    \geq
    1 + \min\Bigl(
      S_j,\,
      \comp{\bracket{M}{2\symTotGames+\delta}{2x}{\frac{y}{2^{j+1}}}}
    \Bigr)
    \qquad\text{for } 0 \leq j \leq 2.
  \]
  Applying Lemma~\ref{lem:TupleCorollary} with
  $\symGamePartSize = \symTotGames+\delta$ gives
  \[
    \comp{\bracket{M}{2\symTotGames+\delta}{2x}{\frac{y}{8}}}
    \geq
    \comp{\bracket{M}{\symTotGames+\delta}{2x}{\left(\frac{y}{8}\right)^{\frac{\symTotGames+\delta}{2\symTotGames+\delta}}}}.
  \]
  Since $\frac{\symTotGames+\delta}{2\symTotGames+\delta} \leq \frac{2}{3}$,
  we have
  \[
    \left(\frac{y}{8}\right)^{\frac{\symTotGames+\delta}{2\symTotGames+\delta}}
    \geq \frac{y}{4}.
  \]
  Therefore another application of Lemma~\ref{lem:mono} gives
  \[
    \comp{\bracket{M}{2\symTotGames+\delta}{2x}{\frac{y}{8}}} \geq S_2.
  \]
  Working upwards, we obtain
  \[
    R_2 \geq 1 + S_2,
    \qquad
    R_1 \geq 1 + \min\left(S_1,\, 1 + S_2\right),
  \]
  and
  \[
    R_0 \geq 1 + \min\left(S_0,\, 1 + S_1,\, 2 + S_2\right),
  \]
  which is exactly \eqref{eq:partition-three-rung-row}.

  For the column-partition cascade, define
  \[
    \begin{aligned}
      T_j &= \comp{\bracket{M}{2\symTotGames}{2x}{\frac{y}{2^j}}},\\
      U_j &= \comp{\bracket{M}{\symTotGames}{x}{\frac{u}{2^j}}},\\
      V_j &= \comp{\bracket{M}{\lfloor \symTotGames(1-\tau)\rfloor+1}{x}{\frac{v}{2^j}}},
    \end{aligned}
    \qquad 0 \leq j \leq 2.
  \]
  The same bottom-rung assumption gives
  \[
    T_0 \geq 1,
    \qquad
    T_1 \geq 1,
    \qquad\text{and}\qquad
    T_2 \geq 1
  \]
  by Lemma~\ref{lem:mono}.
  Applying Lemma~\ref{lem:partition-recurrence} with the density parameter set
  to $\frac{y}{2^j}$ gives
  \[
    T_j \geq 1 + \min\left(
      \comp{\bracket{M}{\symTotGames}{x}{\left(\frac{y}{2^j}\right)^{\frac{1}{1+\tau}}}},
      \comp{\bracket{M}{\lfloor \symTotGames(1-\tau)\rfloor+1}{x}{\left(\frac{y}{2^j}\right)^{\frac{\tau}{1+\tau}}}},
      T_{j+1}
    \right).
  \]
  Since
  \[
    \left(\frac{y}{2^j}\right)^{\frac{1}{1+\tau}} \geq \frac{u}{2^j}
    \qquad\text{and}\qquad
    \left(\frac{y}{2^j}\right)^{\frac{\tau}{1+\tau}} \geq \frac{v}{2^j},
  \]
  Lemma~\ref{lem:mono} gives
  \[
    T_j \geq 1 + \min\left(U_j,\, V_j,\, T_{j+1}\right)
    \qquad\text{for } 0 \leq j \leq 2.
  \]
  Applying Lemma~\ref{lem:TupleCorollary} with
  $\symGamePartSize = \symTotGames$ yields
  \[
    T_3 = \comp{\bracket{M}{2\symTotGames}{2x}{\frac{y}{8}}}
    \geq \comp{\bracket{M}{\symTotGames}{2x}{\sqrt{\frac{y}{8}}}}.
  \]
  Because $0 < y \leq 1$ and $0 < \tau \leq 1$, we have
  \[
    \frac{u}{4} = \frac{y^{\frac{1}{1+\tau}}}{4}
    \leq \frac{\sqrt{y}}{4}
    \leq \sqrt{\frac{y}{8}},
  \]
  and also $\frac{u}{4} \leq \frac{v}{4}$. Hence
  \[
    T_3 \geq U_2 \geq \min(U_2,V_2)
  \]
  by Lemma~\ref{lem:mono}. Working upwards again, we obtain
  \[
    T_2 \geq 1 + \min(U_2,V_2),
  \]
  \[
    T_1 \geq 1 + \min\left(U_1,\, V_1,\, 1 + \min(U_2,V_2)\right),
  \]
  and
  \[
    T_0 \geq 1 + \min\left(U_0,\, V_0,\, 1 + \min(U_1,V_1),\, 2 + \min(U_2,V_2)\right),
  \]
  which is equivalent to \eqref{eq:partition-three-rung-col}.
\end{proof}

For later theorem-level bookkeeping, it is convenient to package these three
densities into the quantity
\[
  \Lambda_M(p,x,y)
  :=
  \min_{0 \leq j < 3}
  \left(
    j + \comp{\bracket{M}{p}{x}{\frac{y}{2^j}}}
  \right).
\]
By Lemma~\ref{lem:mono}, this quantity is monotone in the same direction: if
$1 \leq p' \leq p$, $0 < x' \leq x \leq 1$, and $0 < y' \leq y \leq 1$, then
\[
  \Lambda_M(p',x',y') \leq \Lambda_M(p,x,y).
\]

\begin{lemma}[Row ladder step]\label{lem:lambda-row-step}
  Let $M$ be a matrix, let $\symTotGames \in \mathbb{N}$ with
  $\symTotGames \geq 1$, let $\delta \in \{0,1\}$, and let
  $0 < x \leq \frac12$ and $0 < y \leq 1$. Assume that
  \[
    \comp{\bracket{M}{2\symTotGames}{2x}{\frac{y}{4}}} \geq 1.
  \]
  Then
  \[
    \Lambda_M(2\symTotGames+\delta,2x,y)
    \geq
    1 + \Lambda_M(\symTotGames+\delta,x,y).
  \]
\end{lemma}
\begin{proof}
  In the notation of the proof of Lemma~\ref{lem:partition}, write
  \[
    R_j = \comp{\bracket{M}{2\symTotGames+\delta}{2x}{\frac{y}{2^j}}}
    \qquad\text{and}\qquad
    S_j = \comp{\bracket{M}{\symTotGames+\delta}{x}{\frac{y}{2^j}}}
  \]
  for $0 \leq j \leq 2$. That proof established
  \[
    R_2 \geq 1 + S_2,
    \qquad
    R_1 \geq 1 + \min\left(S_1,\,1 + S_2\right),
  \]
  and
  \[
    R_0 \geq 1 + \min\left(S_0,\,1 + S_1,\,2 + S_2\right).
  \]
  Therefore
  \begin{align*}
    \Lambda_M(2\symTotGames+\delta,2x,y)
    &= \min\left(R_0,\,1 + R_1,\,2 + R_2\right) \\
    &\geq 1 + \min\left(S_0,\,1 + S_1,\,2 + S_2\right) \\
    &= 1 + \Lambda_M(\symTotGames+\delta,x,y),
  \end{align*}
  which is the claim.
\end{proof}

\begin{lemma}[Column ladder step]\label{lem:lambda-col-step}
  Let $M$ be a matrix, let $\symTotGames \in \mathbb{N}$ with
  $\symTotGames \geq 1$, let $0 < x \leq \frac12$, let $0 < y \leq 1$, and
  let $0 < \tau \leq 1$. Write
  \[
    u = y^{\frac{1}{1+\tau}}
    \qquad\text{and}\qquad
    v = y^{\frac{\tau}{1+\tau}}.
  \]
  Assume that
  \[
    \comp{\bracket{M}{2\symTotGames}{2x}{\frac{y}{4}}} \geq 1.
  \]
  Then
  \[
    \Lambda_M(2\symTotGames,2x,y)
    \geq
    1 + \min\Bigl(
      \Lambda_M(\symTotGames,x,u),
      \Lambda_M(\lfloor \symTotGames(1-\tau)\rfloor+1,x,v)
    \Bigr).
  \]
\end{lemma}
\begin{proof}
  In the notation of the proof of Lemma~\ref{lem:partition}, write
  \[
    T_j = \comp{\bracket{M}{2\symTotGames}{2x}{\frac{y}{2^j}}},
    \qquad
    U_j = \comp{\bracket{M}{\symTotGames}{x}{\frac{u}{2^j}}},
    \qquad
    V_j = \comp{\bracket{M}{\lfloor \symTotGames(1-\tau)\rfloor+1}{x}{\frac{v}{2^j}}}
  \]
  for $0 \leq j \leq 2$. That proof established
  \[
    T_j \geq 1 + \min\left(U_j,\,V_j,\,T_{j+1}\right)
    \qquad\text{for } 0 \leq j \leq 2,
  \]
  and
  \[
    T_3 \geq \min(U_2,V_2).
  \]
  Consequently,
  \begin{align*}
    T_2 &\geq 1 + \min(U_2,V_2), \\
    T_1 &\geq 1 + \min\left(U_1,\,V_1,\,1 + \min(U_2,V_2)\right), \\
    T_0 &\geq 1 + \min\left(U_0,\,V_0,\,1 + \min(U_1,V_1),\,2 + \min(U_2,V_2)\right).
  \end{align*}
  Therefore
  \begin{align*}
    \Lambda_M(2\symTotGames,2x,y)
    &= \min\left(T_0,\,1 + T_1,\,2 + T_2\right) \\
    &\geq 1 + \min\Bigl(
      \min(U_0,\,1 + U_1,\,2 + U_2),
      \min(V_0,\,1 + V_1,\,2 + V_2)
    \Bigr) \\
    &= 1 + \min\Bigl(
      \Lambda_M(\symTotGames,x,u),
      \Lambda_M(\lfloor \symTotGames(1-\tau)\rfloor+1,x,v)
    \Bigr),
  \end{align*}
  which is the claim.
\end{proof}

\subsection{Iterated Partition Lemma}\label{app:iterated-partition}

\begin{proof}[Proof of Lemma~\ref{lem:new-partition}]
  Write
  \[
    \tau:=\frac{1}{\rho-1},
    \qquad
    \Lambda_0:=\Lambda_M(p,x,y),
    \qquad
    \beta=\frac{1}{1-\tau}=\frac{\rho-1}{\rho-2}.
  \]
  For integers $t,r$ with
  \[
    0\le t\le s,
    \qquad
    0\le r\le k-s+t,
  \]
  we call $(t,r)$ an \emph{admissible pair}. The index $r$ measures row depth:
  after $r$ row-doubling steps, the row parameter has grown from $x$ to
  $2^r x$. The index $t$ measures the current density-amplification stage.

  For each admissible pair $(t,r)$, define
  \[
    P_{t,r}:=\left\lfloor 2^r\beta^t p \right\rfloor.
  \]
  This is the copy count that we target at that grid point.

  Next define the column exponents. At stage $0$, there has been no density
  amplification yet, so we set
  \[
    E_{0,r}:=1
    \qquad\text{for } 0\le r\le k-s.
  \]
  For stages $1\le t\le s$, define
  \[
    E_{t,r}
    :=
    \rho^t\left(\frac{\rho-1}{\rho}\right)^{k-r-(s-t)}
    \qquad\text{for } 0\le r\le k-s+t.
  \]
  Finally, for each admissible pair $(t,r)$, write
  \[
    F_{t,r}
    :=
    \Lambda_M(P_{t,r},2^r x,y^{E_{t,r}}).
  \]
  The endpoint of the theorem is the terminal admissible pair $(s,k)$:
  \[
    F_{s,k}
    =
    \Lambda_M\!\left(
      \left\lfloor 2^k\beta^s p \right\rfloor,
      2^k x,
      y^{\rho^s}
    \right).
  \]
  Thus the desired inequality \eqref{eq:iterated-partition-bundled} will
  follow once we prove that
  \[
    F_{t,r}\ge r+\Lambda_0
    \qquad\text{for every admissible pair }(t,r).
  \]

  \paragraph{Bridge inequalities.}
  The induction moves across the grid by two elementary floor inequalities. If
  $(t,r)$ is admissible and $r<k-s+t$, then
  \[
    P_{t,r+1}
    =
    \left\lfloor 2^{r+1}\beta^t p\right\rfloor
    \geq
    2\left\lfloor 2^r\beta^t p\right\rfloor
    =
    2P_{t,r}.
  \]
  This is the inequality needed for the row ladder step.

  If $1\le t\le s$, then, since $1-\tau=1/\beta$,
  \[
    \lfloor P_{t,r}(1-\tau)\rfloor+1
    =
    \left\lfloor \frac{P_{t,r}}{\beta}\right\rfloor+1.
  \]
  Also $P_{t,r}\ge 2^r\beta^t p-1$, so
  \[
    \frac{P_{t,r}}{\beta}
    \ge
    2^r\beta^{t-1}p-\frac1\beta.
  \]
  Because $0<1/\beta<1$ and $\lfloor u-c\rfloor+1\ge \lfloor u\rfloor$ for
  every real $u$ and every $0<c<1$, we obtain
  \[
    \lfloor P_{t,r}(1-\tau)\rfloor+1
    \geq
    \lfloor 2^r\beta^{t-1}p\rfloor
    =
    P_{t-1,r}.
  \]
  This is the inequality needed for the column ladder step.

  \paragraph{Side-condition propagation.}
  We also need to know that the weak side condition at the seed point
  propagates across the whole grid. For $1\le t\le s$,
  \[
    E_{t,0}
    =
    \rho^t\left(\frac{\rho-1}{\rho}\right)^{k-s+t}
    =
    \frac{(\rho-1)^{k-s+t}}{\rho^{k-s}}
    \leq
    \frac{(\rho-1)^k}{\rho^{k-s}}
    \leq
    1,
  \]
  because $t\le s$ and $(\rho-1)^k\le \rho^{k-s}$ by hypothesis. Also,
  \[
    E_{t,r+1}
    =
    \frac{\rho}{\rho-1}E_{t,r}
    <
    2E_{t,r}
    \qquad (1\le t\le s),
  \]
  since $\rho>2$. Therefore $E_{t,r}\le 2^r$ for every admissible pair
  $(t,r)$; for stage $0$ this is immediate from $E_{0,r}=1$.

  Since $\beta>1$, we have $P_{t,r}\ge 2^r p$, hence
  \[
    \frac{E_{t,r+1}p}{2P_{t,r}} \leq 1
    \qquad\text{whenever } r<k-s+t.
  \]
  Applying Lemma~\ref{lem:TupleCorollary} with projection parameter $p$ gives
  \[
    \comp{\bracket{M}{2P_{t,r}}{2^{r+1}x}{y^{E_{t,r+1}}/4}}
    \geq
    \comp{\bracket{M}{p}{2^{r+1}x}{(y^{E_{t,r+1}}/4)^{p/(2P_{t,r})}}}.
  \]
  Because $0<y\le 1$, $\frac{E_{t,r+1}p}{2P_{t,r}}\le 1$, and
  $\frac{p}{2P_{t,r}}\le 1$, we have
  \[
    (y^{E_{t,r+1}}/4)^{p/(2P_{t,r})}
    =
    y^{E_{t,r+1}p/(2P_{t,r})}4^{-p/(2P_{t,r})}
    \geq
    y\cdot 4^{-1}
    =
    \frac{y}{4}.
  \]
  Since $x\le 2^{r+1}x$, monotonicity and the hypothesis
  $\comp{\bracket{M}{p}{x}{y/4}}\ge 1$ imply
  \begin{equation}\label{eq:grid-side-condition}
    \comp{\bracket{M}{2P_{t,r}}{2^{r+1}x}{y^{E_{t,r+1}}/4}} \geq 1
    \qquad\text{for every admissible }(t,r)\text{ with }r<k-s+t.
  \end{equation}

  We now prove by induction on $r$ that
  \[
    F_{t,r}\ge r+\Lambda_0
    \qquad\text{for every admissible pair }(t,r).
  \]

  \paragraph{Base case $r=0$.}
  At depth $0$ there is no row amplification yet. For $t=0$, this is exactly
  $F_{0,0}=\Lambda_0$. For $1\le t\le s$, we have $P_{t,0}\ge p$ and
  $E_{t,0}\le 1$, so $y^{E_{t,0}}\ge y$. Monotonicity therefore gives
  \[
    F_{t,0}
    =
    \Lambda_M(P_{t,0},x,y^{E_{t,0}})
    \geq
    \Lambda_M(p,x,y)
    =
    \Lambda_0.
  \]

  \paragraph{Inductive step.}
  Fix $0\le r<k$ and assume that $F_{t,r}\ge r+\Lambda_0$ for every stage $t$
  with $r\le k-s+t$. Let $t$ satisfy $r+1\le k-s+t$. Since $x\le 2^{-k}$ and
  $r<k$, we have
  \[
    2^r x \le 2^{r-k} \le \frac12,
  \]
  so both ladder lemmas are applicable with local row density $2^r x$.

  \paragraph{Case $t=0$.}
  We are still in the pure row-amplification strip. Using the bridge inequality
  $P_{0,r+1}\ge 2P_{0,r}$, monotonicity in the first parameter,
  \eqref{eq:grid-side-condition}, and Lemma~\ref{lem:lambda-row-step}, we get
  \[
    F_{0,r+1}
    =
    \Lambda_M(P_{0,r+1},2^{r+1}x,y)
    \geq
    \Lambda_M(2P_{0,r},2^{r+1}x,y)
    \geq
    1+\Lambda_M(P_{0,r},2^r x,y)
    =
    1+F_{0,r}.
  \]
  Hence $F_{0,r+1}\ge (r+1)+\Lambda_0$.

  \paragraph{Cases $1\le t\le s$.}
  Here one application of the column ladder step either stays on stage $t$ or
  decreases to stage $t-1$. Since
  \[
    E_{t,r+1}\frac{\rho-1}{\rho}=E_{t,r},
  \]
  Lemma~\ref{lem:lambda-col-step} with parameter $\tau=1/(\rho-1)$ and side
  condition \eqref{eq:grid-side-condition} gives
  \begin{align*}
    \Lambda_M(2P_{t,r},2^{r+1}x,y^{E_{t,r+1}})
    \geq
    1+\min\Bigl(
      &\Lambda_M(P_{t,r},2^r x,y^{E_{t,r}}),\\
      &\Lambda_M\!\left(
        \lfloor P_{t,r}(1-\tau)\rfloor+1,
        2^r x,
        y^{E_{t,r+1}/\rho}
      \right)
    \Bigr).
  \end{align*}
  Since $P_{t,r+1}\ge 2P_{t,r}$, monotonicity in the first parameter gives
  $F_{t,r+1}\ge$ the left-hand side.

  \paragraph{Case $t=1$.}
  The second branch reduces to stage $0$. Indeed,
  \[
    \frac{E_{1,r+1}}{\rho}
    =
    \left(\frac{\rho-1}{\rho}\right)^{k-r-s}
    \leq
    1,
  \]
  and the second bridge inequality gives
  $\lfloor P_{1,r}(1-\tau)\rfloor+1\ge P_{0,r}$. Hence monotonicity yields
  \[
    \Lambda_M\!\left(
      \lfloor P_{1,r}(1-\tau)\rfloor+1,
      2^r x,
      y^{E_{1,r+1}/\rho}
    \right)
    \geq
    \Lambda_M(P_{0,r},2^r x,y)
    =
    F_{0,r}.
  \]
  Therefore
  \[
    F_{1,r+1}\ge 1+\min(F_{1,r},F_{0,r}).
  \]

  \paragraph{Case $2\le t\le s$.}
  The second branch decreases by exactly one stage, because
  \[
    \frac{E_{t,r+1}}{\rho}=E_{t-1,r}.
  \]
  Using again $\lfloor P_{t,r}(1-\tau)\rfloor+1\ge P_{t-1,r}$, monotonicity
  gives
  \[
    \Lambda_M\!\left(
      \lfloor P_{t,r}(1-\tau)\rfloor+1,
      2^r x,
      y^{E_{t,r+1}/\rho}
    \right)
    \geq
    \Lambda_M(P_{t-1,r},2^r x,y^{E_{t-1,r}})
    =
    F_{t-1,r}.
  \]
  Therefore
  \[
    F_{t,r+1}\ge 1+\min(F_{t,r},F_{t-1,r}).
  \]

  In all three cases, the induction hypothesis gives both terms inside the
  minimum at least $r+\Lambda_0$, so
  \[
    F_{t,r+1}\ge (r+1)+\Lambda_0.
  \]
  This completes the induction.

  Taking $(t,r)=(s,k)$ yields
  \[
    \Lambda_M\!\left(
      \lfloor 2^k\beta^s p\rfloor,
      2^k x,
      y^{\rho^s}
    \right)
    =
    F_{s,k}
    \geq
    k+\Lambda_0,
  \]
  which is \eqref{eq:iterated-partition-bundled}. Finally, the minimum
  defining $\Lambda_M$ contains the $j=0$ term, so
  \[
    \comp{\bracket{M}{\lfloor 2^k\beta^s p\rfloor}{2^k x}{y^{\rho^s}}}
    \geq
    \Lambda_M\!\left(
      \lfloor 2^k\beta^s p\rfloor,
      2^k x,
      y^{\rho^s}
    \right).
  \]
  Combining this with \eqref{eq:iterated-partition-bundled} gives
  \eqref{eq:iterated-partition-scalar}.
\end{proof}

\printbibliography

\end{document}